\documentclass[journal,onecolumn]{IEEEtran}
\makeatletter
\long\def\@makecaption#1#2{\ifx\@captype\@IEEEtablestring%
\footnotesize\begin{center}{\normalfont\footnotesize #1}\\
{\normalfont\footnotesize\scshape #2}\end{center}%
\@IEEEtablecaptionsepspace
\else
\@IEEEfigurecaptionsepspace
\setbox\@tempboxa\hbox{\normalfont\footnotesize {#1.}~~ #2}%
\ifdim \wd\@tempboxa >\hsize%
\setbox\@tempboxa\hbox{\normalfont\footnotesize {#1.}~~ }%
\parbox[t]{\hsize}{\normalfont\footnotesize \noindent\unhbox\@tempboxa#2}%
\else
\hbox to\hsize{\normalfont\footnotesize\hfil\box\@tempboxa\hfil}\fi\fi}
\makeatother

\usepackage{cite}

\ifCLASSINFOpdf
\else
\fi
\usepackage{tikz}
\usetikzlibrary{calc,arrows.meta,decorations.pathreplacing}

\usepackage{amsmath}
\usepackage{amsfonts}
\usepackage{amssymb}
\usepackage{amsthm}
\usepackage{bm}
\usepackage{mathtools}
\usepackage{dsfont}
\usepackage{thmtools}

\usepackage{algorithm}
\usepackage{algorithmic}

\usepackage{array}
\usepackage{url}
\usepackage[bookmarks=false]{hyperref} 
\usepackage{cleveref} 

\usepackage{xcolor}
\usepackage{units}
\usepackage{booktabs}

\newtheorem{theorem}{Theorem}
\crefname{theorem}{Theorem}{Theorem}
\newtheorem{lemma}{Lemma}
\crefname{lemma}{Lemma}{Lemmas}
\newtheorem{example}{Example}
\declaretheorem[name=Proposition]{proposition}
\crefname{proposition}{Proposition}{Proposition}
\newtheorem{definition}{Definition}
\newtheorem{construction}{Construction}
\crefname{construction}{Construction}{Construction}
\declaretheorem[name=Corollary]{corollary}
\crefname{corollary}{Corollary}{Corollary}
\newcommand{\Romannum}[1]{\MakeUppercase{\romannumeral #1}}
\DeclareMathOperator*{\argmax}{arg\,max}

\newcommand{\bcom}[1]{{\color{blue}#1}}

\newcommand{\indsize}{\scriptsize}
\newcommand{\colind}[2]{\displaystyle\smash{\mathop{#1}^{\raisebox{.5\normalbaselineskip}{\indsize #2}}}}
\newcommand{\rowind}[1]{\mbox{\indsize #1}}

\begin{document}

\title{Coding for Ordered Composite DNA Sequences}

%

\author{Besart~Dollma, Ohad~Elishco, and~Eitan~Yaakobi
\thanks{The research was funded by the European Union (ERC, DNAStorage, 101045114 and EIC, DiDAX 101115134). Views and opinions expressed are however those of the authors only and do not necessarily reflect those of the European Union or the European Research Council Executive Agency. Neither the European Union nor the granting authority can be held responsible for them. This research was funded in part by the Israel Science Foundation (ISF) under Grant Number 1789/23.}
\thanks{An earlier version of this paper was presented in part at the 2025 IEEE International Symposium on Information Theory (ISIT).
This work has been submitted to the IEEE for possible publication. Copyright may be transferred without notice, after which this version may no longer be accessible.}
\thanks{Besart Dollma and Eitan Yaakobi are with the Department of Computer Science, Technion - Israel Institute of Technology, Israel (e-mail: \{besartdollma, yaakobi\}@cs.technion.ac.il). Ohad Elishco is with the School of Electrical and Computer Engineering, Ben-Gurion University of the Negev, Israel (email: ohadeli@bgu.ac.il)}
}

%
%

\markboth{Coding for Ordered Composite DNA Sequences}%
{Dollma \MakeLowercase{\textit{et al.}}: Coding for Ordered Composite DNA Sequences}
%



\maketitle

\begin{abstract}
  To increase the information capacity of DNA storage, composite DNA letters were introduced.
  We propose a novel channel model for composite DNA in which composite sequences are decomposed into ordered standard non-composite sequences.
  The model is designed to handle any alphabet size and composite resolution parameter. 
  We study the problem of reconstructing composite sequences of arbitrary resolution over the binary alphabet under substitution errors.
  We define two families of error-correcting codes and provide lower and upper bounds on their cardinality.
  In addition, we analyze the case in which a single deletion error occurs in the channel and present a systematic code construction for this setting.
  Finally, we briefly discuss the channel's capacity, which remains an open problem.
\end{abstract}

\begin{IEEEkeywords}
DNA storage, composite DNA, error-correcting codes, substitution errors, deletion errors, channel capacity.
\end{IEEEkeywords}

%
\IEEEpeerreviewmaketitle


\section{Introduction}
%
%
%
%

\IEEEPARstart{T}{he} annual demand for digital data storage is expected to surpass the supply of silicon in 2040, assuming
that all data are stored in flash memory for instant access \cite{Zhirnov}. Considering the exponential growth in
the creation of digital data, the development of an alternative storage system is essential.
The idea of using DNA molecules as a volume for storing data was first introduced in the late 1950s by Richard Feynman
in his lecture “There's plenty of room at the bottom”.

Due to its high information density, long-term stability, and robustness, DNA is a promising alternative to
serve as a digital media storage system. Several studies have demonstrated the use of synthetic DNA
for storing digital information on a megabyte scale, exceeding the physical density of current magnetic-tape based systems
by roughly six orders of magnitude \cite{Church, Goldman}.
The process of storing data in DNA begins with DNA synthesis, where synthetic DNA sequences encoding the digital information are generated.
Current synthesis technologies produce millions of copies of the same DNA sequence in parallel and place them in a storage container \cite{LeProust}.
The data is retrieved through DNA sequencing, in which numerous identical copies of the DNA sequences are read and the original information is decoded \cite{Mcginn}.

The next step towards the practical use of DNA-based data storage is to reduce the cost of storing the data.
The total cost of DNA-based data storage is categorized into the cost of data writing through DNA synthesis and the cost of
data reading through DNA sequencing. Prior work shows that DNA becomes viable for archival storage only if the cost of data writing becomes approximately 100 times less \cite{Goldman}.
Traditional encoding schemes for DNA data storage are limited to $\log_2{4}$ bits per character, reflecting the four DNA bases (A, C, T, G).
Introducing additional encoding characters can increase the information capacity logarithmically, reducing overall storage costs.
A novel approach called \emph{composite DNA letters} introduced in \cite{Anavy, Choi} achieves this by extending the encoding alphabet beyond the standard four DNA bases.
It leverages an inherent property of DNA synthesis, the production in parallel of numerous copies of the DNA sequence encoding the digital information.

A composite DNA letter is a mixture of all four standard DNA bases in a specified pre-determined ratio $\phi = (p_A, p_C, p_T, p_G)$ where $p_A + p_C + p_T + p_G = 1$.
For example, $\left(\nicefrac{1}{2}, 0, \nicefrac{1}{2}, 0\right)$ represents a composite DNA letter in which there is a chance of 50\%, 0\%, 50\% and 0\% of seeing A, C, T and G, respectively. 
A composite DNA letter is said to have \emph{resolution} $k \in \mathbb{N}$ if
$\phi = (\frac{k_A}{k}, \frac{k_C}{k}, \frac{k_T}{k}, \frac{k_G}{k})$ for $k_A, k_C, k_T, k_G \in \mathbb{N}$ and $k_A + k_C + k_T + k_G = k$.
A sequence composed of composite letters is called a composite sequence. If the composite letters have resolution $k$, the sequence is referred to as a $k$-resolution composite sequence.

Composite DNA introduces new coding and algorithmic challenges. Zhang et al. \cite{Zhang} were the first to explore error-correcting codes for composite DNA.
In their study, they propose code constructions for cases in which both the number of errors and the error magnitudes are bounded.
Walter et al. \cite{Walter} examined another model of composite synthesis, focusing on substitutions, strand losses, and deletions. 
Preuss et al. \cite{Preuss} further expanded on the concept of larger alphabets by introducing combinatorial composite synthesis.
This approach employs combinatorial DNA encoding, which utilizes a set of easily distinguishable DNA shortmers (fixed-length sequences) to construct large combinatorial alphabets, where each letter is represented by a subset of shortmers.
Sabary et al. \cite{Sabary} examined scenarios in which one or more shortmers are missing from sequencing reads, modeling these cases as asymmetric errors.
Preuss et al. \cite{Preuss2} analyzed the sequencing coverage depth problem for combinatorial DNA encoding by modeling the reconstruction of a single combinatorial letter as a variant of the coupon collector's problem.
Sokolovskii et al. \cite{Sokolovskii} studied the capacity of the combinatorial composite DNA channel and proposed error-correcting codes for this channel.
The authors of \cite{Cohen} gave the expected number of reads required to reconstruct information for composite DNA.
Kobovich et al. \cite{Kobovich} studied how to choose the probabilities of the composite letters to maximize the composite DNA channel capacity.

In the composite DNA channel, any of the potential standard (non-composite) DNA sequences derivable from the composite sequence could serve as channel input.
The number of such sequences grows exponentially with sequence length, creating uncertainty that necessitates performing many sequencing reads to accurately reconstruct the original composite sequence.
This inherent ambiguity also complicates the design of error-correcting codes tailored to the channel.

In this paper, we introduce the \emph{ordered composite DNA channel}, a new channel model for DNA data storage based on composite DNA letters.
In this model, a $k$-resolution composite sequence $\bm{s}$ is deterministically decomposed into $k$ ordered standard sequences $\bm{s}_0, \ldots, \bm{s}_{k-1}$.
Each standard sequence is then transmitted through an independent noisy channel subject to substitution or deletion errors.
The ordered composite DNA channel model assumes that when synthesizing a composite DNA letter $\phi = (\frac{k_A}{k}, \frac{k_C}{k}, \frac{k_T}{k}, \frac{k_G}{k})$ of resolution $k$, 
the multiple copies of DNA sequences produced during synthesis can be partitioned into $k$ groups, with the bases distributed in order across the groups, A in the first $k_A$ groups, C in the next $k_C$, T in the following $k_T$, and G in the remaining $k_G$.
This assumption requires the synthesis process to be aware of the partitioning into $k$ groups, 
and the ordering of the bases within each composite letter.
Photolithographic DNA synthesis \cite{Antkowiak, Somoza} can realize this requirement by enabling parallel and independent synthesis of the ordered sequences, 
and other synthesis approaches may potentially achieve the same. 

In contrast to the regular (non-ordered) composite DNA channel, in the new ordered composite DNA channel model, the number of standard DNA sequences that can be synthesized from a $k$-resolution composite sequence is exactly $k$, 
regardless of the sequence length. This approach reduces uncertainty and may lower the number of reads needed for accurate reconstruction.
Furthermore, the deterministic decomposition of composite sequences into ordered standard sequences enables the design of error-correcting codes tailored to this channel model.

The rest of the paper is organized as follows. 
In Section~\ref{sec:preliminaries} we define composite letters and alphabets, introduce the ordered composite DNA channel, and formulate the problem of reconstructing the original composite sequence from the noisy channel outputs.
We then define two families of error-correcting codes for the case where the channels introduce substitution errors, focusing on the binary alphabet, and present some preliminary results.
In Section~\ref{sec:upper-bounds} we derive upper bounds on the cardinality of these codes.
In Section~\ref{sec:lower-bounds} we establish lower bounds on the cardinality of the proposed codes by presenting explicit code constructions or by relating them to known codes.
In Section~\ref{sec:deletions} we extend the model to deletion errors, deriving upper and lower bounds on the code cardinality, restricted to the case of a single deletion. 
We then present systematic code constructions for this setting, addressing both the known and unknown erroneous channel cases.
Finally, in Section~\ref{sec:conclusion} we conclude the paper and outline directions for future research.
There, we discuss the capacity of the ordered composite DNA channel when the underlying channels are binary substitution channels with crossover probability $p$.
We provide initial insights, reduce the problem to a single-variable optimization, and compute the capacity numerically.
A closed-form expression, however, remains unknown and is left for future work.

\section{Problem Formulation and Preliminary Results} \label{sec:preliminaries}

Let $\Sigma_q = \{ 0, 1, \ldots, q-1 \}$ be a finite alphabet. We assume the natural order on $\Sigma_q$.
Denote by $\Sigma_q^\ell$ the set of all sequences of length $\ell$ over $\Sigma_q$. Denote by $\Sigma_q^{m \times n}$ the set of
$m \times n$ matrices whose components are letters in $\Sigma_q$. For a sequence $\bm{s} \in \Sigma_q^\ell$ and $1 \leq i \leq \ell$, 
$\bm{s}[i]$ represents the letter at position $i$ in $\bm{s}$. For a sequence $\bm{s} \in \Sigma_q^\ell$, $\#_\sigma(\bm{s})$ denotes the number of occurrences of the letter 
$\sigma \in \Sigma_q$ in $\bm{s}$, that is, $\#_\sigma(\bm{s}) = |\{ j : \bm{s}[j] = \sigma \}|$.

A \emph{composite letter $\phi$} over $\Sigma_q$ is a mixture of all the letters in $\Sigma_q$ in a specified predefined ratio. It is represented by a vector of probabilities $\phi = (p_0, p_1, \ldots, p_{q-1}) \in [0, 1]^q$ where $\sum_{i=0}^{q-1} p_i = 1$
and is observed as the letter $i \in \Sigma_q$ with probability $p_i$.
For example, $\phi = (\nicefrac{1}{4}, \nicefrac{1}{4}, \nicefrac{1}{2}, 0)$ represents a composite letter over $\Sigma_4$ which is observed as the letters $0, 1, 2, 3$
with probability $\nicefrac{1}{4}$, $\nicefrac{1}{4}$, $\nicefrac{1}{2}$, and $0$, respectively.
A special family of composite letters are the composite letters of \emph{resolution parameter $k \in \mathbb{N}$} over $\Sigma_q$, where
$\phi = (\frac{k_0}{k}, \frac{k_1}{k}, \ldots, \frac{k_{q-1}}{k})$ for $k_i \in \mathbb{N}$ and $\sum_{i=0}^{q-1} k_i = k$.
The \emph{composite alphabet $\Phi_{q, k}$} is the set of all composite letters of resolution parameter $k$ over $\Sigma_q$, i.e.,
\begin{equation*}
  \Phi_{q, k} \triangleq \left\{ \left( \frac{k_0}{k}, \frac{k_1}{k}, \ldots, \frac{k_{q-1}}{k} \right) \ : \ k_i \in \mathbb{N}, \  \sum_{i=0}^{q-1} k_i = k \right\}.
\end{equation*}

A \emph{$k$-resolution composite sequence} $\bm{s}$ of length $\ell$ is a sequence in a composite alphabet $\Phi_{q, k}^\ell$, that is, $\bm{s} \in \Phi_{q, k}^\ell$.
When the resolution parameter $k$ is clear from the context, we refer to $\bm{s}$ as a \emph{composite sequence}.
A \emph{standard sequence} (or simply a sequence) $\bm{s}$ of length $\ell$ is a sequence in a standard non-composite alphabet $\Sigma_q$, that is, $\bm{s} \in \Sigma_q^\ell$.

A \emph{decomposition} is a mapping $\mathcal{D}: \Phi_{q, k} \to \Sigma_q^{k \times 1}$ such that for a composite letter 
$\phi= (\frac{k_0}{k}, \frac{k_1}{k}, \ldots, \frac{k_{q-1}}{k}) \in \Phi_{q, k}$
\begin{equation*}
  \mathcal{D}(\phi) \triangleq \begin{bmatrix} 0^{k_0} \ 1^{k_1} \ \cdots \ (q-1)^{k_{q-1}}\end{bmatrix}^\intercal,
\end{equation*}
where $i^{k_i}$ indicates that the letter $i$ is repeated $k_i$ times. Since $\Sigma_{i=0}^{q-1} k_i = k$, the decomposition is well-defined,
and the output is a column vector of length $k$ whose components are letters in $\Sigma_q$.

A \emph{reconstruction} is a mapping $\mathcal{R}: \Sigma_q^{k \times 1} \to \Phi_{q,k} \cup \{\mathord{?}\}$ defined as the inverse of the decomposition mapping $\mathcal{D}$, that is,
given a column vector $\bm{v}$ of length $k$ whose components are letters in $\Sigma_q$,
\begin{equation*}
\mathcal{R}\left(\bm{v}\right) \triangleq \begin{cases}
  \phi & \text{if } \mathcal{D}(\phi) = \bm{v} \\
  \mathord{?} & \text{otherwise}
\end{cases}.
\end{equation*}
The symbol "$\mathord{?}$" represents that the reconstruction is not possible for the given input to a valid composite letter.

The decomposition mapping can be naturally extended to receive as input a $k$-resolution composite sequence of length $\ell$ and output a $ k \times \ell$ matrix, i.e., 
$ \mathcal{D}: \Phi_{q, k}^\ell \to \Sigma_q^{k \times \ell}$ by applying the mapping to each letter in the sequence separately.
Given a $k$-resolution composite sequence $\bm{s} \in \Phi_{q, k}^\ell$, we \emph{decompose} it into $k$ ordered standard sequences,
$\bm{s}_0, \ldots, \bm{s}_{k-1} \in \Sigma_q^\ell$, such that $\bm{s}_j$ is the $j$-th row of $\mathcal{D}(\bm{s})$.
We write the $k$ rows of the matrix as the tuple of standard sequences $(\bm{s}_0, \ldots, \bm{s}_{k-1})$ and denote this decomposition as $\mathcal{D}(\bm{s}) = (\bm{s}_0, \ldots, \bm{s}_{k-1})$.

Similarly, the reconstruction mapping can be extended to receive as input a $k \times \ell$ matrix and output a $k$-resolution composite sequence of length $\ell$, i.e.,
$\mathcal{R}: \Sigma_q^{k \times \ell} \to (\Phi_{q, k} \cup \{\mathord{?}\})^\ell$, by applying the mapping to each column of the matrix separately.
Given $k$ ordered standard sequences, $\bm{y}_0, \ldots, \bm{y}_{k-1} \in \Sigma_q^\ell$, that represent the rows of a $k \times \ell$ matrix,
we \emph{reconstruct} the $k$-resolution composite sequence $\bm{y} \in (\Phi_{q, k} \cup \{\mathord{?}\})^\ell$ using the extended reconstruction mapping $\mathcal{R}$.
In the same manner, we write the $k$ ordered standard sequences as the tuple $(\bm{y}_0, \ldots, \bm{y}_{k-1})$ and denote this reconstruction as $\mathcal{R}(\bm{y}_0, \ldots, \bm{y}_{k-1}) = \bm{y}$.

\begin{example}\label{ex:basic}
  Let $\Sigma_3 = \left\{0, 1, 2\right\}$ and 
  \begin{equation*}
    \begin{split}
      \Phi_{3, 2} = \left\{\phi_0 = \left( \frac{2}{2}, \frac{0}{2}, \frac{0}{2} \right), \phi_1 = \left( \frac{0}{2}, \frac{2}{2}, \frac{0}{2}\right), \phi_2 = \left( \frac{0}{2}, \frac{0}{2}, \frac{2}{2}\right), 
       \phi_3 = \left( \frac{1}{2}, \frac{1}{2}, \frac{0}{2}\right), \phi_4 = \left( \frac{1}{2}, \frac{0}{2}, \frac{1}{2}\right), \phi_5 = \left( \frac{0}{2}, \frac{1}{2}, \frac{1}{2}\right) \right\}.
    \end{split}
  \end{equation*} 
  Then,
  \begin{equation*}
    \begin{split}
    \mathcal{D}(\phi_0) = \begin{bmatrix}
      0 \\
      0
    \end{bmatrix} \quad \mathcal{D}(\phi_1) = \begin{bmatrix}
      1 \\
      1
    \end{bmatrix} \quad \mathcal{D}(\phi_2) = \begin{bmatrix}
      2 \\
      2
    \end{bmatrix} \quad \mathcal{D}(\phi_3) = \begin{bmatrix}
      0 \\
      1
    \end{bmatrix} \quad \mathcal{D}(\phi_4) = \begin{bmatrix}
      0 \\
      2
    \end{bmatrix} \quad \mathcal{D}(\phi_5) = \begin{bmatrix}
      1 \\
      2
    \end{bmatrix}.
    \end{split}
  \end{equation*}
  The reconstruction mapping is the inverse of the decomposition mapping, i.e., $\mathcal{R}\left(\mathcal{D}(\phi_i)\right) = \phi_i$ for all $i$,
  and for any other input $\mathcal{R}$ outputs $\mathord{?}$, that is,
  \begin{equation*}
    \mathcal{R}\left(\begin{bmatrix}
    1 \\
    0
  \end{bmatrix}\right) = \mathcal{R}\left(\begin{bmatrix}
    2 \\
    0 \end{bmatrix}\right) = \mathcal{R}\left(\begin{bmatrix}
    2 \\
    1
    \end{bmatrix}\right) = \mathord{?}.
  \end{equation*}
  Let $\bm{s} = \phi_0\phi_1\phi_2\phi_3\phi_4\phi_5 \in \Phi_{3, 2}^6$ be a $2$-resolution composite sequence. Then,
  \begin{equation*}
    \mathcal{D}(\bm{s}) = \begin{bmatrix}
      0 & 1 & 2 & 0 & 0 & 1 \\
      0 & 1 & 2 & 1 & 2 & 2
    \end{bmatrix}.
  \end{equation*}
  We decompose $\bm{s}$ into two sequences, $\bm{s}_0 = 012001 \in \Sigma_3^6$ and $\bm{s}_1 = 012122 \in \Sigma_3^6$ which are the rows of the matrix.
  We write the two sequences as the tuple $(\bm{s}_0, \bm{s}_1)$ and denote the decomposition as $\mathcal{D}(\bm{s}) = (\bm{s}_0, \bm{s}_1)$ and the reconstruction as $\mathcal{R}(\bm{s}_0, \bm{s}_1) = \bm{s}$.
\end{example}

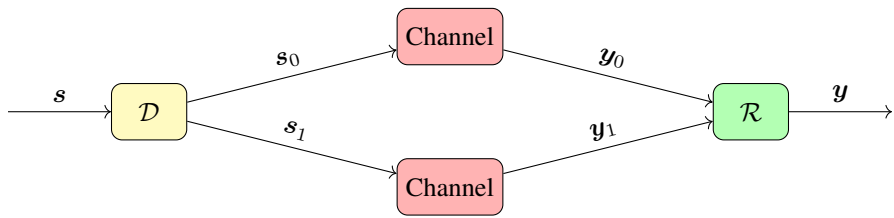
\begin{figure}[htbp]
  \centering
  \begin{tikzpicture}[node distance=2cm]
    \tikzstyle{decompose} = [rectangle, rounded corners, minimum width=1cm, minimum height=0.75cm,text centered, draw=black, fill=yellow!30]
    \tikzstyle{channel} = [rectangle, rounded corners, minimum width=1cm, minimum height=0.75cm,text centered, draw=black, fill=red!30]
    \tikzstyle{reconstruct} = [rectangle, rounded corners, minimum width=1cm, minimum height=0.75cm,text centered, draw=black, fill=green!30]
    \node (S) at (-6, 0) {};
    \node (D) at (-4, 0) [decompose] {$\mathcal{D}$};
    \node (C0) at (0, 1) [channel] {Channel};
    \node (C1) at (0, -1) [channel] {Channel};
    \node (R) at (4, 0) [reconstruct] {$\mathcal{R}$};
    \node (E) at (6, 0) {};

    \draw [->] (S) -- node[above] {$\bm{s}$} (D);
    \draw [->] (D) -- node[above, sloped] {$\bm{s}_0$} (C0);
    \draw [->] (D) -- node[above, sloped] {$\bm{s}_1$} (C1);
    \draw [->] (C0) -- node[above, sloped] {$\bm{y}_0$} (R);
    \draw [->] (C1) -- node[above, sloped] {$\bm{y}_1$} (R);
    \draw [->] (R) -- node[above] {$\bm{y}$} (E);

  \end{tikzpicture}
  \caption{Ordered composite DNA channel for resolution $k=2$.}
  \label{fig:channel}
\end{figure}

We now present the \emph{ordered composite DNA channel}. Let $\bm{s} \in \Phi_{q, k}^\ell$ be a $k$-resolution composite sequence of length $\ell$. 
Let $\bm{s}_0, \bm{s}_1, \ldots, \bm{s}_{k-1} \in \Sigma_q^\ell$ be the ordered decomposed sequences of $\bm{s}$, i.e., $\mathcal{D}(\bm{s}) = (\bm{s}_0, \bm{s}_1, \ldots, \bm{s}_{k-1})$. 
Each of the sequences $\bm{s}_i$ is sent through a separate noisy channel $i$ that may introduce errors.
We denote the received sequence of channel $i$ by $\bm{y}_i \in \Sigma_q^{\ell '}$.
Note that the length $\ell'$ of the received sequence may differ from the original length $\ell$ depending on the type of errors introduced by the channel.
Given the received sequences $\bm{y}_{i}$, and the index of the channel $i$ on which each sequence is received,
we aim to reconstruct the $k$-resolution composite sequence $\bm{y} \in (\Phi_{q, k} \cup \{\mathord{?}\})^\ell$, with the goal of having $\bm{y} = \bm{s}$.
For $k=2$, the model is depicted in Figure \ref{fig:channel}.

As illustrated in Example~\ref{ex:basic}, for a resolution parameter $k$ and composite alphabet $\Phi_{q, k}$, the image of the decomposition mapping $\mathcal{D}$ consists of all non-decreasing column vectors in $\Sigma_q^{k \times 1}$. 
Under this characterization, the ordered composite DNA channel can be viewed as a scheme in which the input is a matrix in $\Sigma_q^{k \times \ell}$, with each column constrained to be non-decreasing. 
Each row of the matrix is transmitted through a separate, independent noisy channel, and the objective is to reconstruct the original matrix from the possibly corrupted rows and their corresponding indices.

For the remainder of the paper, we assume that the noisy channels introduce only substitution errors in all sections except Section~\ref{sec:deletions}, where deletion errors are considered, and we work exclusively in the binary setting where $q = 2$, so the composite and standard alphabets are $\Phi_{2, k}$ and $\Sigma_2$, respectively.
The composite alphabet $\Phi_{2, k}$ definition is then simplified to
\begin{equation*}
  \Phi_{2, k} \triangleq \left\{ \left( \frac{k_0}{k}, \frac{k_1}{k} \right) : k_0 + k_1 = k, \  k_0, k_1 \in \mathbb{N} \right\},
\end{equation*}
and has cardinality $|\Phi_{2, k}| = k + 1$. 
We enumerate the letters of the alphabet $\Phi_{2, k}$ using the notation $\phi_i$, defined for each integer $i \in \left\{0, 1, \ldots, k\right\}$ as
\begin{equation*}
\phi_i \triangleq \left( \frac{k - i}{k}, \frac{i}{k} \right).
\end{equation*}

\begin{example}\label{ex:binary}
  Let $\Sigma_2 = \{0, 1\}$ and $k=4$. Then
  \begin{equation*}
    \Phi_{2, 4} = \left\{ \phi_0 = \left(\frac{4}{4}, \frac{0}{4}\right),  \phi_1 = \left(\frac{3}{4}, \frac{1}{4}\right), \phi_2 = \left(\frac{2}{4}, \frac{2}{4}\right), \phi_3 = \left(\frac{1}{4}, \frac{3}{4}\right), \phi_4 = \left(\frac{0}{4}, \frac{4}{4}\right)\right\}.
  \end{equation*}
  We can view $\Phi_{2, 4}$ as the quinary alphabet $\Sigma_5 = \{0, 1, 2, 3, 4\}$ by mapping $\phi_i \mapsto i$.
  The mappings for decomposition and reconstruction are
  \begin{equation*}
    \begin{split}
      \mathcal{D}(0) & = \begin{bmatrix}
        0 \\
        0 \\ 
        0 \\
        0
      \end{bmatrix} \quad \mathcal{D}(1) = \begin{bmatrix}
        0 \\
        0 \\
        0 \\
        1
      \end{bmatrix} \quad \mathcal{D}(2) = \begin{bmatrix}
        0 \\
        0 \\
        1 \\
        1
      \end{bmatrix} \quad \mathcal{D}(3) = \begin{bmatrix}
        0 \\
        1 \\
        1 \\
        1
      \end{bmatrix} \quad \mathcal{D}(4) = \begin{bmatrix}
        1 \\
        1 \\
        1 \\
        1
      \end{bmatrix},
    \end{split}
  \end{equation*}

  \begin{equation*}
      \mathcal{R}\left(\begin{bmatrix}
        0 \\
        0 \\ 
        0 \\
        0
      \end{bmatrix} \right) = 0 \quad \mathcal{R}\left(\begin{bmatrix}
        0 \\
        0 \\
        0 \\
        1
      \end{bmatrix} \right) = 1 \quad \mathcal{R}\left(\begin{bmatrix}
        0 \\
        0 \\
        1 \\
        1
      \end{bmatrix} \right) = 2 \quad \mathcal{R}\left(\begin{bmatrix}
        0 \\
        1 \\
        1 \\
        1
      \end{bmatrix} \right) = 3 \quad \mathcal{R}\left(\begin{bmatrix}
        1 \\
        1 \\
        1 \\
        1
      \end{bmatrix} \right) = 4,
  \end{equation*}
  and for every other binary column vector $\bm{v} \in \Sigma_2^{4 \times 1}$, we have $\mathcal{R}(\bm{v}) = \mathord{?}$.
  Let $\bm{s}=012340$ be a composite sequence over $\Phi_{2, 4}$, represented as a quinary sequence. Then,
  \begin{equation*}
    \mathcal{D}(\bm{s}) = \begin{bmatrix}
      0 & 0 & 0 & 0 & 1 & 0 \\
      0 & 0 & 0 & 1 & 1 & 0 \\
      0 & 0 & 1 & 1 & 1 & 0 \\
      0 & 1 & 1 & 1 & 1 & 0
    \end{bmatrix}.
  \end{equation*}
  We decompose $\bm{s}$ into four binary sequences, $\bm{s}_0=000010, \bm{s}_1=000110, \bm{s}_2=001110, \bm{s}_3=011110$, which correspond to the rows of the matrix,
  and write $\mathcal{D}(\bm{s}) = \left(\bm{s}_0, \bm{s}_1, \bm{s}_2, \bm{s}_3\right)$.
  We then transmit each sequence $\bm{s}_i, i \in \left\{0, 1, 2, 3\right\}$ through separate independent binary substitution channels.
  Suppose the third channel introduced a substitution error in the second bit of $\bm{s}_2$ and the fourth channel introduced a substitution error in the third bit of $\bm{s}_3$.
  The received sequences then become $\bm{y}_0=000010, \bm{y}_1=000110, \bm{y}_2=011110, \bm{y}_3=010110$, and their reconstruction is
  \begin{equation*}
    \mathcal{R}\left(\begin{bmatrix}
      0 & 0 & 0 & 0 & 1 & 0 \\
      0 & 0 & 0 & 1 & 1 & 0 \\
      0 & 1 & 1 & 1 & 1 & 0 \\
      0 & 1 & 0 & 1 & 1 & 0
    \end{bmatrix} \right) = 02\mathord{?}340.
  \end{equation*}
  We write the received sequences as the tuple $(\bm{y}_0, \bm{y}_1, \bm{y}_2, \bm{y}_3)$ and denote the reconstruction as $\mathcal{R}(\bm{y}_0, \bm{y}_1, \bm{y}_2, \bm{y}_3) = \bm{y}$.
\end{example}

The composite alphabet $\Phi_{2, k}$ can be naturally associated with the alphabet $\Sigma_{k+1}$ via the mapping $\phi_i \mapsto i$, as described in Example~\ref{ex:binary}. 
Under this association, the decomposition mapping becomes $\mathcal{D} : \Sigma_{k+1} \to \Sigma_2^{k \times 1}$, defined by $\mathcal{D}(i) = \begin{bmatrix} 0^{k-i} \ 1^{i} \end{bmatrix}^\intercal$. 
In other words, each letter $i \in \Sigma_{k+1}$ is mapped to a binary column vector of length $k$ consisting of $k - i$ zeros followed by $i$ ones.
To emphasize the binary setting, we refer to a $k$-resolution composite sequence over $\Phi_{2, k}$ as a \emph{$k$-resolution composite binary sequence}. 
When the resolution parameter $k$ is clear from context, we refer to it simply as a \emph{composite binary sequence}.
Following the association of $\Phi_{2, k}$ with $\Sigma_{k+1}$, it is convenient to represent a $k$-resolution composite binary sequence $\bm{s}$ as a sequence over the $(k+1)$-ary alphabet.
Accordingly, we write $\bm{s} \in \Phi_{2, k}^\ell$ as $\bm{s} \in \Sigma_{k+1}^\ell$.


We define two families of error-correcting codes for the ordered composite DNA channel. 
In the first family, each channel is allowed a fixed number of substitution errors, under the assumption that the error budget per channel is known in advance. 
This corresponds to assigning a separate error budget to each channel. 
In the second family, the codes correct a fixed total number of substitution errors, regardless of how the errors are distributed across the channels. 
Here, the error budget is shared collectively among all channels. 

\begin{definition}
  An $(e_0, e_1, \ldots, e_{k-1})$-composite-error-correcting code ($(e_0, e_1, \ldots, e_{k-1})$-CECC) $\mathcal{C}$ is a code that can correct up to $e_i$ substitution errors in $\bm{s}_i$,
  introduced by the $i$-th channel, for each $i \in \left\{0, 1, \ldots, k-1\right\}$.
\end{definition}

\begin{definition}
  A $k$-resolution $e$-composite-error-correcting code ($k$-resolution $e$-CECC) $\mathcal{C}$ is a code that can correct up to $e$ substitution errors in total, introduced collectively by all $k$ channels.
\end{definition}

Let $\mathcal{S}_{k}\left(n; (e_0, e_1, \ldots, e_{k-1})\right)$ denote the largest cardinality of an $(e_0, e_1, \ldots, e_{k-1})$-CECC of length $n$, and let $\mathcal{S}_k\left(n; e\right)$ denote the largest cardinality of a $k$-resolution $e$-CECC of length $n$.
An $(e_0, e_1, \ldots, e_{k-1})$-CECC is called \emph{optimal} if its size equals $\mathcal{S}_{k}\left(n; (e_0, e_1, \ldots, e_{k-1})\right)$.
Similarly, a $k$-resolution $e$-CECC is called \emph{optimal} if its size equals $\mathcal{S}_k\left(n; e\right)$.
We denote by $\mathcal{A}_q(n; e)$ the largest cardinality of a $q$-ary $e$-error-correcting code of length $n$.
Throughout the paper we assume that the number of errors is independent of the length of the sequence, i.e., $e_i, e$ are constants and $ e_i, e \ll n$, for $0 \leq i \leq k-1$.
We now present several immediate propositions that follow directly from the code definitions, with proofs provided in Appendix~\ref{appendix:preliminaries}.


\begin{restatable}{proposition}{kAryEcc}\label{prop:k-ary}
  A $(k+1)$-ary $e$-error-correcting code is also a $k$-resolution $e$-CECC, i.e., $\mathcal{A}_{k+1}(n; e) \leq \mathcal{S}_k(n; e).$
\end{restatable}

However, the converse does not hold: if two channels introduce a substitution error at the same position of the sequence, correcting the errors requires a $k$-resolution 2-CECC, 
even though a $(k+1)$-ary single-error-correcting code would suffice.


\begin{restatable}{proposition}{connection}\label{prop:connection}
  For any $e \in \mathbb{N}^{+}$, a $k$-resolution $e$-CECC is also an $(e_0, e_1, \ldots, e_{k-1})$-CECC for all tuples $(e_0, e_1, \ldots, e_{k-1}) \in \mathbb{N}^k$ satisfying $\sum_{i=0}^{k-1} e_i \leq e$. That is,
  \begin{equation*}
  \mathcal{S}_k\left(n; \sum_{i=0}^{k-1} e_i\right) \leq \mathcal{S}_{k}\left(n; (e_0, e_1, \ldots, e_{k-1})\right).
  \end{equation*}
\end{restatable}


\begin{restatable}{proposition}{reversal}\label{prop:reversal}
  For any tuple $(e_0, e_1, \ldots, e_{k-2}, e_{k-1}) \in \mathbb{N}^k$ and any code length $n$, it holds that
  \begin{equation*}
  \mathcal{S}_{k}\left(n; (e_0, e_1, \ldots, e_{k-2}, e_{k-1})\right) = \mathcal{S}_{k}\left(n; (e_{k-1}, e_{k-2}, \ldots, e_{1}, e_0)\right).
  \end{equation*}
\end{restatable}

A natural question is whether this proposition extends to arbitrary permutations of the tuple $(e_0, e_1, \ldots, e_{k-1})$, specifically, whether
\begin{equation*}
\mathcal{S}_{k}\left(n; (e_0, e_1, \ldots, e_{k-1})\right) \overset{?}{=} \mathcal{S}_{k}\left(n; (e_{\pi(0)}, e_{\pi(1)}, \ldots, e_{\pi(k-1)})\right)
\end{equation*} holds for any permutation $\pi$ of the indices?
At present, the general case remains open, and there is no clear reason to expect the equality to hold in full generality.
However, the following proposition shows that in the case of a single error, the equality does hold.


\begin{restatable}{proposition}{swapSingleError}\label{prop:swap-single-error}
  Let $\bm{e}_i = (0, \ldots ,0, 1, 0, \ldots, 0)$ be the $i$-th unit vector in $\mathbb{N}^k$, where the $1$ is in the $i$-th position.
  Then for any code length $n$ and any $0 \leq i, j \leq k-1$ 
  \begin{equation*}
    \mathcal{S}_k(n; \bm{e}_i) = \mathcal{S}_k(n; \bm{e}_j).
  \end{equation*}
\end{restatable}

Proposition~\ref{prop:swap-single-error} allows us to reduce the analysis to the case of $\bm{e}_0 = (1, 0, \ldots, 0)$-CECCs, corresponding to a single substitution error in the first channel,
without having to consider each individual channel separately.

\section{Upper Bounds} \label{sec:upper-bounds}

In this section, we derive upper bounds on the cardinality of the proposed code families using sphere packing arguments.
For each family, we define a composite error ball centered at a $k$-resolution composite binary sequence, consisting of all valid $k$-resolution sequences obtainable under that family's substitution error constraints.
The main challenge is that these composite error balls are non-uniform in size.
We first consider an arbitrary number of errors with resolution restricted to $k=2$, which allows us to compute the minimum ball size and apply the sphere packing bound.
We then refine this bound through an asymptotic analysis following the approach of Levenshtein \cite{Levenshtein}.
Next, we address a limited number of errors by applying the generalized sphere packing bound (GSPB) \cite{gspb} to derive improved non-asymptotic bounds for a single error with arbitrary resolution $k$ and for two errors with resolution $k=2$.

For any two binary sequences $\bm{x}, \bm{y} \in \{0,1\}^n$, let $d(\bm{x}, \bm{y})$ denote their Hamming distance, and define $\mathcal{X}_{k}^{n} \triangleq \Sigma_{k+1}^n$ as the set of all $k$-resolution composite binary sequences of length $n$.  
Given $\bm{s} \in \mathcal{X}_{k}^{n}$ with decomposition $\mathcal{D}(\bm{s}) = (\bm{s}_0, \dots, \bm{s}_{k-1})$, we define two types of composite error balls centered at $\bm{s}$, corresponding to the two families of composite-error-correcting codes.  
The first, 
\begin{equation*}
\mathcal{B}_{k, (e_0, e_1, \ldots, e_{k-1})}(\bm{s}) \triangleq \left\{ \mathcal{R}(\bm{y}_0, \bm{y}_1, \ldots, \bm{y}_{k-1}) \cap \mathcal{X}_{k}^{n} \ : \ \bm{y}_i \in \{0,1\}^n, \ d(\bm{s}_i, \bm{y}_i) \leq e_i, \ 0 \leq i \leq k-1 \right\},
\end{equation*}
contains all sequences obtainable from $\bm{s}$ by introducing at most $e_i$ substitution errors in the $i$-th channel, while the second,
\begin{equation*}
\mathcal{B}_{k, e}(\bm{s}) \triangleq \left\{ \mathcal{R}(\bm{y}_0, \bm{y}_1, \ldots, \bm{y}_{k-1}) \cap \mathcal{X}_{k}^{n} \ : \ \bm{y}_i \in \{0,1\}^n, \ \sum_{i=0}^{k-1} d(\bm{s}_i, \bm{y}_i) \leq e \right\},
\end{equation*}
contains all sequences obtainable from $\bm{s}$ with at most $e$ total substitution errors.
The intersection with $\mathcal{X}_{k}^{n}$ ensures only valid reconstructions are included, excluding sequences containing the symbol $\mathord{?}$.
$\mathcal{B}_{k, (e_0, e_1, \ldots, e_{k-1})}(\bm{s})$ and $\mathcal{B}_{k, e}(\bm{s})$ are referred to as the \emph{$k$-resolution composite error balls of radius $(e_0, \dots, e_{k-1})$ and $e$}, respectively, or simply \emph{composite error balls} when $k$ is clear from context.
A code $\mathcal{C} \subseteq \mathcal{X}_{k}^{n}$ is an $(e_0, e_1, \ldots, e_{k-1})$-CECC if the composite error balls centered at any two distinct codewords are disjoint, that is, for all distinct codewords $\bm{c}, \bm{c}' \in \mathcal{C}$, 
\begin{equation*}
  \mathcal{B}_{k, (e_0, e_1, \ldots, e_{k-1})}(\bm{c}) \cap \mathcal{B}_{k, (e_0, e_1, \ldots, e_{k-1})}(\bm{c}') = \emptyset.
\end{equation*}
Similarly, a code $\mathcal{C} \subseteq \mathcal{X}_{k}^{n}$ is a $k$-resolution $e$-CECC if for all distinct codewords $\bm{c}, \bm{c}' \in \mathcal{C}$, it holds
\begin{equation*}
  \mathcal{B}_{k, e}(\bm{c}) \cap \mathcal{B}_{k, e}(\bm{c}') = \emptyset.
\end{equation*}


\subsection{Arbitrary Error Parameters}

The non-uniform size of composite error balls complicates the cardinality estimates for arbitrary error parameters.
We therefore restrict our analysis to the resolution $k=2$ case.
Our approach has two parts. First we apply the classical sphere packing bound with the minimum ball size in order to establish a baseline.
Second we obtain a tighter result by employing an asymptotic analysis that follows Levenshtein~\cite{Levenshtein}.
A summary of the upper bounds derived in this section is given in Table~\ref{tab:upper-bounds-1}.

{
\renewcommand{\arraystretch}{2} 
\begin{table}[htbp]
  \caption{Upper bounds on the cardinality of composite error correcting codes for resolution $k=2$ and arbitrary error parameters.}
  \label{tab:upper-bounds-1}
  \centering
  \begin{tabular}{c c c}
  \toprule
  \textbf{Code Family}                             & \textbf{Sphere Packing Bound}           & \textbf{Asymptotic Bound} \\
  \midrule
  $\mathcal{S}_2\left(n; (e_0, e_1)\right)$   & $\frac{3^n}{\binom{n}{\min\{e_0, e_1\}}}$     & $\frac{3^n}{\left(\frac{n}{3}\right)^{e_0+e_1}}\cdot e_0^{e_0} \cdot e_1^{e_1}$ \\ 
  $\mathcal{S}_2\left(n; e\right)$            & $\frac{3^n}{\binom{n}{e}}$                    & $\frac{3^n}{(\frac{4n}{3e})^e}$   \\
  \bottomrule
  \end{tabular}
\end{table}
}

Recall that a $2$-resolution composite binary sequence $\bm{s} \in \mathcal{X}_2^n$ is represented as a ternary sequence over $\Sigma_3^n$, and the decomposition mapping for ternary letters is given by
\begin{equation*}
   \mathcal{D}(0) = \begin{bmatrix} 0 \\ 0 \end{bmatrix}, \quad 
   \mathcal{D}(1) = \begin{bmatrix} 0 \\ 1 \end{bmatrix}, \quad 
   \mathcal{D}(2) = \begin{bmatrix} 1 \\ 1 \end{bmatrix}.
\end{equation*}

\begin{theorem}\label{th:up_bound1}
For any positive integers $e_0, e_1, e$ and code length $n$, the cardinalities of $(e_0, e_1)$-CECCs and $2$-resolution $e$-CECCs are upper bounded by
\begin{equation*}
\mathcal{S}_2\left(n; (e_0, e_1)\right) \leq \frac{3^n}{\binom{n}{\min\{e_0, e_1\}}} \quad \text{and} \quad \mathcal{S}_2\left(n; e\right) \leq \frac{3^n}{\binom{n}{e}}.
\end{equation*}
\end{theorem}

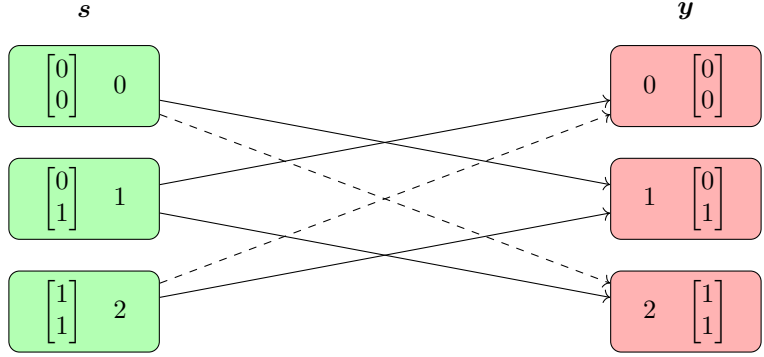
\begin{figure}[htbp]
    \centering
    \begin{tikzpicture}[node distance=1cm]

      \tikzstyle{ng} = [rectangle, rounded corners, minimum width=2cm, minimum height=0.5cm,text centered, draw=black, fill=green!30]
      \tikzstyle{nr} = [rectangle, rounded corners, minimum width=2cm, minimum height=0.5cm,text centered, draw=black, fill=red!30]

      \node (s) at (-4, 2.5) {$\bm{s}$};
      \node (y) at (4, 2.5) {$\bm{y}$};
      \node (A0) at (-4, 1.5) [ng] {$\begin{bmatrix}0 \\0\end{bmatrix} \quad 0$};
      \node (A1) at (-4, 0) [ng] {$\begin{bmatrix}0 \\1\end{bmatrix} \quad 1$};
      \node (A2) at (-4, -1.5) [ng] {$\begin{bmatrix}1 \\1\end{bmatrix} \quad 2$};
      \node (B0) at (4, 1.5) [nr] {$0 \quad \begin{bmatrix}0 \\0\end{bmatrix} $};
      \node (B1) at (4, 0) [nr] {$1 \quad \begin{bmatrix}0 \\1\end{bmatrix} $};
      \node (B2) at (4, -1.5) [nr] {$2 \quad \begin{bmatrix}1 \\1\end{bmatrix} $};

      \draw [->] (A0) edge (B1) (A1) edge (B0) (A1) edge (B2) (A2) edge (B1);
      \draw [->, dashed] (A0) edge (B2) (A2) edge (B0); 

    \end{tikzpicture}
    \caption{Transformations resulting from channel errors in $2$-resolution $e$-CECCs. Dashed edges indicate transformations requiring both channels to err at the same position.}
      \label{fig:error-channel-e-cecc}
  \end{figure}

\begin{IEEEproof}
The proof uses the sphere packing bound, which requires establishing a minimum size for the composite error balls.
For any sequence $\bm{s} \in \mathcal{X}_2^n$, we show that the sizes of the composite error balls satisfy the following lower bounds
\begin{equation*}
|\mathcal{B}_{2, (e_0, e_1)}(\bm{s})| \geq \binom{n}{\min\{e_0, e_1\}} \quad \text{and} \quad |\mathcal{B}_{2, e}(\bm{s})| \geq \binom{n}{e}.
\end{equation*}
Applying the sphere packing bound with these lower bounds yields the upper bounds stated in the theorem.
\begin{itemize}
    \item \textbf{$(e_0, e_1)$-CECC}: Assume without loss of generality that $e_0 \geq e_1$.
    If $\sigma \in \{ 0, 2\}$, we can introduce an error in both channels at the same position and still obtain a valid letter, as illustrated by the dashed arrows in Figure \ref{fig:error-channel-e-cecc}.
    If $\sigma = 1$, introducing a single error in any of the channels will result in a valid letter.
    In either case, we can select any $e_1$ letters in $\bm{s}$ to introduce one or two errors, resulting in a valid $\bm{y} \in \mathcal{B}_{2, (e_0, e_1)}(\bm{s})$.
    Furthermore, note that for $\bm{s} = \bm{0}$, the bound is strict, as we cannot introduce an error in the first channel without also introducing
    one in the second channel to obtain a valid sequence. This shows that $|\mathcal{B}_{2, (e_0, e_1)}(\bm{s})| \geq \binom{n}{e_1}$.

    \item \textbf{$2$-resolution $e$-CECC}: For each letter in $\bm{s}$ there is at least one way to transform it into another letter in the reconstructed sequence $\bm{y}$ by introducing exactly one error, as illustrated in Figure \ref{fig:error-channel-e-cecc}.
  Therefore we can select any $e$ letters in $\bm{s}$ to introduce an error, resulting in a valid $\bm{y} \in \mathcal{B}_{2, e}(\bm{s})$, i.e., $|\mathcal{B}_{2, e}(\bm{s})| \geq \binom{n}{e}$.
\end{itemize}

\end{IEEEproof} 

The upper bounds in Theorem~\ref{th:up_bound1} are loose because the sizes of the $2$-resolution composite error balls, which are provided in \cref{prop:ball-arbitrary} from Appendix \ref{appendix:upper}, vary significantly with the center sequence $\bm{s}$.
To quantify this looseness we consider $n=30$ and CECCs with parameters $(1, 0)$ which correspond to a single substitution error in the first channel.
For the all zero sequence $\bm{s} = \bm{0}$ no letter can be transformed by an error in the first channel into a valid ternary letter.
Hence $|\mathcal{B}_{2, (1, 0)}(\bm{0})| = |\{\bm{0}\}| = 1$.
In contrast for the alternating sequence $\bm{s} = 012\cdots012$ there are $20$ nonzero letters each of which can be transformed by an error in the first channel into another valid letter, so $|\mathcal{B}_{2, (1, 0)}(\bm{s})| = 21$.
This large discrepancy highlights the imprecision of the bound.


To obtain a tighter upper bound on the cardinality of the code we apply an asymptotic method inspired by Levenshtein's work on insertion and deletion errors~\cite{Levenshtein}.
The main idea is to partition the codewords into \textbf{typical} and \textbf{atypical} subsets.
We then prove that the typical subset asymptotically dominates the size of the code and that the atypical subset becomes negligible.
For this analysis we use the notation $f(n) \lesssim g(n)$ which means that $\lim_{n \to \infty} \frac{f(n)}{g(n)} \leq 1$.


\begin{theorem}\label{th:asymptotic}
    For any positive integers $e_0,e_1>0$, it holds that
    \begin{equation*}
    \mathcal{S}_2\left(n; (e_0, e_1)\right) \lesssim \frac{3^n}{\left(\frac{n}{3}\right)^{e_0+e_1}}\cdot e_0^{e_0} \cdot e_1^{e_1}.
    \end{equation*}
    If, in addition, $0 < e_1 \leq e_0 \leq 2e_1$, then
    \begin{equation*}
    \mathcal{S}_2\left(n; (e_0, e_1)\right) \lesssim \frac{3^n}{\left(\frac{2n}{3}\right)^{e_0+e_1}}\cdot e_0^{e_0} \cdot e_1^{e_1}.
    \end{equation*}
    Moreover, for any positive even integer $e > 0$
    \begin{equation*}
    \mathcal{S}_2(n; e) \lesssim \frac{3^n}{(\frac{4n}{3e})^e}.
    \end{equation*}
\end{theorem}

\begin{IEEEproof}
  Let $\mathcal{C}$ be an optimal $(e_0, e_1)$-CECC, that is, $|\mathcal{C}| = \mathcal{S}_2\left(n; (e_0, e_1)\right)$.
  Denote by $\Delta = \frac{n}{3} - \sqrt{(e_0+e_1) n \ln n}$.
  Let $\mathcal{C}_0 \subseteq \mathcal{C}$ be the subset of codewords $\bm{c} \in \mathcal{C}$ such that the number of ones in $\bm{c}$ 
  is least $\Delta$, i.e., $$ \mathcal{C}_0 = \left\{ \bm{c} \in \mathcal{C} \ : \ \#_1(\bm{c}) \geq \Delta \right\}.$$
  Note that for any such $\bm{c}$ it holds that
  \begin{equation*}
    \begin{split}
      |\mathcal{B}_{2, (e_0, e_1)}(\bm{c})| \geq \binom{ \#_1(\bm{c})}{e_0}  \binom{ \#_1(\bm{c})-e_0}{e_1} \geq \binom{\Delta}{e_0} \binom{\Delta - e_0}{e_1}
      \geq \left( \frac{\Delta}{e_0} \right)^{e_0} \left( \frac{\Delta-e_0}{e_1} \right)^{e_1},
    \end{split}
  \end{equation*}
  where the last inequality is due to the fact that $\binom{a}{b} \geq \left( \frac{a}{b} \right)^b$.
  Since $\mathcal{C}$ is a code, then the composite error balls $\mathcal{B}_{2, (e_0, e_1)}(\bm{c})$ are disjoint for all distinct $\bm{c} \in \mathcal{C}_0$, yielding
  \begin{equation*}
      |\mathcal{C}_0| \leq \frac{3^n}{\left( \frac{\Delta}{e_0} \right)^{e_0} \left( \frac{\Delta-e_0}{e_1} \right)^{e_1}} = \frac{3^n}{\left( \frac{\frac{n}{3} - \sqrt{(e_0+e_1) n \ln n}}{e_0} \right)^{e_0} \left( \frac{\frac{n}{3} - \sqrt{(e_0+e_1) n \ln n}-e_0}{e_1} \right)^{e_1}}
      \lesssim \frac{3^n}{\left(\frac{n}{3}\right)^{e_0+e_1}}\cdot e_0^{e_0} \cdot e_1^{e_1}.
  \end{equation*}
  Let $\mathcal{C}_1 = \mathcal{C} \setminus \mathcal{C}_0$. The size of $\mathcal{C}_1$ is constrained by the number of codewords with fewer than 
  $\Delta$ ones, that is,
  \begin{equation*}
    |\mathcal{C}_1| \leq \sum_{m=0}^{\Delta} \binom {n}{m} 2^{n-m} = \sum_{m=0}^{\frac{n}{3} - \sqrt{(e_0 + e_1) n \ln n}} \binom {n}{m} 2^{n-m}.
  \end{equation*}
  We can apply \cref{lemma:asymp} from Appendix~\ref{appendix:upper} for $t=e_0+e_1$ to get an asymptotic upper bound on the size of $\mathcal{C}_1$, namely,
  \begin{equation*}
    |\mathcal{C}_1| \leq \sum_{m=0}^{\frac{n}{3} - \sqrt{(e_0 + e_1) n \ln n}} \binom {n}{m} 2^{n-m} \overset{(\ref{lemma:asymp})}{\lesssim} \frac{3^n}{n^{\frac{9{(e_0 + e_1)}}{4}}}
  \end{equation*}
  This shows that the upper bound on $\mathcal{C}_1$ is negligible compared to that on $\mathcal{C}_0$.
  Therefore we have that
  \begin{equation*}
    \mathcal{S}_2\left(n; (e_0, e_1)\right) = |\mathcal{C}| \simeq |\mathcal{C}_0| \lesssim \frac{3^n}{\left(\frac{n}{3}\right)^{e_0+e_1}}\cdot e_0^{e_0} \cdot e_1^{e_1}.
  \end{equation*}

  We now proceed to show the second part of the theorem. Let $\mathcal{C}$ be an optimal $(e_0, e_1)$-CECC, that is, $|\mathcal{C}| = \mathcal{S}_2\left(n; (e_0, e_1)\right)$.
  Let $\mathcal{C}_0 \subseteq \mathcal{C}$ be the subset of codewords $\bm{c} \in \mathcal{C}$ such that
  \begin{equation*}
      \#_0(\bm{c}) \leq \frac{n}{3} + \sqrt{e_0 n \ln n} \qquad \text{and} \qquad  \#_2(\bm{c}) \leq \frac{n}{3} + \sqrt{e_1 n \ln n}.
  \end{equation*}
  Note that for any such $\bm{c}$ it holds that
  \begin{equation*}
    \begin{split}
      |\mathcal{B}_{2, (e_0, e_1)}(\bm{c})| & \geq \binom{\#_1(\bm{c}) + \#_2(\bm{c})}{e_0} \binom{\#_1(\bm{c}) + \#_0(\bm{c}) - e_0}{e_1} = \binom{n-\#_0(\bm{c})}{e_0} \binom{n-\#_2(\bm{c})-e_0}{e_1} \\
      & \geq \binom{\frac{2n}{3} - \sqrt{e_0 n \ln n}}{e_0} \binom{\frac{2n}{3} - \sqrt{e_1 n \ln n} - e_0}{e_1} 
      \geq \left( \frac{\frac{2n}{3} - \sqrt{e_0 n \ln n}}{e_0} \right)^{e_0} \left( \frac{\frac{2n}{3} - \sqrt{e_1 n \ln n} - e_0}{e_1} \right)^{e_1},
    \end{split}
  \end{equation*}
  where the last inequality is due to the fact that $\binom{a}{b} \geq \left( \frac{a}{b} \right)^b$.
  Since $\mathcal{C}$ is a code, then the composite error balls $\mathcal{B}_{2, (e_0, e_1)}(\bm{c})$ are disjoint for all distinct $\bm{c} \in \mathcal{C}_0$, yielding
  \begin{equation*}
      |\mathcal{C}_0| \leq \frac{3^n}{\left( \frac{\frac{2n}{3} - \sqrt{e_0 n \ln n}}{e_0} \right)^{e_0} \left( \frac{\frac{2n}{3} - \sqrt{e_1 n \ln n} - e_0}{e_1} \right)^{e_1}} \lesssim \frac{3^n}{\left(\frac{2n}{3}\right)^{e_0+e_1}}\cdot e_0^{e_0} \cdot e_1^{e_1}.
  \end{equation*}
  Define $\mathcal{C}_1 = \mathcal{C} \setminus \mathcal{C}_0$. First we provide an upper bound on the number of codewords
  which have more than $\frac{n}{3} + \sqrt{e_0 n \ln n}$ zeroes or more than $\frac{n}{3} + \sqrt{e_1 n \ln n}$ twos, i.e., 
  \begin{equation*}
    \sum_{j=\frac{n}{3} + \sqrt{e_0 n \ln n}}^{n} \binom {n}{j} 2^{n-j} + \sum_{\ell =\frac{n}{3} + \sqrt{e_1 n \ln n}}^{n} \binom {n}{\ell} 2^{n-\ell}.
  \end{equation*}
  This sum has double counting, however the bound is enough.
  If we apply the second inequality of Lemma \ref{lemma:asymp} from Appendix~\ref{appendix:upper} to each of these summations, and remember that $e_1 \leq e_0$, we get that the number of such codewords is asymptotically bounded by
  \begin{equation*}
    \frac{3^n}{n^{\frac{9{e_0}}{4}}} + \frac{3^n}{n^{\frac{9{e_1}}{4}}} \leq 2 \cdot \frac{3^n}{n^{\frac{9{e_1}}{4}}}.
  \end{equation*}
  Next, for each $\bm{c} \in \mathcal{C}_1$, remember that
  \begin{equation*}
    |\mathcal{B}_{2, (e_0, e_1)}(\bm{c})| \geq \binom{n}{\min\{e_0, e_1\}} = \binom{n}{e_1} \geq \left(\frac{n}{e_1}\right)^{e_1},
  \end{equation*}
  and since these are still codewords, the composite error balls $\mathcal{B}_{2, (e_0, e_1)}(\bm{c})$ are disjoint for all distinct $\bm{c} \in \mathcal{C}_1$, therefore
  \begin{equation*}
    |\mathcal{C}_1| \lesssim \frac{2  \cdot \frac{3^n}{n^{\frac{9{e_1}}{4}}}}{(\frac{n}{e_1})^{e_1}} = \frac{3^n}{n^{\frac{13e_1}{4}}} \cdot 2 \cdot e_1^{e_1}.
  \end{equation*}
  Since $e_0, e_1 \ll n $, if $e_0 + e_1 < \frac{13e_1}{4}$ then $|\mathcal{C}_1|$ is negligible compared to $|\mathcal{C}_0|$, and this is indeed 
  the case because we assumed $e_0 \leq 2e_1$.
  Therefore
  \begin{equation*}
    \mathcal{S}_2\left(n; (e_0, e_1)\right) = |\mathcal{C}| \simeq |\mathcal{C}_0| \lesssim \frac{3^n}{\left(\frac{2n}{3}\right)^{e_0+e_1}}\cdot e_0^{e_0} \cdot e_1^{e_1}.
  \end{equation*}

  Finally, we prove the last part of the theorem to establish an upper bound on $\mathcal{S}_2(n; e)$ for any positive even integer $e > 0$.
  By Proposition \ref{prop:connection}, and since $e$ is even, it holds that $\mathcal{S}_2\left(n; e\right) \leq \mathcal{S}_2\left(n; \left(\frac{e}{2}, \frac{e}{2}\right)\right)$.
  By using the second part of this theorem with $e_0 = e_1 = \frac{e}{2}$, we obtain
  \begin{equation*}
      \mathcal{S}_2\left(n; e \right) \leq \mathcal{S}_2\left(n; \left(\frac{e}{2}, \frac{e}{2}\right)\right) \lesssim \frac{3^n}{\left(\frac{2n}{3}\right)^{e}} \cdot \left(\frac{e}{2}\right)^e = \frac{3^n}{\left(\frac{4n}{3e}\right)^e}.
  \end{equation*}
\end{IEEEproof}

\subsection{Limited Error Parameters}

We now focus on several scenarios in which the number of errors is limited.
First, we analyze the scenario of a single substitution error for both code families, for arbitrary resolution parameter $k$.
When the erroneous channel is known, Proposition~\ref{prop:swap-single-error} implies that it suffices to consider the scenario in which the error occurs in the first channel.
Next, we consider the scenario of two substitution errors, focusing on the special case of resolution $k = 2$.
Throughout this analysis, we employ the generalized sphere packing bound (GSPB) framework introduced in \cite{gspb}, which enables the derivation of nontrivial, non-asymptotic upper bounds.
These bounds improve upon the sphere packing bound for the arbitrary error parameters established in the previous section and, in certain cases, also surpass the corresponding asymptotic bounds.
We additionally compute the average sizes of the respective composite error balls together with the average sphere packing value defined in this framework,  which serve as intuitive indicators of the expected upper bounds but do not provide formal guarantees.
Table~\ref{tab:upper-bounds-2} summarizes the results of this section for a single substitution error with arbitrary resolution, 
whereas Table~\ref{tab:upper-bounds-3} presents the results for two substitution errors with resolution $k=2$ and includes a comparison with the asymptotic bounds from \cref{th:asymptotic}.

{
\renewcommand{\arraystretch}{2} 
\begin{table}[htbp]
  \caption{Upper bounds and values for composite error correcting codes with a single error and arbitrary resolution.}
  \label{tab:upper-bounds-2}
  \centering
    \begin{tabular}{c c c}
    \toprule
    \textbf{Code Family}                               &   \textbf{Generalized Sphere Packing Bound }     &   \textbf{Average Sphere Packing Value} \\
    \midrule
    $\mathcal{S}_k\left(n; (1, 0, \ldots, 0)\right)$   & $\frac{(k+1)^{n+1} - (k-1)^{n+1}}{2(n+1)}$       &   $\frac{(k+1)^n}{\frac{2n}{k+1} + 1}$ \\
    $\mathcal{S}_k\left(n; 1\right)$                   & $\frac{(k+1)^n}{\frac{2kn}{k+1}-1}$              &   $\frac{(k+1)^n}{\frac{2kn}{k+1}+1}$ \\
    \bottomrule
    \end{tabular}
\end{table}
}

{
\renewcommand{\arraystretch}{2} 
\begin{table}[htbp]
  \caption{Upper bounds and values for composite error correcting codes with two errors and resolution $k=2$.}
  \label{tab:upper-bounds-3}
  \centering
    \begin{tabular}{c c c c}
    \toprule
    \textbf{Code Family}                         &   \textbf{Generalized Sphere Packing Bound }     &   \textbf{Average Sphere Packing Value}               & \textbf{Asymptotic Bound} \\
    \midrule
    $\mathcal{S}_{2}\left(n; (1, 1)\right)$      & $\frac{3^n}{\frac{(n-3)^2}{6}}$                  &   $\frac{3^n}{\frac{4n^2}{9} + \frac{14n}{9} + 1}$    & $\frac{3^n}{\frac{4n^2}{9}}$ \\
    $\mathcal{S}_{2}\left(n; 2\right)$           & $\frac{\sqrt{\frac{8n}{6}}}{\sqrt{\frac{8n}{6}}-1} \cdot \frac{3^n}{\frac{8n^2}{9}-\frac{2n (\sqrt{\frac{8n}{6}})}{3}}$  & $ \frac{3^n}{\frac{8n^2}{9} + \frac{10n}{9} + 1}$ & $\frac{3^n}{\frac{4n^2}{9}}$  \\
    \bottomrule
    \end{tabular}
\end{table}
}

Let $\mathcal{H} = (\mathcal{X}, \mathcal{E})$ be a hypergraph with vertex set $\mathcal{X} = \left\{x_1, \ldots, x_N\right\}$ and hyperedge set $\mathcal{E} = \left\{E_1, \ldots, E_M\right\}$. 
Let $A \in \left\{0,1\right\}^{N \times M}$ denote the incidence matrix of $\mathcal{H}$, where $A_{i,j} = 1$ if $x_i \in E_j$, and $A_{i,j} = 0$ otherwise.
The \emph{relaxed transversal number} of $\mathcal{H}$ is defined as
\begin{equation*}
  \tau^{*}\left(\mathcal{H}\right) \triangleq \min \left\{ \sum_{i=1}^{N} w_i \ : \ A^\intercal \cdot \bm{w} \geq \bm{1}, \ \bm{w} \in [0, 1]^N \right\}.
\end{equation*}
A \emph{fractional transversal} is any vector $\bm{w} \in [0, 1]^N$ assigning weights to the vertices such that $ A^\intercal \cdot \bm{w} \geq \bm{1}$.
For any such $\bm{w}$, it holds that
\begin{equation*}
\tau^*\left({\mathcal{H}}\right) \leq \sum_{i=1}^{N} w_i.
\end{equation*}

We define a hypergraph corresponding to each family of composite-error-correcting codes.
In both cases, the vertex set is the set of all $k$-resolution composite binary sequences of length $n$, that is $\mathcal{X} \triangleq \mathcal{X}_{k}^{n}$.
The hyperedges are determined by the associated $k$-resolution composite error balls.
Formally, for any tuple $(e_0, e_1, \ldots, e_{k-1}) \in \mathbb{N}^k$ and any positive integer $e$, let
\begin{equation*}
  \begin{split}
  \mathcal{H}_k(e_0, e_1, \ldots, e_{k-1}) & \triangleq \left( \mathcal{X}, \left\{ \mathcal{B}_{k, (e_0, e_1, \ldots, e_{k-1})}(\bm{s}) \ : \ \bm{s} \in \mathcal{X} \right\} \right), \\
  \mathcal{H}_k(e) & \triangleq \left( \mathcal{X}, \left\{ \mathcal{B}_{k, e}(\bm{s}) \ : \ \bm{s} \in \mathcal{X} \right\} \right).
  \end{split}
\end{equation*}
The results from \cite{gspb} state that
\begin{equation*}
    \mathcal{S}_k\left(n; (e_0, e_1, \ldots, e_{k-1})\right) \leq \tau^*\left(\mathcal{H}_k(e_0, e_1, \ldots, e_{k-1})\right) \quad \text{and} \quad \mathcal{S}_k\left(n; e\right) \leq \tau^*\left(\mathcal{H}_k(e)\right).
\end{equation*}
This upper bound is referred to as the \emph{generalized sphere packing bound}.
Lastly, for a $k$-resolution composite binary sequence $\bm{s} \in \mathcal{X}$, we denote by $\mathcal{B}_{k, (e_0, e_1, \ldots, e_{k-1})}^{in}(\bm{s})$ the set of vertices in $\mathcal{X}$ that can reach $\bm{s}$ via at most $e_i$ substitution errors in the $i$-th channel, for all $0 \leq i \leq k-1$.
Similarly, we denote by $\mathcal{B}_{k, e}^{in}(\bm{s})$ the set of vertices in $\mathcal{X}$ that can reach $\bm{s}$ via at most $e$ substitution errors, regardless of how the errors are distributed across the $k$ channels. That is,
\begin{equation*}
  \begin{split}
    \mathcal{B}_{k, (e_0, e_1, \ldots, e_{k-1})}^{in}(\bm{s}) & \triangleq \left\{ \bm{y} \in \mathcal{X} \ : \  d(\bm{y}_i, \bm{s}_i) \leq e_i, \ 0 \leq i \leq k-1\right\}, \\
    \mathcal{B}_{k, e}^{in}(\bm{s}) & \triangleq \left\{ \bm{y} \in \mathcal{X} \ : \ \sum_{i=0}^{k-1} d(\bm{y}_i, \ \bm{s}_i) \leq e \right\},
  \end{split}
\end{equation*}
where $\mathcal{D}(\bm{y}) = (\bm{y}_0, \bm{y}_1, \ldots, \bm{y}_{k-1})$ and $\mathcal{D}(\bm{s}) = (\bm{s}_0, \bm{s}_1, \ldots, \bm{s}_{k-1})$.
It is further shown in \cite{gspb} that
\begin{equation} \label{eq:fractra}
  w_i = \frac{1}{\min_{\bm{x} \in \mathcal{B}^{in}(\bm{x}_i)} |\mathcal{B}(\bm{x})|}
\end{equation}
defines a valid fractional transversal.
We use this result together with the generalized sphere packing bound to provide improved non-asymptotic upper bounds for the scenarios of up to two errors.


\subsubsection{$(1, 0, \ldots, 0)$-composite-error-correcting codes}

We begin by analyzing the scenario of a single substitution error under the assumption that the erroneous channel is known.
As noted in Proposition~\ref{prop:swap-single-error}, it suffices to consider the case where the error occurs in the first channel.

\begin{proposition} \label{prop:ball-1-0-general}
  Let $\bm{s} \in \mathcal{X}$ be a $k$-resolution composite binary sequence of length $n$. Denote by $m \triangleq \#_{k-1}(\bm{s}) + \#_{k}(\bm{s})$.
  Then 
  \begin{equation*}
    |\mathcal{B}_{k, (1, 0, \ldots, 0)}(\bm{s})| = 1 + m.
  \end{equation*}
\end{proposition}

\begin{IEEEproof}
    Figure \ref{fig:error-channel-1-0-general} depicts the transformations that a letter in the reconstructed sequence $\bm{y} \triangleq \mathcal{R}(\bm{y}_0, \bm{y}_1, \ldots, \bm{y}_{k-1})$ can undergo due to an error in the first channel, where $\bm{y}_i$ denotes the sequence received from the $i$-th channel.
    According to this mapping, the following cases arise.
    \begin{itemize}
      \item No error is introduced. In this case, the output is exactly $\bm{s}$, which belongs to $\mathcal{B}_{k, (1, 0, \ldots, 0)}(\bm{s})$.
      \item Invalid reconstructions (dashed arrows). This happens when the error occurs at a position in which $\bm{s}$ takes a value $ i \in \left\{0, 1, \ldots, k-2\right\}$.
      In the binary column vector representation, such letters have $0$s in their first two entries.
      An error in the first channel flips the first bit, resulting in $10\ldots$, which does not correspond to any valid letter in $\Sigma_{k+1}$.
      Since the composite error ball includes only valid reconstructions, this case does not contribute to $\mathcal{B}_{k, (1, 0, \ldots, 0)}(\bm{s})$.
      \item An error occurs at a position in which $\bm{s}$ takes the value $k-1$ or $k$. Since there are $m$ such positions in $\bm{s}$, this case contributes $m$ sequences to $\mathcal{B}_{k, (1, 0, \ldots, 0)}(\bm{s})$.
    \end{itemize}
    Hence, we conclude that
    \begin{equation*}
      |\mathcal{B}_{k, (1,0, \ldots, 0)}(\bm{s})|  = 1 + m.
    \end{equation*}

    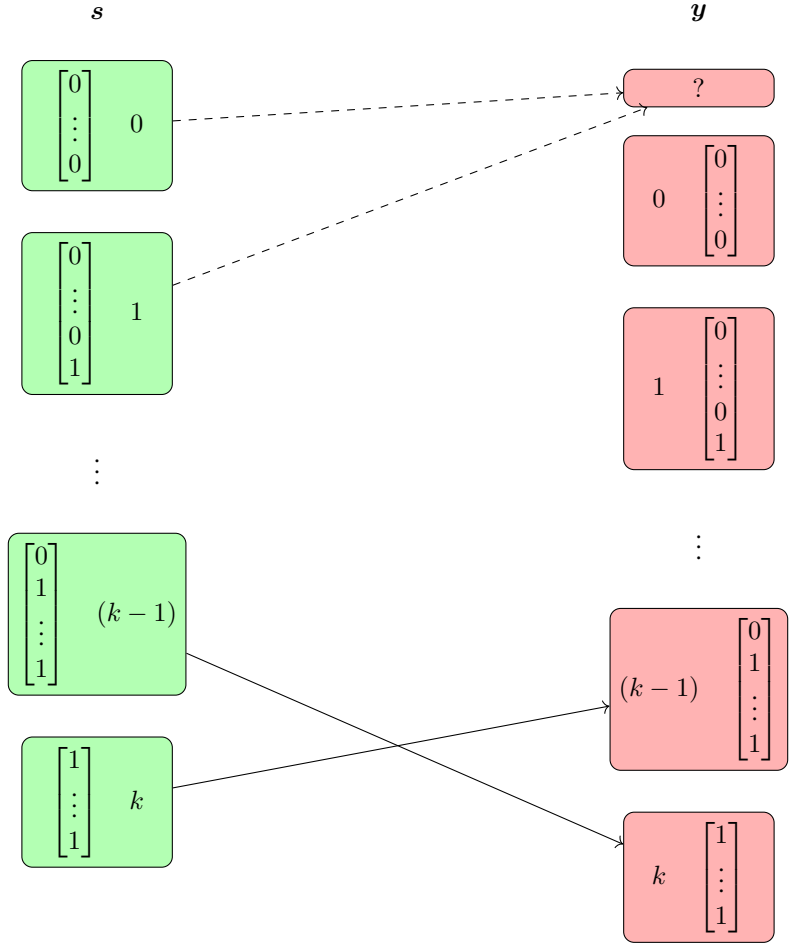
\begin{figure}[htbp]
    \centering
    \begin{tikzpicture}[node distance=1cm]

      \tikzstyle{ng} = [rectangle, rounded corners, minimum width=2cm, minimum height=0.5cm,text centered, draw=black, fill=green!30]
      \tikzstyle{nr} = [rectangle, rounded corners, minimum width=2cm, minimum height=0.5cm,text centered, draw=black, fill=red!30]

      \node (s) at (-4, 6) {$\bm{s}$};
      
      \node (A0) at (-4, 4.5) [ng] {$\begin{bmatrix}0 \\ \vdots \\ 0 \end{bmatrix} \quad 0$};
      \node (A1) at (-4, 2) [ng] {$\begin{bmatrix}0 \\ \vdots \\ 0 \\ 1\end{bmatrix} \quad 1$};
      \node (0) at (-4 , 0) {$\vdots$};
      \node (AK-1) at (-4, -2) [ng] {$\begin{bmatrix} 0 \\ 1 \\ \vdots \\ 1\end{bmatrix} \quad (k-1)$};
      \node (AK) at (-4, -4.5) [ng] {$\begin{bmatrix}1 \\ \vdots \\ 1 \end{bmatrix} \quad k$};

      \node (y) at (4, 6) {$\bm{y}$};
      \node (?) at (4, 5) [nr] {$\mathord{?} $};
      \node (B0) at (4, 3.5) [nr] {$0 \quad \begin{bmatrix}0 \\ \vdots \\ 0 \end{bmatrix} $};
      \node (B1) at (4, 1) [nr] {$1 \quad \begin{bmatrix}0 \\ \vdots \\ 0 \\ 1\end{bmatrix} $};
      \node (1) at (4 , -1) {$\vdots$};
      \node (BK-1) at (4, -3) [nr] {$(k-1) \quad \begin{bmatrix} 0 \\ 1 \\ \vdots \\ 1\end{bmatrix} $};
      \node (BK) at (4, -5.5) [nr] {$k \quad \begin{bmatrix} 1 \\ \vdots \\ 1\end{bmatrix} $};

      \draw [->] (AK-1) edge (BK) (AK) edge (BK-1);
      \draw [->, dashed] (A0) edge (?) (A1) edge (?); 
    \end{tikzpicture}
    \caption{Transformations resulting from channel errors in $(1, 0, \ldots, 0)$-CECCs. Dashed arrows represent transformations to the invalid symbol.}
    \label{fig:error-channel-1-0-general}
  \end{figure}
\end{IEEEproof}

For a $k$-resolution composite binary sequence $\bm{s}$ as in the proposition, any $\bm{x} \in \mathcal{B}_{k, (1, 0, \ldots, 0)}^{in}(\bm{s})$ has the same value $m$ as $\bm{s}$, as illustrated in Figure \ref{fig:error-channel-1-0-general}.
Therefore, the corresponding weight of the fractional transversal $w_i$ from \cref{eq:fractra} is $w_i = \frac{1}{1 +m}$.

\begin{theorem}\label{thm:gspb-1-0-general}
  For any $n \geq 1$
  \begin{equation*}
     \mathcal{S}_k\left(n; (1, 0, \ldots, 0)\right) \leq \tau^{*}\left(\mathcal{H}_k(1, 0, \ldots, 0)\right) \leq \sum_{i=1}^{N} w_i \leq \frac{(k+1)^{(n+1)} - (k-1)^{(n+1)}}{2(n+1)}.
  \end{equation*}
\end{theorem}

\begin{IEEEproof}
    We iterate over the fractional transversal weights $w_i$ based on the value of $m$ in the $k$-resolution composite binary sequence $\bm{s} \in \mathcal{X}$.
    The number of $k$-resolution composite binary sequences in $\mathcal{X}$ with exactly $m$ symbols equal to $k-1$ or $k$, and the remaining $n - m$ symbols drawn from $\left\{0, 1, \ldots, k-2\right\}$ is $\binom{n}{m} 2^m (k - 1)^{n - m}$.
    \begin{equation*}
      \begin{split}
      \sum_{i=1}^{N} w_i &= \sum_{m=0}^n \binom{n}{m} 2^{m} (k-1)^{n-m} \frac{1}{1+m}
      = (k-1)^{n} \sum_{m=0}^n \binom{n}{m} \left(\frac{2}{k-1}\right)^{m} \frac{1}{1+m} \\
      & = (k-1)^{n} \left( \frac{k-1}{2} \right) \sum_{m=0}^n \binom{n}{m} \left(\frac{2}{k-1}\right)^{m+1} \frac{1}{1+m} \\
      & \overset{\text{(BI)}}{=} \frac{(k-1)^{n+1}}{2} \cdot \left( \frac{\left(\frac{k+1}{k-1}\right)^{n+1} - 1}{n+1} \right)
      = \frac{(k+1)^{n+1} - (k-1)^{n+1}}{2(n+1)},
      \end{split}
    \end{equation*}
    where step $\overset{\text{(BI)}}{=}$ follows by applying a binomial identity from Appendix~\ref{appendix:binom-identities} at $x = \frac{2}{k-1}$.
\end{IEEEproof}


\subsubsection{$k$-resolution single-composite-error-correcting codes}
We now turn our attention to the scenario of $k$-resolution $1$-CECCs. In this case one error may be introduced in any of the $k$ channels and the erroneous channel is unknown.

\begin{proposition} \label{prop:ball-1-cecc-general}
 Let $\bm{s} \in \mathcal{X}$ be a $k$-resolution composite binary sequence of length $n$. Denote by $m \triangleq \sum_{i=1}^{k-1} \#_{i}(\bm{s})$.
 Then 
 \begin{equation*}
  |\mathcal{B}_{k, 1}(\bm{s})| = 1 + n + m.
 \end{equation*}
\end{proposition}

\begin{IEEEproof}
  Figure \ref{fig:error-channel-1-cecc-general} depicts the valid transformations that a letter in the reconstructed sequence $\bm{y} \triangleq \mathcal{R}(\bm{y}_0, \bm{y}_1, \ldots, \bm{y}_{k-1})$ can undergo due to an error in any of the channels, where $\bm{y}_i$ denotes the sequence received from the $i$-th channel.
  In this case, we intentionally leave out the transformations to the invalid symbol, but from every letter in $\Sigma_{k+1}$ we can obtain an invalid reconstruction.
  The following cases arise.
  \begin{itemize}
    \item No error is introduced. In this case, the output is exactly $\bm{s}$, which belongs to $\mathcal{B}_{k, 1}(\bm{s})$.
    \item An error occurs at a position in which $\bm{s}$ takes a value $i \in \left\{1, \ldots, k-1\right\}$. Each such letter can be transformed to either $i-1$ or $i+1$ via a single substitution in one of the channels.
    Since there are $m$ such positions in $\bm{s}$, this case contributes $2m$ sequences to $\mathcal{B}_{k, 1}(\bm{s})$.
    \item An error occurs at a position in which $\bm{s}$ takes the value $0$ or $k$. $0$ can only be transformed to $1$ by an error in the last channel, and $k$ can only be transformed to $k-1$ by an error in the first channel.
    Since there are $n - m$ such positions in $\bm{s}$, this case contributes $(n - m)$ sequences to $\mathcal{B}_{k, 1}(\bm{s})$.
  \end{itemize}
  Hence, we conclude that
  \begin{equation*}
    |\mathcal{B}_{k, 1}(\bm{s})| = 1 + 2m + (n - m) = 1 + n + m.
  \end{equation*}

    \begin{figure}[htbp]
    \centering
    \begin{tikzpicture}[node distance=1cm]

      \tikzstyle{ng} = [rectangle, rounded corners, minimum width=2cm, minimum height=0.5cm,text centered, draw=black, fill=green!30]
      \tikzstyle{nr} = [rectangle, rounded corners, minimum width=2cm, minimum height=0.5cm,text centered, draw=black, fill=red!30]

      \node (s) at (-4, 4) {$\bm{s}$};
      \node (A0) at (-4, 3) [ng] {$0$};
      \node (A1) at (-4, 2) [ng] {$1$};
      \node (A2) at (-4, 1) [ng] {$2$};
      \node (0) at (-4 , 0) {$\vdots$};
      \node (AK-2) at (-4, -1) [ng] {$k-2$};
      \node (AK-1) at (-4, -2) [ng] {$k-1$};
      \node (AK) at (-4, -3) [ng] {$k$};

      \node (y) at (4, 4) {$\bm{y}$};
      \node (B0) at (4, 3) [nr] {$0$};
      \node (B1) at (4, 2) [nr] {$1$};
      \node (B2) at (4, 1) [nr] {$2$};
      \node (1) at (4 , 0) {$\vdots$};
      \node (BK-2) at (4, -1) [nr] {$k-2$};
      \node (BK-1) at (4, -2) [nr] {$k-1$};
      \node (BK) at (4, -3) [nr] {$k $};

      \draw [->] (A0) edge (B1) (A1) edge (B0) (A1) edge (B2) (A2) edge (B1) (A2) edge (1) (AK-2) edge (1) (AK-2) edge (BK-1) (AK-1) edge (BK-2) (AK-1) edge (BK) (AK) edge (BK-1);
    \end{tikzpicture}
    \caption{Transformations resulting from channel errors in $k$-resolution $1$-CECCs. Transformations to the invalid symbol are omitted.}
    \label{fig:error-channel-1-cecc-general}
  \end{figure}
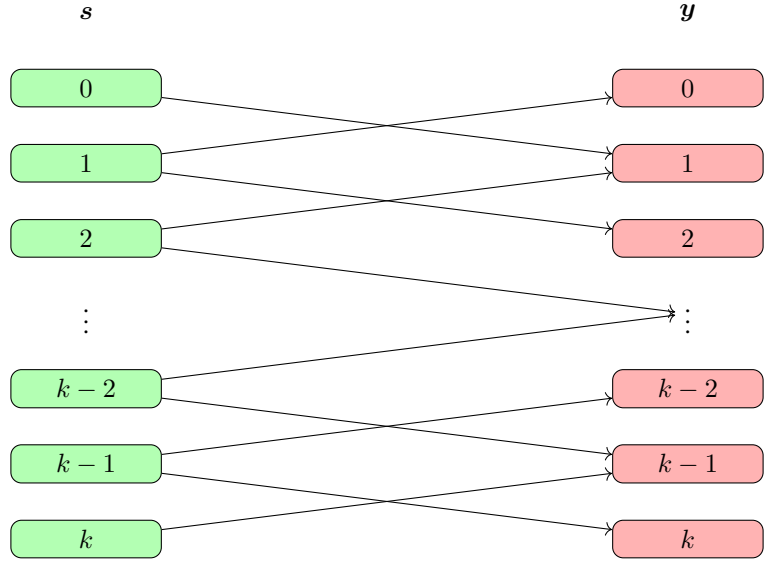
\end{IEEEproof}

For a $k$-resolution composite binary sequence $\bm{s}$ as in the proposition, each $\bm{x} \in \mathcal{B}_{k, 1}^{in}(\bm{s})$ contains between $m-1$ and $m+1$ letters from the set $\left\{1, \ldots, k-1\right\}$, as illustrated in Figure \ref{fig:error-channel-1-cecc-general}.
Therefore the value of the fractional transversal $w_i$ from \cref{eq:fractra} is given by $w_i = \frac{1}{n + m}$.

\begin{theorem}\label{thm:gspb-1-cecc-general}
For any $ n \geq 1$
\begin{equation*}
  \mathcal{S}_k\left(n; 1\right) \leq \tau^{*}\left(\mathcal{H}_k(1)\right) \leq \sum_{i=1}^N w_i \leq \frac{(k+1)^{n}}{\frac{2kn}{k+1} - 1}.
\end{equation*}
\end{theorem}

\begin{IEEEproof}
  We iterate over the fractional transversal weights $w_i$ based on the value of $m$ in the $k$-resolution composite binary sequence $\bm{s} \in \mathcal{X}$.
  The number of $k$-resolution composite binary sequences in $\mathcal{X}$ with exactly $m$ letters in the set $\left\{1, \ldots, k-1\right\}$ is $\binom{n}{m} (k-1)^{m} 2^{n-m}$.
  We then make use of Lemma~\ref{lemma:k-resolution-1-cecc} from Appendix \ref{appendix:upper} for the inequality in the following equation.

  \begin{equation*}
    \sum_{i=1}^N w_i = \sum_{m=0}^n \binom{n}{m} (k-1)^{m} 2^{n-m} \frac{1}{n+m}
    = 2^{n} \sum_{m=0}^n \binom{n}{m} \left(\frac{k-1}{2}\right)^{m} \frac{1}{n+m}
    \overset{(\ref{lemma:k-resolution-1-cecc})}{\leq} 2^{n} \frac{\left( \frac{k+1}{2} \right)^{n}}{\frac{2kn}{k+1} - 1} 
    = \frac{(k+1)^{n}}{\frac{2kn}{k+1} - 1}.
  \end{equation*}
\end{IEEEproof}


This concludes the analysis for the scenarios of a single substitution error.
We now consider scenarios involving two substitution errors with the resolution parameter restricted to $k = 2$.
The general approach remains similar, and the proofs are deferred to Appendix~\ref{appendix:upper}.
Specifically, we examine the following two scenarios.
\begin{itemize}
  \item $(1,1)$-CECCs, where each of the two channels may introduce at most one error.
  \item $2$-resolution $2$-CECCs, where up to two errors may occur across the two channels without any constraint on their distribution.
\end{itemize}


\subsubsection{$(1, 1)$-composite-error-correcting codes}
In this scenario, any of the two channels may introduce at most one error.

\begin{restatable}{proposition}{ballOneOne} \label{prop:ball-1-1}
  Let $\bm{s} \in \mathcal{X}$ be a $2$-resolution composite binary sequence of length $n$ with $j$ zeroes and $m$ ones. Then
  \begin{equation*}
    |\mathcal{B}_{2,(1, 1)}(\bm{s})| = 2n + 1 + m (n-1) + j (n-m-j).
  \end{equation*}
\end{restatable}

For a $2$-resolution composite binary sequence $\bm{s}$ as in the proposition, the minimal $|\mathcal{B}_{2, (1, 1)}(\bm{x})|$ for $\bm{x} \in \mathcal{B}_{2, (1, 1)}^{in}(\bm{s})$ is received for a sequence $\bm{x}$ with $m-2$ ones.
Since one error must be introduced in each channel, then we have a transformation of the type $0 \to 1 $ and another of the type $2 \to 1$, and therefore the value of $w_i$ from \cref{eq:fractra} is upper bounded by $\frac{1}{m(n-1) + (j+1) (n - m - j + 1)}$ as shown below

\begin{equation*}
  \begin{split}
    w_i &= \frac{1}{\min_{\bm{x} \in \mathcal{B}_{2, (1, 1)}^{in}(\bm{s})} |\mathcal{B}_{2, (1,1)}(\bm{x})|} 
    = \frac{1}{2n + 1 + (m-2) \cdot (n-1) + (j+1) \cdot (n - m - j + 1)} \\
    & \leq \frac{1}{m(n-1) + (j+1) \cdot (n - m - j + 1)}.
  \end{split}
\end{equation*}

\begin{restatable}{theorem}{gspbOneOne}\label{thm:gspb-1-1}
  For $n \geq 4$
  \begin{equation*}
      \mathcal{S}_2\left(n; (1, 1)\right) \leq \tau^{*}\left(\mathcal{H}_2(1, 1)\right) \leq \sum _{i=1}^N w_i \leq \frac{3^n}{\frac{(n-3)^2}{6}}.
  \end{equation*}
\end{restatable}


\subsubsection{$2$-resolution $2$-composite-error-correcting codes}
In this scenario at most two errors may be introduced, and the distribution of the errors to the two channels is not restricted.

\begin{restatable}{proposition}{ballTwo}\label{prop:ball-2}
  Let $\bm{s} \in \mathcal{X}$ be a $2$-resolution composite binary sequence of length $n$ with $m$ ones. Then
  \begin{equation*}
    |\mathcal{B}_{2, 2}(\bm{s})| = \frac{n^2}{2} + \frac{3n}{2} + 1 + m(n-1) + \frac{m^2-m}{2}.
  \end{equation*}
\end{restatable}

For a $2$-resolution composite binary sequence $\bm{s}$ as in the proposition, the minimal $|\mathcal{B}_{2, 2}(\bm{x})|$ for $\bm{x} \in \mathcal{B}_{2, 2}^{in}(\bm{s})$ is received for a
sequence $\bm{x}$ with $m-2$ ones, and therefore the value of $w_i$ from \cref{eq:fractra} is upper bounded by $\frac{2}{(n+m)^2 - n - 7m}$, as shown below

\begin{equation*}
  \begin{split}
    w_i &= \frac{1}{\min_{\bm{x} \in \mathcal{B}_{2, 2}^{in}(\bm{s})} |\mathcal{B}_{2, 2}(\bm{x})|} 
    = \frac{2}{n^2 + 3n + 2 + 2(m-2)(n-1) + (m-2)^2 + (m-2)}
    = \frac{2}{(n+m)^2 - n - 7m + 12} \\
    & \leq \frac{2}{(n+m)^2 - n - 7m}.
  \end{split}
\end{equation*}

\begin{restatable}{theorem}{gspbTwo}\label{thm:gspb-2}
  For $n \geq 48$
  \begin{equation*}
    \mathcal{S}_{2}\left(n; 2\right) \leq \tau^{*}\left(\mathcal{H}_2(2)\right) \leq \sum _{i=1}^N w_i \leq \frac{\sqrt{\frac{8n}{6}}}{\sqrt{\frac{8n}{6}}-1} \cdot \frac{3^n}{\frac{8n^2}{9}-\frac{2n (\sqrt{\frac{8n}{6}})}{3}}.
  \end{equation*}
\end{restatable}


As previously noted, the size of a composite error ball depends on the specific composite binary sequence.
An additional pair of important notions introduced in \cite{gspb} are the average ball size and the average sphere packing value.
The \emph{average size} of a $k$-resolution composite error ball of radius $(e_0, e_1, \ldots, e_{k-1})$ and of radius $e$ are defined, respectively, as
\begin{equation*}
    \bar{\Delta}_{k, (e_0, e_1, \ldots, e_{k-1})} \triangleq \frac{1}{|\mathcal{X}|} \sum_{\bm{s} \in \mathcal{X}} |\mathcal{B}_{k, (e_0, e_1, \ldots, e_{k-1})}(\bm{s})| \qquad \text{and} \qquad
    \bar{\Delta}_{k, e} \triangleq \frac{1}{|\mathcal{X}|} \sum_{\bm{s} \in \mathcal{X}} |\mathcal{B}_{k, e}(\bm{s})|.
\end{equation*}
The corresponding \emph{average sphere packing values} are given by
\begin{equation*}
  \text{ASPV}_k(e_0, e_1, \ldots e_{k-1}) \triangleq \frac{|\mathcal{X}|}{\bar{\Delta}_{k, (e_0, e_1, \ldots, e_{k-1})}} \qquad \text{and} \qquad \text{ASPV}_k(e) \triangleq \frac{|\mathcal{X}|}{\bar{\Delta}_{k, e}}.
\end{equation*}
Although these quantities do not, in general, constitute valid upper bounds on the code cardinality, they serve as useful benchmarks for comparison. 
Next, we compute the average composite error ball sizes for the four scenarios analyzed above.
The following theorem gives the main result, and its proof is provided in Appendix~\ref{appendix:upper}.

\begin{restatable}{theorem}{averageBallSize}
  The average sizes of the $k$-resolution composite error balls with radii $(1, 0, \ldots, 0)$ and $1$ are given by
  \begin{equation*}
    \bar{\Delta}_{k, (1, 0, \ldots, 0)} = \frac{2n}{k+1}+1 \quad \text{and} \quad \bar{\Delta}_{k, 1} = \frac{2kn}{k+1}+1,
   \end{equation*}
  respectively. The average sizes of the $2$-resolution composite error balls with radii $(1, 1)$ and $2$ are given by
  \begin{equation*}
    \bar{\Delta}_{2, (1, 1)} = \frac{4n^2}{9} + \frac{14n}{9} + 1 \quad \text{and} \quad \bar{\Delta}_{2, 2} = \frac{8n^2}{9} + \frac{10n}{9} + 1,
  \end{equation*}
  respectively.
\end{restatable}

The corresponding average sphere packing values for the $k$-resolution composite error balls with radii $(1, 0, \ldots, 0)$ and $1$ are given by
\begin{equation*}
    \text{ASPV}_k(1, 0, \ldots, 0) = \frac{(k+1)^n}{\frac{2n}{k+1}+1} \qquad \text{and} \qquad
    \text{ASPV}_k(1)  = \frac{(k+1)^n}{\frac{2kn}{k+1}+1},
\end{equation*}
which as shown in Table~\ref{tab:upper-bounds-2} closely resemble the upper bounds on code cardinality derived in \cref{thm:gspb-1-0-general} and \cref{thm:gspb-1-cecc-general}.
The corresponding average sphere packing values for the $2$-resolution composite error balls with radii $(1, 1)$ and $2$ are given by
\begin{equation*}
    \text{ASPV}_2(1, 1) = \frac{3^n}{\frac{4n^2}{9} + \frac{14n}{9} + 1} \qquad \text{and} \qquad
    \text{ASPV}_2(2) = \frac{3^n}{\frac{8n^2}{9} + \frac{10n}{9} + 1}.
\end{equation*}
The key distinction to the GSPB, as shown in Table~\ref{tab:upper-bounds-3}, appears in the case of $(1,1)$-CECCs, which stems from the application of the inequality
$\frac{1}{x + y} \leq \frac{1}{2} \left( \frac{1}{x} + \frac{1}{y} \right)$, for $x, y > 0$ in the proof of \cref{thm:gspb-1-1}.
As a result, a gap emerges between the average sphere packing value and the upper bound on the code cardinality derived in this case.

\section{Constructions and Lower Bounds} \label{sec:lower-bounds}

We begin this section by presenting a basic lower bound on the cardinality of $k$-resolution $e$-CECCs. 
This bound, when combined with Proposition~\ref{prop:connection}, yields a corresponding lower bound on the cardinality of $(e_0, e_1, \ldots, e_{k-1})$-CECCs.
However, this method results in a relatively weak result.
To improve upon it, we propose a general construction that leads to a stronger lower bound for the cardinality of $(e_0, e_1, \ldots, e_{k-1})$-CECCs under arbitrary error parameters.
We then restrict our attention to the case of a single substitution error, considering both families of codes, where in one the erroneous channel is known and in the other it is unknown.
As in the previous section, when the erroneous channel is known, Proposition~\ref{prop:swap-single-error} permits the assumption that the substitution occurs in the first channel.
When the erroneous channel is unknown, we show that any code capable of correcting a single symmetric error of limited magnitude one, or equivalently, any code in the Lee metric with Lee distance at least three, can be used as a $k$-resolution $1$-CECC.

We have already established a lower bound on the cardinality of $k$-resolution $e$-CECCs.
Proposition~\ref{prop:k-ary} states that
\begin{equation*}
  \mathcal{S}_{k}\left(n; e\right) \geq \mathcal{A}_{k+1}(n; e).
\end{equation*}
This is the strongest lower bound on the cardinality of $k$-resolution $e$-CECCs that we are aware of.
The exact value of $\mathcal{A}_{k+1}(n; e)$ is not known for arbitrary values of $n$ and $e$.
However, when $q = k + 1$ is a prime power, we can use BCH codes to obtain a lower bound on $\mathcal{A}_{k+1}(n;e)$, as stated in the following corollary with proof in Appendix~\ref{appendix:lower-bounds}.

\begin{restatable}{corollary}{lowerBoundBCH}\label{cor:e-cecc-lower-bch}
  For any resolution parameter $k$ such that $k+1$ is a prime power, number of errors $e > 0$ and code length $n$,
  \begin{equation*}
    \mathcal{S}_{k}\left(n; e\right) \geq \mathcal{A}_{k+1}(n; e) \geq \frac{\left(k+1\right)^n}{\left(k+1\right) ^ {\lceil\log_{k+1} (n+1) \rceil \cdot \lceil \frac{k(2e-1)}{k+1}\rceil + 1}} .
  \end{equation*}
\end{restatable}

As previously noted, in the case of $(e_0, e_1, \ldots, e_{k-1})$-CECCs, a straightforward lower bound can be obtained by combining Proposition~\ref{prop:connection} with Corollary~\ref{cor:e-cecc-lower-bch}, namely,
\begin{equation*}
    \mathcal{S}_{k}\left(n; (e_0, e_1, \ldots, e_{k-1})\right) \geq \mathcal{S}_{k}\left(n; \sum_{i=0}^{k-1}e_i \right) \geq \frac{\left(k+1\right)^n}{\left(k+1\right) ^ {\lceil\log_{k+1} (n+1) \rceil \cdot \lceil \frac{k(2 \sum_{i=0}^{k-1}e_i -1)}{k+1}\rceil + 1}} .
\end{equation*}


To obtain a tighter bound, we now introduce the following construction, which is designed to improve upon the previously straightforward estimate.
This construction is natural and requires that if the underlying channel $0 \leq i \leq k-1$ may introduce up to $e_i$ substitution errors, 
then the sequences transmitted through this channel belong to a binary error-correcting code capable of correcting up to $e_i$ substitution errors.

\begin{construction}\label{const:e0e1-general}
  Let $\mathcal{C}_i$ be a binary $e_i$-error-correcting code of length $n$. Let $ \mathcal{C}_{\Romannum{1}}\left( \mathcal{C}_0, \ldots, \mathcal{C}_{k-1} \right)$ be the code
  \begin{equation*}
    \mathcal{C}_{\Romannum{1}}\left( \mathcal{C}_0, \ldots, \mathcal{C}_{k-1} \right) \triangleq \left\{ \bm{c} \in \Sigma_{k+1}^n \ : \ \bm{c}_i \in \mathcal{C}_i, \ 0 \leq i \leq k-1 \right\},
  \end{equation*}
  where $\mathcal{D}(\bm{c}) = (\bm{c}_0, \bm{c}_1, \ldots, \bm{c}_{k-1})$ is the decomposition of the codeword $\bm{c}$.
\end{construction}

In the following theorem we establish the correctness of the construction and provide a formal proof, even though its validity may already appear intuitive.
\begin{theorem}\label{thm:e0e1}
  The code $ \mathcal{C}_{\Romannum{1}}\left( \mathcal{C}_0, \ldots, \mathcal{C}_{k-1} \right)$ is an $(e_0, e_1, \ldots, e_{k-1})$-CECC.
\end{theorem}

\begin{IEEEproof}
  Let $\bm{c}$ denote the transmitted codeword. Let $(\bm{c}_0, \bm{c}_1, \ldots, \bm{c}_{k-1})$ be the binary sequences obtained by applying the decomposition mapping to $\bm{c}$, so that $\mathcal{D}(\bm{c}) = (\bm{c}_0, \bm{c}_1, \ldots, \bm{c}_{k-1})$.
  For each index $i$ with $0 \leq i \leq k-1$, let $\bm{y}_i$ be the output of the $i$-th channel.
  By assumption, $\bm{y}_i$ differs from $\bm{c}_i$ in at most $e_i$ positions due to substitution errors.
  Since $\bm{c}_i$ belongs to the binary code $\mathcal{C}_i$, which is capable of correcting up to $e_i$ substitution errors, we can recover $\bm{c}_i$ from $\bm{y}_i$.
  After recovering all the binary sequences $\bm{c}_0, \bm{c}_1, \ldots, \bm{c}_{k-1}$, we apply the reconstruction mapping to obtain the original codeword, that is,
  $\mathcal{R}(\bm{c}_0, \bm{c}_1, \ldots, \bm{c}_{k-1}) = \bm{c}$.
\end{IEEEproof}

The improved lower bound deriving from this construction is given in the following corollary, and its proof can be found in Appendix~\ref{appendix:lower-bounds}.
\begin{restatable}{corollary}{lowerBoundCosets}\label{cor:e0e1-lower-bound}
  For any tuple $(e_0, e_1, \ldots, e_{k-1}) \in \mathbb{N}^k$ and code length $n$,
  \begin{equation*}
    \mathcal{S}_{k}\left(n; (e_0, e_1, \ldots, e_{k-1})\right) \geq \frac{(k+1)^n}{2 ^ { \lceil \log_2(n+1) \rceil \cdot  \sum_{i=0}^{k-1}e_i}}.
  \end{equation*}
\end{restatable}


We now consider the cases of a single substitution error.
We begin with the setting in which the erroneous channel is known.
By \cref{prop:swap-single-error}, it is sufficient to focus on the case where the substitution occurs in the first channel.
As discussed in Section~\ref{sec:upper-bounds}, and illustrated in Figure \ref{fig:error-channel-1-0-general}, under the assumption of a single substitution error occurring in the first channel, only the following transformations are possible in a $k$-resolution composite binary sequence $\bm{s} \in \Sigma_{k+1}^n$.
\begin{itemize}
  \item The letter $k-1$ may be transformed to the letter $k$.
  \item The letter $k$ may be transformed to the letter $k-1$.
  \item A letter $\sigma \in \left\{0, 1, \ldots, k-2\right\}$ may be transformed into the symbol $\mathord{?}$.
\end{itemize}
The invalid symbol $\mathord{?}$ can be detected and corrected without the need for any error-correcting code. 
In this case, the symbol $\mathord{?}$ corresponds to the vector
\begin{equation*}
  \begin{bmatrix}
    1 \ 0^{k-\sigma-1} \ 1 ^\sigma
  \end{bmatrix}^\intercal,
\end{equation*}
which arises when a bit flip from $0$ to $1$ occurs in the first channel.
Reverting this bit from $1$ back to $0$ restores the original vector representation of the letter $\sigma$.
Therefore, our goal is to construct a code capable of correcting the two valid types of transformations that may occur under a single substitution error in the first channel.

For a $k$-resolution composite binary sequence $\bm{s} \in \Sigma_{k+1}^n$, let $v(\bm{s}) = \#_{k-1}(\bm{s}) + \#_{k}(\bm{s})$.
This function can be used to partition $\Sigma_{k+1}^n$ into equivalence classes, where two sequences $\bm{s}, \bm{t} \in \Sigma_{k+1}^n$ are considered equivalent if and only if $v(\bm{s}) = v(\bm{t})$.
We denote these equivalence classes by
\begin{equation*}
  \Sigma_{k+1}^n(\ell) \triangleq \left\{ \bm{s} \in \Sigma_{k+1}^n \ : \ v(\bm{s}) = \ell \right\}.
\end{equation*}
The cardinality of each equivalence class is then given by $\binom{n}{\ell} \cdot 2^\ell \cdot (k-1)^{n-\ell}$.

We construct a function $\mathcal{F}: \Sigma_{k+1}^n(\ell) \to \left\{0, 1\right\}^\ell$ that maps a $k$-resolution composite binary sequence $\bm{s} \in \Sigma_{k+1}^n(\ell)$ to a binary sequence of length $\ell$.
The function $\mathcal{F}$ removes all letters in $\bm{s}$ that are not in the set $\left\{k-1, k\right\}$, and replaces each occurrence of $k-1$ with $0$, and each occurrence of $k$ with $1$.
For example, if $k=4$ and $\bm{s} = 1324403$, then $v(\bm{s}) = 4$ and $\mathcal{F}(\bm{s}) = 0110$.
To define $\mathcal{F}$ formally, we first introduce the following notation.
For a $k$-resolution composite binary sequence $\bm{s} \in \Sigma_{k+1}^n(\ell)$, let $\mathcal{J}(\bm{s})$ denote the set of positions where $\bm{s}$ takes a value from the set $\{k-1, k\}$, that is,
\begin{equation*}
  \mathcal{J}(\bm{s}) \triangleq \left\{ 1 \leq j \leq n \ : \ \bm{s}[j] \in \left\{k-1, k\right\} \right\}.
\end{equation*}
The size of this set satisfies $|\mathcal{J}(\bm{s})| = v(\bm{s}) = \ell$. 
We write the elements of $\mathcal{J}(\bm{s})$  in increasing order as $\mathcal{J}(\bm{s}) = \left\{ j_1, j_2, \ldots, j_\ell \right\}$, where $j_1 < j_2 < \ldots < j_\ell$.
Then $\mathcal{F}(\bm{s})$ is defined as the following concatenation
\begin{equation*}
  \mathcal{F} (\bm{s}) \triangleq \left(\bm{s}[j_1] - (k-1) \right) \circ \left(\bm{s}[j_2] - (k-1) \right) \circ \ldots \circ \left(\bm{s}[j_\ell] - (k-1) \right).
\end{equation*}

We are now ready to construct a code $\mathcal{C} \subseteq \Sigma_{k+1}^n$ that is capable of correcting a single substitution error in the first channel.
\begin{construction}
  For each $0 \leq \ell \leq n$, let $\mathcal{C}(\ell)$ be a binary single-error-correcting code of length $\ell$. Let the code $\mathcal{C}_{\Romannum{2}}$ be defined as
  \begin{equation*}
    \mathcal{C}_{\Romannum{2}} \triangleq \bigcup_{\ell=0}^{n} \left\{ \bm{c} \in \Sigma_{k+1}^n(\ell) \ : \ \mathcal{F}(\bm{c}) \in \mathcal{C}(\ell) \right\}.
  \end{equation*}
\end{construction}

For $\ell = 0$, the sequence $\mathcal{F}(\bm{c})$ is empty, that is, $\mathcal{F}(\bm{c}) = \epsilon$.
We let $\mathcal{C}(0) = \left\{\epsilon\right\}$, so that the condition $\mathcal{F}(\bm{c}) \in \mathcal{C}(0)$ holds.

\begin{theorem}
  The code $\mathcal{C}_{\Romannum{2}}$ is a $(1, 0, \ldots, 0)$-CECC of length $n$.
\end{theorem}

\begin{IEEEproof}
  Let $\bm{c}$ be the transmitted codeword. Let $\bm{y}_0, \bm{y}_1, \ldots, \bm{y}_{k-1}$ be the received sequences from the $k$ channels, and
  let $\bm{y} = \mathcal{R}(\bm{y}_0, \bm{y}_1, \ldots, \bm{y}_{k-1})$ be the reconstructed sequence.
  If $\bm{y}$ contains the symbol $\mathord{?}$ at some position $j$, then this indicates that $\bm{y}_0[j]$ has flipped from $0$ to $1$.
  We can revert this bit to $0$, reconstruct the sequence again, and recover $\bm{c}$.
  Otherwise, $\bm{y}$ consists only of valid letters from $\Sigma_{k+1}$, and one of the following cases must have occurred.
  \begin{enumerate}
    \item A letter $k-1$ in $\bm{c}$ was changed to $k$ in $\bm{y}$. This corresponds to a $0$ in $\mathcal{F}(\bm{c})$ being flipped to a $1$ in $\mathcal{F}(\bm{y})$.
    \item A letter $k$ in $\bm{c}$ was changed to $k-1$ in $\bm{y}$. This corresponds to a $1$ in $\mathcal{F}(\bm{c})$ being flipped to a $0$ in $\mathcal{F}(\bm{y})$.
    \item No error occurred, so $\bm{y} = \bm{c}$, and therefore $\mathcal{F}(\bm{y}) = \mathcal{F}(\bm{c})$.
  \end{enumerate}
  In all cases, the number of letters in $\bm{y}$ from the set $\{k-1, k\}$ is the same as in $\bm{c}$, i.e., $v(\bm{y}) = v(\bm{c})$.
  Let $\ell = v (\bm{y})$. Then we can apply the decoder of $\mathcal{C}(\ell)$ to $\mathcal{F}(\bm{y})$ to recover $\mathcal{F}(\bm{c})$.
  If $\mathcal{F}(\bm{y}) = \mathcal{F}(\bm{c})$, then we are in case (3), and we immediately conclude that $\bm{y} = \bm{c}$.
  Otherwise, $\mathcal{F}(\bm{y})$ and $\mathcal{F}(\bm{c})$ differ at exactly one position, say position $j$.
  We can determine whether this is case (1) or (2) by comparing the bits at position $j$ in $\mathcal{F}(\bm{y})$ and $\mathcal{F}(\bm{c})$.
  Finally, we identify the $j$-th occurrence of a letter from the set $\{k-1, k\}$ in $\bm{y}$, modify it according to the difference between $\mathcal{F}(\bm{y})[j]$ and $\mathcal{F}(\bm{c})[j]$, 
  and thereby recover the original codeword $\bm{c}$.
\end{IEEEproof}

\begin{corollary}
  The cardinality of the code $\mathcal{C}_{\Romannum{2}}$ is given by
    \begin{equation*}
      |\mathcal{C}_{\Romannum{2}}| = \sum_{\ell=0}^{n} \binom{n}{\ell} (k-1)^{n-\ell} |\mathcal{C}(\ell)| = (k-1)^{n} \sum_{\ell=0}^{n} \binom{n}{\ell} \left(\frac{1}{k-1}\right)^{\ell} |\mathcal{C}(\ell)|.
    \end{equation*}
\end{corollary}

\begin{IEEEproof}
  For each $0 \leq \ell \leq n$, we can choose the $\ell$ positions in the codeword $\bm{c}$ where the letters $k-1$ or $k$ will appear.
  We can additionally choose the letters in the remaining $n - \ell$ positions from the set $\left\{0, 1, \ldots, k-2\right\}$.
  The number of such choices is given by $\binom{n}{\ell} \cdot (k-1)^{n-\ell}$.
\end{IEEEproof}

\begin{corollary}\label{cor:1-0-cecc-lower-bound}
  For binary single-error-correcting codes $\mathcal{C}(\ell)$ of length $\ell$ with size $|\mathcal{C}(\ell)| = 2^{\ell - \lceil\log_2(\ell + 1)\rceil}$, it holds that
  \begin{equation*}
    \mathcal{S}_{k}\left(n; (1, 0, \ldots, 0)\right) \geq |\mathcal{C}_{\Romannum{2}}| = (k-1)^{n} \sum_{\ell=0}^{n} \binom{n}{\ell} \left(\frac{1}{k-1}\right)^{\ell} 2^{\ell - \lceil\log_2(\ell + 1)\rceil}.
  \end{equation*}
\end{corollary}
To better understand the lower bound obtained in \cref{cor:1-0-cecc-lower-bound}, observe that $\frac{2^\ell}{2 (\ell + 1)} \leq 2^{\ell - \lceil\log_2(\ell + 1)\rceil} \leq \frac{2^\ell}{(\ell + 1)}$.
Note that
\begin{equation*}
  \begin{split}
    & (k-1)^{n} \sum_{\ell=0}^{n} \binom{n}{\ell} \left(\frac{1}{k-1}\right)^{\ell} \frac{2^{\ell}}{\ell+1} 
    = \frac{(k-1)^{n+1}}{2} \sum_{\ell=0}^{n} \binom{n}{\ell} \left(\frac{2}{k-1}\right)^{\ell+1} \frac{1}{\ell+1} \\
    & \overset{\text{(BI)}}{=} \frac{(k-1)^{n+1}}{2} \cdot \frac{\left( \left(\frac{k+1}{k-1}\right)^{n+1} - 1 \right)}{(n+1)}
    = \frac{(k+1)^{n+1} - (k-1)^{n+1}}{2(n+1)},
  \end{split}
\end{equation*}
where step $\overset{\text{(BI)}}{=}$ follows by applying a binomial identity from Appendix~\ref{appendix:binom-identities} at $x = \frac{2}{k-1}$.
Therefore,
\begin{equation*}
  \frac{(k+1)^{n+1} - (k-1)^{n+1}}{4(n+1)} \leq (k-1)^{n} \sum_{\ell=0}^{n} \binom{n}{\ell} \left(\frac{1}{k-1}\right)^{\ell} 2^{\ell - \lceil\log_2(\ell + 1)\rceil} \leq \frac{(k+1)^{n+1} - (k-1)^{n+1}}{2(n+1)}.
\end{equation*}

Note that the expression on the right-hand side coincides with the upper bound on the cardinality of $(1, 0, \ldots, 0)$-CECCs given in~\cref{thm:gspb-1-0-general}.
Remarkably, this construction is optimal.
As shown in the next theorem, when choosing optimal binary single-error-correcting codes $\mathcal{C}(\ell)$ we obtain
$\mathcal{S}_{k}\left(n; (1, 0, \ldots, 0)\right) = |\mathcal{C}_{\Romannum{2}}|$.

\begin{theorem} \label{thm:1-0-cecc-optimal}
  For optimal binary single-error-correcting codes $\mathcal{C}(\ell)$ of length $\ell$, it holds that
  \begin{equation*}
    \mathcal{S}_{k}\left(n; (1, 0, \ldots, 0)\right) = |\mathcal{C}_{\Romannum{2}}|.
  \end{equation*}
\end{theorem}

\begin{IEEEproof}
  Let $[ n ] \triangleq \{1, 2, \ldots, n\}$. For $\mathcal{J} \subseteq [n]$, let $\overline{\mathcal{J}} = [n]\setminus \mathcal{J}$ denote its complement.
  For a sequence $\bm{s}$ and a set of positions $\mathcal{J}$, denote by $\bm{s}_{\mathcal{J}}$ the restriction of $\bm{s}$ to the positions in $\mathcal{J}$.
  For any $\mathcal{J}\subseteq [n]$ of size $|\mathcal{J}|=\ell$ and $\bm{a}\in \Sigma_{k-1}^{n-\ell}$, define the fiber
  \begin{equation*}
    \text{Fib}(\mathcal{J}, \bm{a}) \triangleq \left\{ \bm{s} \in \Sigma_{k+1}^n \ : \ \mathcal{J}(\bm{s}) = \mathcal{J}, \ \bm{s}_{\overline{\mathcal{J}}} = \bm{a} \right\}.
  \end{equation*}
  This fiber fixes the positions of the letters from $\{k-1, k\}$ to the set $\mathcal{J}$, while the entries at the remaining positions $\overline{\mathcal{J}}$ are fixed to $\bm{a}$.
  For any fixed fiber $\text{Fib}(\mathcal{J},\bm{a})$ with $|\mathcal{J}|=\ell$, the map $\mathcal{F}$ restricted to this fiber is a bijection onto $\{0,1\}^\ell$.
  A single error in the first channel only toggles $k-1 \leftrightarrow k$ at a single position.
  Hence both $\mathcal{J}$ and $\bm{a}$ are invariants under these errors, and the error corresponds exactly to a single bit flip in $\{0,1\}^\ell$ under the bijection $\mathcal{F}$.
  Thus, if $\mathcal{C}$ is a $(1,0,\ldots,0)$-CECC, then $\mathcal{C}\cap \text{Fib}(\mathcal{J},\bm{a})$ must be a binary single-error-correcting code of length $\ell$. 
  By the optimality of $\mathcal{C}(\ell)$, we therefore have
  \begin{equation*}
    | \mathcal{C} \cap \text{Fib}(\mathcal{J}, \bm{a}) | \leq |\mathcal{C}(\ell)|.
  \end{equation*}
  Summing over all fibers yields
  \begin{equation*}
    |\mathcal{C}| \leq \sum_{\ell=0}^n \sum_{\substack{\mathcal{J} \subseteq [ n ] \\ |\mathcal{J}| = \ell}} \sum_{\bm{a} \in \Sigma_{k-1}^{n-\ell}} | \mathcal{C} \cap \text{Fib}(\mathcal{J}, \bm{a}) | \leq \sum_{\ell=0}^n \sum_{\substack{\mathcal{J} \subseteq [ n ] \\ |\mathcal{J}| = \ell}} \sum_{\bm{a} \in \Sigma_{k-1}^{n-\ell}} |\mathcal{C}(\ell)|.
  \end{equation*}
  In each fiber, the construction $\mathcal{C}_\Romannum{2}$ consists exactly of the words that map under $\mathcal{F}$ to an optimal binary single-error-correcting code $\mathcal{C}(\ell)$.
  The construction ranges over all fibers, namely over every $\ell$, every $\mathcal{J}\subseteq [n]$ with $|\mathcal{J}|=\ell$, and every $\bm{a}\in \Sigma_{k-1}^{n-\ell}$, so no fiber is omitted.
  Therefore \(|\mathcal{C}_{\Romannum{2}}|\) attains the bound with equality.
\end{IEEEproof}


We now turn our attention to the case of $k$-resolution $1$-CECCs.
As illustrated in Figure~\ref{fig:error-channel-1-cecc-general}, under the assumption of a single substitution error occurring in any of the $k$ channels, the following transformations are possible.
\begin{itemize}
  \item A letter $\sigma \in \left\{1, \ldots, k-1\right\}$ may be transformed to $\sigma \pm 1$.
  \item The letter $0$ may be transformed to the letter $1$, or the letter $k$ may be transformed to the letter $k-1$.
  \item Any letter may be transformed into the invalid symbol $\mathord{?}$.
\end{itemize}
Unlike the case of $(1, 0, \ldots, 0)$-CECCs, where the invalid symbol $\mathord{?}$ can always be corrected without any error-correcting code,
in the case of $k$-resolution $1$-CECCs, the symbol $\mathord{?}$ can only be corrected without coding in certain specific cases.

The first two types of transformations resemble symmetric limited magnitude errors of magnitude $\ell = 1$.
A \emph{$q$-ary symmetric single-limited-magnitude-error-correcting code of magnitude $\ell=1$} is a code that can correct a single substitution error where a letter $\sigma \in \Sigma_q$ may be altered to $\sigma \pm 1 \mod q$.
These codes are equivalent to codes in the Lee metric with minimum distance $d_{\mathcal{L}} = 3$.
The key distinction, however, is that $k$-resolution $1$-CECCs do not permit circular transformations: the letter $0$ cannot be changed to the letter $k$, nor can the letter $k$ be changed to the letter $0$.
In contrast, the symmetric limited-magnitude-error-correcting codes and Lee metric codes allow such wrap-around errors.
The following theorem demonstrates that it is possible to address the invalid symbol $\mathord{?}$ and still use a symmetric single-limited-magnitude-error-correcting code of magnitude $\ell = 1$, 
or alternatively a code with Lee distance at least 3, to correct a single substitution error occurring in any of the $k$ channels.

\begin{theorem}
  Let $\mathcal{C} \subseteq \Sigma_{k+1}^n$ be a $(k+1)$-ary symmetric single-limited-magnitude-error-correcting code of magnitude $\ell = 1$ of length $n$. Then $\mathcal{C}$ is a $k$-resolution $1$-CECC.
\end{theorem}

\begin{IEEEproof}
  By definition, a $(k+1)$-ary symmetric single-limited-magnitude-error-correcting code of magnitude $\ell = 1$ can correct a substitution of a letter $\sigma \in \Sigma_{k+1}$ to $\sigma \pm 1$.
  Therefore, $\mathcal{C}$ can correct the first two types of transformations described earlier.
  It remains to show how the invalid symbol $\mathord{?}$ can be handled.

  Suppose that a letter $\sigma$ is transformed into $\mathord{?}$. Let $\bm{v}$ be the binary column vector representation of this $\mathord{?}$.
  By the structure of the decomposition mapping, $\bm{v}$ must contain a $1$ in some row $i$ and a $0$ in row $i+1$, for some $0 \leq i < k-1$, namely, $\bm{v} = [\cdots 1 0 \cdots]^\intercal$.
  This implies a violation of the non-decreasing property, and we must determine whether the error occurred in row $i$ (flipping a $0$ to $1$) or in row $i+1$ (flipping a $1$ to $0$).
  Since a single substitution error occurred, we distinguish the cases as follows.
  \begin{itemize}
    \item  If $i > 0$ and the bit in row $i-1$ of $\bm{v}$ is $1$, then the bits in rows $i-1$, $i$, and $i+1$ form the pattern $110$, that is $\bm{v} = [\cdots 1 1 0 \cdots]^\intercal$.
    In this case, the only valid explanation is that the bit in row $i+1$ flipped from $1$ to $0$. We revert it to $1$ to yield $[\cdots 1 1 1 \cdots]^\intercal$ and this allows to recover the original letter $\sigma$.
    \item If $i < k-2$ and the bit in row $i+2$ of $\bm{v}$ is $0$, then the bits in rows $i$, $i+1$, and $i+2$ form the pattern $100$, that is $\bm{v} = [\cdots 1 0 0 \cdots]^\intercal$.
    This implies that the bit in row $i$ flipped from $0$ to $1$. We can then revert it to $0$ to yield $[\cdots 0 0 0 \cdots]^\intercal$ and this allows to recover the original letter $\sigma$.
    \item Otherwise $\bm{v} = [0^{i}101^{k-i-2}]$. We can swap the bits in rows $i$ and $i+1$ of $\bm{v}$, changing the pattern $10$ to $01$, and denote the resulting vector by $\bm{w} = [0^{i}011^{k-i-2}]$.
    The vector $\bm{w}$ corresponds to a valid decomposition of a letter in $\Sigma_{k+1}$, as the non-decreasing violation is resolved.
    Furthermore, observe that the vector $\bm{w}$ must represent either $\sigma-1$ (if the error occurred in row $i+1$ and the original pattern was $11$) or $\sigma+1$ (if the error occurred in row $i$ and the original pattern was $00$).
    Therefore, the symmetric single-limited-magnitude-error-correcting code $\mathcal{C}$ can be used to recover the original letter $\sigma$.
  \end{itemize}
\end{IEEEproof}

In the following corollary, we provide a lower bound on the cardinality of $k$-resolution $1$-CECCs, obtained from the cardinality of a Lee metric code with distance at least 3.
\begin{corollary}
  For any code length $n$ and even resolution parameter $k$, it holds that
  \begin{equation*}
    \mathcal{S}_k\left(n; 1\right) \geq \frac{(k+1)^n}{\left(k+1\right)^{\lceil \log_{k+1} (2n +1) \rceil}}.
  \end{equation*}
\end{corollary}

\begin{IEEEproof}
  By the previous theorem, we may use any error-correcting code with Lee distance $d_{\mathcal{L}} \geq 3$.
  A well-known example is the Berlekamp code~\cite{Berlekamp}.
  A more general construction appears in Problem 10.13 of~\cite{Roth}, which applies to alphabets of arbitrary odd size.
  Since $k$ is even, the alphabet size $k+1 > 2$ is odd.
  According to this construction, for $m = \lceil \log_{k+1}(2n + 1) \rceil$, there exists a code $\mathcal{C}$ of length
  $\ell = \frac{1}{2} \left( (k+1)^m - 1 \right)$
  with Lee distance at least 3 and redundancy $m$. By shortening this code to length $n$, we obtain a code $\mathcal{C}' \subseteq \Sigma_{k+1}^n$ of length $n$, with the same Lee distance and redundancy.
  The cardinality of $\mathcal{C}'$ provides a lower bound on the cardinality of $k$-resolution $1$-CECCs.
  \begin{equation*}
    \mathcal{S}_k\left(n; 1\right) \geq |\mathcal{C}'| = (k+1)^{n-m} = \frac{(k+1)^n}{\left(k+1\right)^{\lceil \log_{k+1} (2n +1) \rceil}}.
  \end{equation*}
\end{IEEEproof}

Finally, observe that when $n = \frac{1}{2} \left( (k+1)^m - 1 \right)$, the code $\mathcal{C}'$ satisfies
\begin{equation*}
  |\mathcal{C}'| = \frac{(k+1)^n}{(2n+1)},
\end{equation*}
while the upper bound on the cardinality of $k$-resolution $1$-CECCs is given in Theorem~\ref{thm:gspb-1-cecc-general} as
\begin{equation*}
   \frac{(k+1)^n}{\frac{2kn}{k+1} - 1}.
\end{equation*}

This gap can be attributed to the structural difference in the error models. In the Lee metric, each letter $\sigma \in \Sigma_{k+1}$ can undergo exactly two transformations, to $\sigma \pm 1$ respectively, with wrap-around at the boundaries.
In contrast, under the $k$-resolution $1$-CECC model, each letter $\sigma \in \{1, \ldots, k-1\}$ also allows two transformations to $\sigma \pm 1$, but the boundary letters $0$ and $k$ allow only a single transformation each (to $1$ and $k-1$, respectively).
Therefore, the average number of allowed error transformations per letter in $\Sigma_{k+1}$ is $\frac{2(k-1) + 1 + 1}{k+1} = \frac{2k}{k+1}$, explaining the gap.


\section{Deletions} \label{sec:deletions}

Unlike conventional storage mediums, which primarily suffer from substitution and erasure errors, DNA data storage is also prone to insertion and deletion errors.
In this section, we examine the ordered composite DNA channel over the binary alphabet, i.e., $q=2$ with composite letters of resolution $k=2$, focusing on deletions errors.
The channel model remains as illustrated in Figure \ref{fig:channel}, but the underlying channels are now binary deletion channels rather than binary substitution channels.
An immediate complication is that the received sequences $\bm{y}_0$ and $\bm{y}_1$ may have different lengths. 
As a result, the reconstruction mapping $\mathcal{R}$ is not well-defined.
To address this, we consider error correction prior to reconstruction, that is, we first correct the errors in $\bm{y}_0$ and $\bm{y}_1$ before reconstructing the sequence $\bm{y}$.
Additionally, we define respective composite error balls to account for deletions.
Before doing so, let us revisit the definitions and propositions introduced in the Section~\ref{sec:preliminaries} and examine how they can be adapted to the case of deletions.
For this section only, we adopt the following definitions of composite-deletion-correcting codes.
Since $k = 2$ is assumed throughout this section, we simplify the notation whenever possible.

\begin{definition}
  An $(e_0, e_1)$-composite-deletion-correcting code ($(e_0, e_1)$-CDCC) $\mathcal{C}$ is a code that can correct up to $e_0$ deletion errors in $\bm{s}_0$
  (introduced by the first channel) and up to $e_1$ deletion errors in $\bm{s}_1$ (introduced by the second channel).
\end{definition}

\begin{definition}
  An $e$-composite-deletion-correcting code ($e$-CDCC) $\mathcal{C}$ is a code that can correct up to $e$ deletion errors in total, introduced by the two channels together.
\end{definition}

We denote the largest cardinality of these codes as $\mathcal{S}_{\mathsf{D}}\left(n; (e_0, e_1)\right)$ and $\mathcal{S}_{\mathsf{D}}\left(n; e\right)$, to distinguish the deletion case from the substitution case.
As in the preceding parts of the paper, a $2$-resolution composite binary sequence $\bm{s}$ of length $\ell$, or simply a composite binary sequence of length $\ell$, is represented as a ternary sequence $\bm{s} \in \mathcal{X}_2^\ell \triangleq \Sigma_3^\ell$.

We now revisit the propositions from Section~\ref{sec:preliminaries} and examine whether they apply to the case of deletions.
\cref{prop:k-ary} states that a ternary error-correcting code capable of correcting $e$ substitution errors would also be a $2$-resolution $e$-CECC.
This, however, does not carry over to deletions.
Unlike in the substitution case, a single deletion can lead to multiple errors in the reconstructed ternary sequence when reconstruction is attempted from two sequences of different lengths.
\cref{prop:connection} establishes a relation between $\mathcal{S}_2\left(n; (e_0, e_1)\right)$ and $\mathcal{S}_2\left(n; e\right)$, stating that for any non-negative integers
$e_0, e_1$, we have $\mathcal{S}_2\left(n; e_0 + e_1\right) \leq \mathcal{S}_2\left(n; (e_0, e_1)\right)$. 
This inequality holds independently of the error model. 
A code that corrects $e_0+e_1$ deletions in total also corrects the case where the deletions are distributed as $(e_0, e_1)$. Therefore 
$\mathcal{S}_{\mathsf{D}}\left(n; e_0+e_1\right) \leq \mathcal{S}_{\mathsf{D}}\left(n; (e_0, e_1)\right)$.
For resolution parameter $k=2$, note that \cref{prop:reversal} also implies \cref{prop:swap-single-error}.
\cref{prop:reversal} states that $\mathcal{S}_2\left(n; (e_0, e_1)\right) = \mathcal{S}_2\left(n; (e_1, e_0)\right)$.
The same holds for deletions, as it follows from the symmetry of the channel model. 
The proof is the same, with substitutions replaced by deletions.
Thus, $\mathcal{S}_{\mathsf{D}}\left(n; (e_0, e_1)\right) = \mathcal{S}_{\mathsf{D}}\left(n; (e_1, e_0)\right)$.

We now limit our focus to the case of a single deletion error. 
Since $\mathcal{S}_{\mathsf{D}}\left(n; (1, 0)\right) = \mathcal{S}_{\mathsf{D}}\left(n; (0, 1)\right)$, it suffices to consider $(1, 0)$-CDCCs and $1$-CDCCs.
We begin by deriving upper bounds on the cardinality of these codes and then proceed to establish lower bounds through explicit constructions.


\subsection{Upper Bounds}

In this section, we establish upper bounds on the cardinality of $(1, 0)$-CDCCs and $1$-CDCCs. As in the case of substitution errors, 
we derive these bounds using the GSPB~\cite{gspb}. 
This approach is necessary because, as we will see, the size of the composite error ball under the deletion model also depends on the center sequence, and the smallest such ball has constant size, 
which renders the standard sphere packing bound ineffective.

We say that a sequence $\bm{x}$ is a \emph{subsequence} of $\bm{y}$ if $\bm{x}$ can be obtained from $\bm{y}$ by deleting zero or more letters.
Correspondingly, we say that $\bm{y}$ is a \emph{supersequence} of $\bm{x}$.
Unlike substitution errors, the length of the channel output sequence is $n$ when no deletion occurs and $n - 1$ when a single deletion error occurs.
This variability complicates the definition of composite error balls under the deletion model.
To simplify the analysis we assume that a deletion error always occurs so the output sequence length is always $n - 1$.

Let $\bm{s} \in \mathcal{X}_2^n$ be a composite binary sequence of length $n$, and let $\bm{s}_0, \bm{s}_1$ be its binary decomposed sequences, that is, $\mathcal{D}(\bm{s}) = (\bm{s}_0, \bm{s}_1)$.
We define $\mathcal{B}_{(1, 0)}^{\mathsf{D}}(\bm{s})$ and $\mathcal{B}_{(0, 1)}^{\mathsf{D}}(\bm{s})$ as the sets of channel outputs obtained from a single deletion error in the first and second channel, respectively. That is,
\begin{equation*}
  \begin{split}
    \mathcal{B}_{(1, 0)}^{\mathsf{D}}(\bm{s}) & \triangleq \left\{ (\bm{y}_0, \bm{s}_1) \ : \ \bm{y}_0 \in \left\{0, 1\right\}^{n-1}, \ \bm{y}_0 \ \text{is a subsequence of} \ \bm{s}_0 \right\}, \\
    \mathcal{B}_{(0, 1)}^{\mathsf{D}}(\bm{s}) & \triangleq \left\{ (\bm{s}_0, \bm{y}_1) \ : \ \bm{y}_1 \in \left\{0, 1\right\}^{n-1}, \ \bm{y}_1 \ \text{is a subsequence of} \ \bm{s}_1 \right\}.
  \end{split}
\end{equation*}

We define $\mathcal{B}_1^{\mathsf{D}}(\bm{s})$ to be the set of channel outputs that can be obtained from a single deletion error occurring in any of the two channels, but not both, i.e.,
\begin{equation*}
    \mathcal{B}_1^{\mathsf{D}}(\bm{s}) \triangleq \mathcal{B}_{(1, 0)}^{\mathsf{D}}(\bm{s}) \cup \mathcal{B}_{(0, 1)}^{\mathsf{D}}(\bm{s}).
\end{equation*}

As mentioned earlier it is sufficient to consider $(1, 0)$-CDCC since $\mathcal{S}_{\mathsf{D}}\left(n; (1, 0)\right) = \mathcal{S}_{\mathsf{D}}\left(n; (0, 1)\right)$.
We therefore restrict our analysis to $(1, 0)$-CDCC, while noting that all subsequent results can be directly adapted to $(0, 1)$-CDCC.
The sets $\mathcal{B}_{(1, 0)}^{\mathsf{D}}(\bm{s})$ and $\mathcal{B}_{1}^{\mathsf{D}}(\bm{s})$ are referred to as the \emph{deletion composite error balls of radius $(1, 0)$ and $1$}, respectively.
A code $\mathcal{C} \subseteq \mathcal{X}_{2}^{n}$ is a $(1, 0)$-CDCC if the deletion composite error balls of radius $(1, 0)$ centered at any two distinct codewords are disjoint, that is, for all distinct codewords $\bm{c}, \bm{c}' \in \mathcal{C}$,
\begin{equation*}
  \mathcal{B}_{(1, 0)}^{\mathsf{D}}(\bm{c}) \cap \mathcal{B}_{(1, 0)}^{\mathsf{D}}(\bm{c}') = \emptyset.
\end{equation*}
Similarly, a code $\mathcal{C} \subseteq \mathcal{X}_{2}^{n}$ is a $1$-CDCC if for all distinct codewords $\bm{c}, \bm{c}' \in \mathcal{C}$, it holds
\begin{equation*}
  \mathcal{B}_{1}^{\mathsf{D}}(\bm{c}) \cap \mathcal{B}_{1}^{\mathsf{D}}(\bm{c}') = \emptyset.
\end{equation*}

To compute the sizes of the deletion composite error balls we first introduce the following notation.
For a binary sequence $\bm{x} \in \{0, 1\}^n$, let $\rho(\bm{x})$ denote the number of runs in $\bm{x}$.
For example, if $\bm{x} = 001010010$, then $\rho(\bm{x}) = 7$.
The size of a deletion composite error ball depends on the composite binary sequence and is stated in the following proposition.

\begin{proposition}\label{prop:deletion-composite-error-balls}
  Let $\bm{s} \in \mathcal{X}_2^n$ be a composite binary sequence with decomposition $\mathcal{D}(\bm{s}) = (\bm{s}_0, \bm{s}_1)$. Then,
  \begin{equation*}
    |\mathcal{B}_{(1, 0)}^{\mathsf{D}}(\bm{s})| = \rho{(\bm{s}_0)} \quad \text{and} \quad |\mathcal{B}_1^{\mathsf{D}}(\bm{s})| = \rho{(\bm{s}_0)} + \rho{(\bm{s}_1)}.
  \end{equation*}
\end{proposition}

\begin{IEEEproof}
  The number of elements in $\mathcal{B}_{(1, 0)}^{\mathsf{D}}(\bm{s})$ equals to the number of subsequences $\bm{y}_0$ of $\bm{s}_0$ of length $n-1$, which is exactly the number of runs in $\bm{s}_0$.
  Hence $|\mathcal{B}_{(1, 0)}^{\mathsf{D}}(\bm{s})| = \rho{(\bm{s}_0)}$. Similarly, $|\mathcal{B}_{(0, 1)}^{\mathsf{D}}(\bm{s})| = \rho(\bm{s}_1)$.
  Finally, note that the sets $\mathcal{B}_{(1, 0)}^{\mathsf{D}}(\bm{s})$ and $\mathcal{B}_{(0, 1)}^{\mathsf{D}}(\bm{s})$ are disjoint, 
  therefore the size of the union is simply the sum of their sizes.
\end{IEEEproof}

Note that for the composite binary sequence $\bm{s} = \bm{0}$ the decomposed sequences are the all-zero sequences $\bm{s}_0 = \bm{s}_1 = \bm{0}$.
In this case $|\mathcal{B}_{(1, 0)}^{\mathsf{D}}(\bm{s})| = \rho(\bm{s}_0) = 1$ and $|\mathcal{B}_1^{\mathsf{D}}(\bm{s})| = 2$.
Hence a direct application of the sphere packing bound based on the minimal size of the deletion composite error ball is not effective.
We thus resort to the GSPB.

The first step is to define the hypergraphs that represent the model.
Following the approach used for substitution errors, we associate a hypergraph with each family of composite-deletion-correcting codes.
The vertex sets consist of all pairs of binary sequences that can occur as channel outputs under the deletion restrictions of the family.
The hyperedges are given by the corresponding deletion composite error balls. Formally,
\begin{equation*}
  \begin{split}
    \mathcal{H}_{\mathsf{D}}(1, 0) & \triangleq \left(\mathcal{X}_{(1, 0)}, \left\{ \mathcal{B}_{(1, 0)}^{\mathsf{D}}(\bm{s}) \ : \ \bm{s} \in \mathcal{X}_2^n \right\} \right), \\
    \mathcal{H}_{\mathsf{D}}(1) & \triangleq \left(\mathcal{X}_1, \left\{ \mathcal{B}_{1}^{\mathsf{D}}(\bm{s}) \ : \ \bm{s} \in \mathcal{X}_2^n \right\} \right), \\
  \end{split}
\end{equation*}
where
\begin{equation*}
  \begin{split}
    \mathcal{X}_{(1, 0)} & \triangleq \left\{ (\bm{y}_0, \bm{s}_1) \in \{0, 1\}^{n-1} \times \{0, 1\}^n \ : \ \exists \bm{s} \in \mathcal{X}_2^n, \ (\bm{y}_0, \bm{s}_1) \in \mathcal{B}_{(1, 0)}^{\mathsf{D}}(\bm{s}) \right\}, \\
    \mathcal{X}_{(0, 1)} & \triangleq \left\{ (\bm{s}_0, \bm{y}_1) \in \{0, 1\}^{n} \times \{0, 1\}^{n-1} \ : \ \exists \bm{s} \in \mathcal{X}_2^n, \ (\bm{s}_0, \bm{y}_1) \in \mathcal{B}_{(0, 1)}^{\mathsf{D}}(\bm{s}) \right\}, \\
    \mathcal{X}_{1} & \triangleq \mathcal{X}_{(1, 0)} \cup \mathcal{X}_{(0, 1)}.
  \end{split}
\end{equation*}

The definitions of the vertex sets are declarative and their cardinalities are not immediately clear.
To gain insight into the structure of $\mathcal{X}_{(1, 0)}$ and enable the computation of GSPB, we introduce a few auxiliary objects.

Given a binary sequence $\bm{y}_0 \in \{0, 1\}^{n-1}$ of Hamming weight $w$, let $\mathcal{V}(n; w)$ denote the number of binary sequences $\bm{s}_1 \in \{0, 1\}^n$
such that $(\bm{y}_0, \bm{s}_1) \in \mathcal{X}_{(1, 0)}$.
\cref{prop:value-v} provides a closed-form expression for $\mathcal{V}(n; w)$ with its proof given in Appendix~\ref{appendix:deletions}.

We remind that the existence of a composite binary sequence $\bm{s} \in \mathcal{X}_2^n$ satisfying $\mathcal{D}(\bm{s}) = (\bm{s}_0, \bm{s}_1)$ is equivalent to the condition $\bm{s}_0 \leq \bm{s}_1$,
where the inequality is taken component-wise. In other words, for a binary sequence $\bm{y}_0 \in \{0, 1\}^{n-1}$ of Hamming weight $w$, 
$\mathcal{V}(n; w)$ counts the number of distinct binary sequences $\bm{s}_1 \in \{0, 1\}^n$ such that $\bm{s}_0 \leq \bm{s}_1$, where $\bm{s}_0$ is a supersequence of $\bm{y}_0$ of length $n$.
Given a binary sequence $\bm{x} \in \{0, 1\}^{n-1}$, let $\mathcal{I}_1(\bm{x})$ denote the set of all supersequences of $\bm{x}$ that can be obtained by inserting a single bit into $\bm{x}$, that is, 
\begin{equation*}
  \mathcal{I}_1(\bm{x}) \triangleq \left\{ \bm{y} \in \{0, 1\}^{n} \ : \ \bm{y} \ \text{is a supersequence of} \ \bm{x} \right\}.
\end{equation*}

\begin{restatable}{proposition}{valueOfV}\label{prop:value-v}
  Let $\bm{y}_0 \in \{0, 1\}^{n-1}$ be a binary sequence of Hamming weight $w$. The number of distinct binary sequences $\bm{s}_1 \in \{0, 1\}^n$ such that 
  there exists $\bm{s}_0 \in \mathcal{I}_1(\bm{y}_0)$ and $\bm{s}_0 \leq \bm{s}_1$ is given by 
  \begin{equation*}
    \mathcal{V}(n; w) = 2^{n-w} + w \cdot 2^{n-w-1}.
  \end{equation*}
\end{restatable}

This result enables the computation of the cardinality of the vertex set $\mathcal{X}_{(1, 0)}$ by iterating over all possible Hamming weights $w$.  
The cardinality is given in the following proposition, with its proof provided in Appendix~\ref{appendix:deletions}.

\begin{restatable}{proposition}{vertexSetSize}\label{prop:vertex-set-size}
  The vertex set $\mathcal{X}_{(1, 0)}$ has cardinality $|\mathcal{X}_{(1, 0)} | = 2 \cdot 3^{n-1} + (n-1) \cdot 3^{n-2}.$
\end{restatable}

Observe that the cardinality of the vertex set $\mathcal{X}_{(0, 1)}$ exceeds the total number of composite binary sequences, $|\mathcal{X}_2^n| = 3^n$, for all $n \geq 4$.
This observation is not immediately evident from the definitions, as each composite binary sequence can correspond to multiple vertices in the hypergraph. 
However, the same vertex can arise from multiple distinct composite binary sequences.

We now construct a fractional transversal in the hypergraph  $\mathcal{H}_{\mathsf{D}}(1, 0)$. 
For each vertex $\left(\bm{y}_0, \bm{s}_1\right) \in \mathcal{X}_{(1, 0)}$, we assign the weight
\begin{equation*}
  w _{\left(\bm{y}_0, \bm{s}_1\right)} \triangleq \frac{1}{\rho(\bm{y}_0)}.
\end{equation*}

We show that the assigned weights constitute a valid fractional transversal.
It suffices to verify that the sum of weights over each hyperedge is at least 1.
Recall that the hyperedges are the deletion composite error balls of radius $(1, 0)$, namely $\mathcal{B}_{(1, 0)}^{\mathsf{D}}(\bm{s})$.
Note that if $\bm{y}_0$ is a subsequence of $\bm{s}_0$ then $\rho(\bm{y}_0) \leq \rho(\bm{s}_0)$, which implies $\frac{1}{\rho(\bm{y}_0)} \geq \frac{1}{\rho(\bm{s}_0)}$.
Therefore, for every hyperedge $\mathcal{B}_{(1, 0)}^{\mathsf{D}}(\bm{s})$, we have
\begin{equation*}
  \sum_{(\bm{y}_0, \bm{s}_1) \in \mathcal{B}_{(1, 0)}^{\mathsf{D}}(\bm{s})} \frac{1}{\rho(\bm{y}_0)} \geq \sum_{(\bm{y}_0, \bm{s}_1) \in \mathcal{B}_{(1, 0)}^{\mathsf{D}}(\bm{s})} \frac{1}{\rho(\bm{s}_0)} = 
  \frac{1}{\rho(\bm{s}_0)} \cdot \sum_{(\bm{y}_0, \bm{s}_1) \in \mathcal{B}_{(1, 0)}^{\mathsf{D}}(\bm{s})} 1 = \frac{1}{\rho(\bm{s}_0)} \cdot \rho(\bm{s}_0) = 1.
\end{equation*}

Since the fractional transversal is dependent on the number of runs in $\bm{y}_0$ for each vertex $\left(\bm{y}_0, \bm{s}_1\right)$, 
we need to be able to iterate the vertices based on the number of runs in $\bm{y}_0$.
The following proposition helps us understand the number of binary sequences of length $n$ with a given number of runs $\rho$ and Hamming weight $w$.
Its proof can be found in Appendix~\ref{appendix:deletions}.

\begin{restatable}{proposition}{numBinarySequences}
  The number of binary sequences of length $n$ with $\rho$ runs and Hamming weight $w$ is given by
  \begin{equation*}
  \mathcal{N}(n; \rho; w) = \begin{cases}
    1 & \text{if } \rho = 1 \ \text{and } \left(w = 0 \ \text{or } w = n \right) \\
    0 & \text{if } \rho = 1 \ \text{and } 0 < w < n \\
    \binom{w-1}{\lceil\frac{\rho}{2}\rceil -1} \binom{n-w-1}{\lfloor \frac{\rho}{2} \rfloor - 1} + \binom{w-1}{\lfloor\frac{\rho}{2}\rfloor -1} \binom{n-w-1}{\lceil \frac{\rho}{2} \rceil - 1} & \text{if } \rho \geq 2  \ \text{and } 0 < w < n \\
  \end{cases}.
  \end{equation*}
\end{restatable}

We are now prepared to formally derive an upper bound on the cardinality of $(1, 0)$-CDCCs using the GSPB.
Recall that the GSPB asserts
  \begin{equation*}
     \mathcal{S}_{\mathsf{D}}\left(n; (1, 0)\right) \leq \tau^{*}\left(\mathcal{H}_{\mathsf{D}}(1, 0)\right) \leq \sum _{(\bm{y}_0, \bm{s}_1) \in \mathcal{X}_{(1, 0)}} w_{(\bm{y}_0, \bm{s}_1)}.
  \end{equation*}

\begin{theorem}\label{thm:del_upper_1-0}
  For any code length $n$, it holds that
  \begin{equation*}
    \mathcal{S}_{\mathsf{D}}\left(n; (1, 0)\right) \leq \sum_{\rho=1} \sum_{w=0}^{n-1} \frac{\mathcal{N}(n-1; \rho; w) \cdot \mathcal{V}(n; w)}{\rho}.
  \end{equation*}
\end{theorem}

\begin{IEEEproof}
  We iterate over all the vertices $(\bm{y}_0, \bm{s}_1) \in \mathcal{X}_{(1, 0)}$ based on the number of runs $\rho$ in $\bm{y}_0$ and the Hamming weight $w$ of $\bm{y}_0$.
  Each such $\bm{y}_0$ is associated with $\mathcal{V}(n; w)$ vertices in $\mathcal{X}_{(1, 0)}$, each contributing a weight of $\frac{1}{\rho}$.
  The total number of such $\bm{y}_0$ sequences is given by $\mathcal{N}(n-1; \rho; w)$, yielding
  \begin{equation*}
    \mathcal{S}_{\mathsf{D}}\left(n; (1, 0)\right) \leq \sum _{(\bm{y}_0, \bm{s}_1) \in \mathcal{X}_{(1, 0)}} w_{(\bm{y}_0, \bm{s}_1)} = \sum _{(\bm{y}_0, \bm{s}_1) \in \mathcal{X}_{(1, 0)}} \frac{1}{\rho(\bm{y}_0)} = \sum_{\rho=1} \sum_{w=0}^{n-1} \frac{\mathcal{N}(n-1; \rho; w) \cdot \mathcal{V}(n; w)}{\rho}.
  \end{equation*}
\end{IEEEproof}

We now turn our attention to the case of $1$-CDCCs.
Owing to the symmetry between $(1, 0)$-CDCC and $(0, 1)$-CDCC, the corresponding vertex sets satisfy $|\mathcal{X}_{(0, 1)}| = |\mathcal{X}_{(1, 0)}|$.
The vertex sets $\mathcal{X}_{(1, 0)}$ and $\mathcal{X}_{(0, 1)}$ are disjoint, therefore $|\mathcal{X}_1| = 2 \cdot |\mathcal{X}_{(1, 0)}|$.
An immediate upper bound on the cardinality of $1$-CDCCs follows from the observation that every $1$-CDCC is also a $(1,0)$-CDCC. Thus,
\begin{equation*}
  \mathcal{S}_{\mathsf{D}}\left(n; 1\right) \leq \mathcal{S}_{\mathsf{D}}\left(n; (1, 0)\right) \leq \sum_{\rho=1} \sum_{w=0}^{n-1} \frac{\mathcal{N}(n-1; \rho; w) \cdot \mathcal{V}(n; w)}{\rho}.
\end{equation*}
At this stage, we have not identified a fractional transversal in the hypergraph $\mathcal{H}_{\mathsf{D}}(1)$ that would yield a tighter upper bound.


As shown in \cref{prop:deletion-composite-error-balls}, the size of a deletion composite error ball depends on its center composite binary sequence.
Analogously to the case of substitution errors, we compute the average deletion composite error ball sizes and the corresponding average sphere packing values.
Although these quantities do not in general provide valid upper bounds on the code cardinality, they serve as useful benchmarks for comparison, particularly when the upper bounds are expressed as summations.

The \emph{average sizes} of a deletion composite error ball of radius $(1, 0)$ and of radius $1$ are defined, respectively, as
\begin{equation*}
  \bar{\Delta}_{(1, 0)}^{\mathsf{D}} \triangleq \frac{1}{|\mathcal{X}_2^n|} \sum_{\bm{s} \in \mathcal{X}_2^n} |\mathcal{B}_{(1, 0)}^{\mathsf{D}} (\bm{s})| \qquad \text{and} \qquad \bar{\Delta}_1^{\mathsf{D}}  \triangleq \frac{1}{|\mathcal{X}_2^n|} \sum_{\bm{s} \in \mathcal{X}_2^n} |\mathcal{B}_1^{\mathsf{D}} (\bm{s})|.
\end{equation*}
The corresponding \emph{average sphere packing values} are given by
\begin{equation*}
  \text{ASPV}_{\mathsf{D}}(1, 0) \triangleq \frac{|\mathcal{X}_2^n|}{\bar{\Delta}_{(1, 0)}^{\mathsf{D}}} \qquad \text{and} \qquad \text{ASPV}_{\mathsf{D}}(1) \triangleq \frac{|\mathcal{X}_2^n|}{\bar{\Delta}_1^{\mathsf{D}}}.
\end{equation*}

The following theorem gives the main result, and its proof is provided in Appendix~\ref{appendix:deletions}.

\begin{restatable}{theorem}{deletionAverageBallSize}
  The average sizes of the deletion composite error balls of radius $(1, 0)$ and $1$ are given by
  \begin{equation*}
    \bar{\Delta}_{(1, 0)}^{\mathsf{D}} = 1 + \frac{4}{9} (n-1) \quad \text{and} \quad \bar{\Delta}_1^{\mathsf{D}} = 2 + \frac{8}{9} (n-1),
  \end{equation*}
  respectively.
\end{restatable}

The corresponding average sphere packing values for the deletion composite error balls with radii $(1, 0)$ and $1$ become
\begin{equation*}
  \text{ASPV}_{\mathsf{D}}(1, 0) = \frac{3^n}{1 + \frac{4}{9} (n-1)} \quad \text{and} \quad \text{ASPV}_{\mathsf{D}}(1) = \frac{3^n}{2 + \frac{8}{9} (n-1)}.
\end{equation*}

Since the derived upper bound for $(1, 0)$-CDCCs and $1$-CDCCs is expressed as a summation, its relationship to the average sphere packing values $\text{ASPV}_{\mathsf{D}}(1, 0)$ and $\text{ASPV}_{\mathsf{D}}(1)$ is not immediately apparent.
To clarify this relationship, we evaluate these quantities for small code lengths $2 \le n \le 10$ and present the results in Table~\ref{tab:deletion-comparison}.
We remind the reader that the average sphere packing values do not constitute valid upper bounds on the code cardinalities.

{
\renewcommand{\arraystretch}{2} 
\begin{table}[htbp]
  \caption{Upper bounds and values for composite deletion correcting codes with a single deletion error and resolution $k=2$ for code length $2 \leq n \leq 10$.}
  \label{tab:deletion-comparison}
  \centering
    \begin{tabular}{c c c c}
    \toprule
    $n$ & $\sum_{\rho=1} \sum_{w=0}^{n-1} \frac{\mathcal{N}(n-1; \rho; w) \cdot \mathcal{V}(n; w)}{\rho}$ &   $\text{ASPV}_{\mathsf{D}}(1, 0)$     &   $\text{ASPV}_{\mathsf{D}}(1)$ \\
    \midrule
    2 & 7 & 6 & 3 \\
    3 & 18 & 14 & 7 \\
    4 & 47 & 34 & 17 \\
    5 & 129 & 87 & 43 \\
    6 & 357 & 226 & 113 \\
    7 & 1001 & 596 & 298 \\
    8 & 2836 & 1595 & 797 \\
    9 & 8106 & 4320 & 2160 \\
    10 & 23329 & 11809 & 5904 \\
    \bottomrule
    \end{tabular}
\end{table}
}

\subsection{Lower Bounds}

\begin{figure}[htbp]
  \centering
  \begin{tikzpicture}[>=Stealth, every node/.style={font=\small}]

  \node[left] (msg) at (0,0.5) {Message $\bm{s} \in \Sigma_q^m$};
  \draw[->, thick] (msg.east) -- (1.5,0.5);

  \foreach \i in {1,...,3} {
      \node[draw, minimum size=1cm] (s\i) at (1.5+1*\i,0.5) {$s_{\i}$};
  }
  \node[minimum size=1cm] (sc) at (5.5,0.5) {$\cdots$};
  \node[draw, minimum size=1cm] (sm-1) at (6.5,0.5) {$s_{m-1}$};
  \node[draw, minimum size=1cm] (sm) at (7.5,0.5) {$s_m$};

  \node[left] (code) at (0,-1.5) {Codeword $\bm{c} \in \Sigma_q^n$};
  \draw[<- , thick] (code.east) -- (1.5,-1.5);

  \foreach \i in {1,...,3} {
      \node[draw, minimum size=1cm] (cx\i) at (1.5+1*\i,-1.5) {$s_{\i}$};
  }

  \node at (5.5,-1.5) {$\cdots$};
  \node[draw, minimum size=1cm] (cxm-1) at (6.5,-1.5) {$s_{m-1}$};
  \node[draw, minimum size=1cm] (cxm) at (7.5,-1.5) {$s_{m}$};

  \node[draw, minimum size=1cm, text=blue] (p1) at (8.5,-1.5) {$p$};
  \node[draw, minimum size=1cm, text=blue] (p2) at (9.5,-1.5) {$p$};

  \node[draw, minimum size=1cm, text=red] (z1) at (10.5,-1.5) {$z_1$};
  \node[draw, minimum size=1cm, text=red] (z2) at (11.5,-1.5) {$z_2$};
  \node[minimum size=1cm, text=red] (zc) at (12.5,-1.5) {$\cdots$};
  \node[draw, minimum size=1cm, text=red] (zt) at (13.5,-1.5) {$z_t$};
  \node[draw, minimum size=1cm, text=red] (zt1) at (14.5,-1.5) {$z_{t+1}$};

  \draw[->, thick] (sm.south) -- (cxm.north) node[midway, left] {$\mathrm{ENC}$};

  \draw[very thick, decorate, decoration={brace, mirror, amplitude=10pt}] (cx1.south west) -- (cxm.south east) node[midway, below=8pt] {Data};
  \draw[green!50!black, very thick, decorate, decoration={brace, mirror, amplitude=10pt}] (p1.south west) -- (zt1.south east) node[midway, below=8pt, text=green!50!black] {Redundancy};

  \node[above=2pt, text=blue] at (p1.north east) {Marker};

  \end{tikzpicture}
  \caption{Systematic encoder $\mathrm{ENC}$ of the Tenengolts $q$-ary single-deletion-correcting code.
  The message $\bm{s} \in \Sigma_q^m$ is encoded into a codeword $\bm{c} \in \Sigma_q^n$.
  Here $t = \lceil \log_q m \rceil$.
  The marker $\bcom{pp}$, where $p \equiv (s_m + 1) \mod{q}$, serves as a separator between the data part and the redundancy part.}
  \label{fig:systematic-code}
\end{figure}
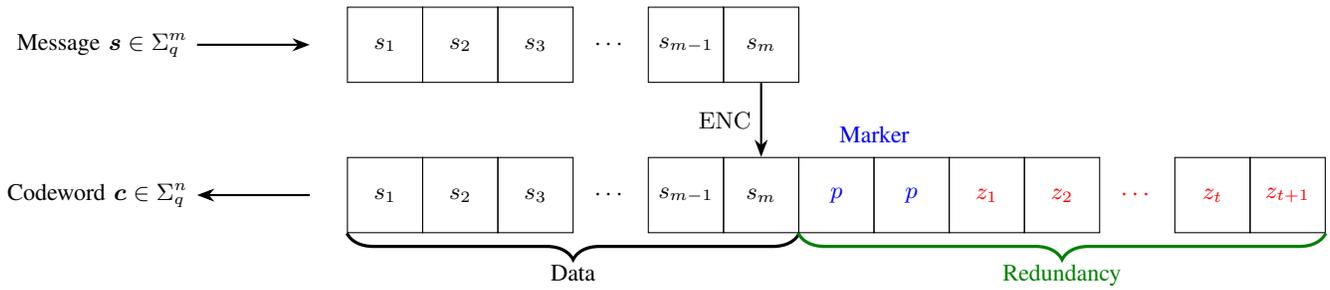

In this section, we provide constructions for $(1, 0)$-CDCCs and $1$-CDCCs, thereby establishing lower bounds on their cardinalities.
Our approach leverages the well-known, nearly optimal Varshamov-Tenengolts (VT) binary single-deletion-correcting codes \cite{VT}.
Levenshtein \cite{Levenshtein} observed that the Varshamov-Tenengolts codes could be used for correcting a single deletion.
For all $0 \leq a \leq n$, the Varshamov-Tenengolts (VT) code is defined as
\begin{equation*}
  VT_a(n) = \left\{ \bm{x} = (x_1, \ldots, x_n) \in \left\{0, 1\right\}^n \ : \ \sum_{i=1}^n ix_i \equiv a \mod{(n+1)} \right\}.
\end{equation*}
We additionally provide systematic constructions based on the Tenengolts $q$-ary single-deletion-correcting code \cite{Tenengolts}.
As illustrated in Figure~\ref{fig:systematic-code}, this $q$-ary systematic single-deletion-correcting code encodes a $q$-ary message $\bm{s} \in \Sigma_q^{m}$ of length $m$ into a codeword of length $n$ as
\begin{equation*}
  \mathrm{ENC}(\bm{s}) = \bm{s} pp \bm{z},
\end{equation*}
where $\bm{z} \in \Sigma_q^{t+1}$ constitutes the redundancy and $t = \lceil \log_q m \rceil$.
The marker $pp$, with $p \triangleq (\bm{s}[m] + 1) \mod{q}$, serves as a separator between the data part and the redundancy part.
The decoder of this code, denoted by $\mathrm{DEC}$, takes as input a subsequence $\bm{x}$ of length $n-1$, obtained from $\mathrm{ENC}(\bm{s})$ for some $\bm{s} \in \Sigma_q^{m}$ by a single deletion, and reconstructs the original message, yielding $\mathrm{DEC}(\bm{x}) = \bm{s}$.

We begin with the case of $(1, 0)$-CDCCs, where a single deletion is introduced by the first channel.
We first construct a code based on the VT code, which yields a lower bound on the cardinality of $(1, 0)$-CDCCs.
We then present a systematic construction based on the Tenengolts ternary single-deletion-correcting code, which incurs slightly more redundancy to achieve a systematic form.

\begin{construction}
  For each $0 \leq a \leq n$, let the code $\mathcal{C}_\Romannum{3}(a)$ be defined as
  \begin{equation*}
    \mathcal{C}_\Romannum{3}(a) \triangleq \left\{ \bm{c} \in \Sigma_3^n \ : \ \bm{c}_0 \in VT_a(n) \right\},
  \end{equation*}
  where $\bm{c}_0$ is the first sequence in the decomposition of $\bm{c}$, that is, $\mathcal{D}(\bm{c}) = (\bm{c}_0, \bm{c}_1)$.
\end{construction}

\begin{theorem}
  For any $0 \leq a \leq n$, the code $\mathcal{C}_\Romannum{3}(a)$ is a $(1, 0)$-CDCC.
\end{theorem}

\begin{IEEEproof}
  Fix some $0 \leq a \leq n$.
  Let $\bm{c} \in \mathcal{C}_\Romannum{3}(a)$ be the transmitted sequence and let $\bm{c}_0, \bm{c}_1$ be the decomposed binary sequences of $\bm{c}$, i.e., $\mathcal{D}(\bm{c}) = (\bm{c}_0, \bm{c}_1)$.
  Let $\bm{y}_0, \bm{y}_1$ be the outputs of the first and second channel, respectively. Since the second channel does not introduce any errors, then $\bm{y}_1 = \bm{c}_1$.
  Use the $VT_a(n)$ code to correct the deletion error in $\bm{y}_0$ and obtain $\bm{c}_0$. Reconstruct $\bm{c} = \mathcal{R}(\bm{c}_0, \bm{c}_1)$.
\end{IEEEproof}

\begin{corollary}
  For any code length $n$, there exists $0 \leq a \leq n$ such that
  \begin{equation*}
    \mathcal{S}_\mathsf{D}\left(n; (1, 0)\right) \geq |\mathcal{C}_\Romannum{3}(a)| \geq \frac{3^n}{n+1}.
  \end{equation*}
\end{corollary}

\begin{IEEEproof}
  The $VT_a(n)$ codes partition the space of composite binary sequences of length $n$ into $n+1$ cosets.
  By the pigeonhole principle, there exists at least one coset of size no smaller than $\frac{3^n}{n+1}$.
\end{IEEEproof}


The codes $\mathcal{C}_\Romannum{3}(a)$ are neither constructive nor systematic.
By slightly increasing the redundancy, we can construct a systematic $(1, 0)$-CDCC, based on the Tenengolts $q$-ary single-deletion-correcting code, for $q=3$.

\begin{construction}
  Let $\bm{s} \in \Sigma_3^m$ be a composite binary sequence of length $m$.
  Let $\bm{s}_0$ and $\bm{s}_1$ be the decomposed binary sequences of $\bm{s}$, so $\mathcal{D}(\bm{s}) = (\bm{s}_0, \bm{s}_1)$.
  Let $\mathrm{ENC}(\bm{s}_0) = \bm{s}_0 pp \bm{z}$ be the codeword obtained by encoding $\bm{s}_0$ with the Tenengolts ternary single-deletion-correcting code.
  Define the code
  \begin{equation*}
    \mathcal{C}_\Romannum{4} \triangleq \left\{ \bm{s} p'p' \bm{z} \ : \ \bm{s} \in \Sigma_3^m \right\},
  \end{equation*}
  where $p' \equiv (p + 1) \mod{3}$.
\end{construction}

\begin{theorem}
  The code $\mathcal{C}_\Romannum{4}$ is a systematic $(1, 0)$-CDCC.
\end{theorem}

\begin{IEEEproof}
  It is immediate by the definition of $\mathcal{C}_\Romannum{4}$ that the code is systematic. 
  We show that it is a $(1, 0)$-CDCC.
  Denote the transmitted codeword $\bm{c} = \bm{s} p'p' \bm{z}$, for some $\bm{s} \in \Sigma_3^m$.
  Let $\bm{c}_0$ and $\bm{c}_1$ be the decomposed binary sequences of $\bm{c}$, so that $\mathcal{D}(\bm{c}) = (\bm{c}_0, \bm{c}_1)$. 
  These sequences are transmitted on the deletion channels.
  Assume that $\mathcal{D}(p'p'\bm{z}) = (p'_0p'_0 \bm{z}_0, p'_1p'_1 \bm{z}_1)$, then $\bm{c}_0 = \bm{s}_0 p'_0p'_0 \bm{z}_0$ and $\bm{c}_1 = \bm{s}_1 p'_1p'_1 \bm{z}_1$.

  We assume that a single deletion occurs in the first channel.
  Let $\bm{y}_0, \bm{y}_1$ denote the received sequences. Since the second channel is error-free, then $\bm{y}_1 = \bm{c}_1$.
  The sequence $\bm{y}_0$ is a subsequence of $\bm{c}_0$ that has suffered a single deletion error.
  It suffices to recover $\bm{s}_0$, as $\bm{s}_1$ is already known as the data part of $\bm{y}_1$, and $\bm{s} = \mathcal{R}(\bm{s}_0, \bm{s}_1)$.

  The ternary letter $p$ is computed by $\mathrm{ENC}(\bm{s}_0)$ as $ p \equiv (\bm{s}_0[m] + 1) \mod{3}$ and $p' \equiv (p + 1) \mod{3}$, hence
  \begin{equation*}
    p' = \begin{cases}
    2 & \text{if } \bm{s}_0[m] = 0 \\
    0 & \text{if } \bm{s}_0[m] = 1 \\
  \end{cases}
  \qquad and  \qquad
    p'_0 = \begin{cases}
    1 & \text{if } \bm{s}_0[m] = 0 \\
    0 & \text{if } \bm{s}_0[m] = 1 \\
  \end{cases}.
  \end{equation*}

  Therefore $\bm{s}_0[m] = \bm{c}_0[m] \neq \bm{c}_0[m+1] = p'_0 $.
  We now show how to recover $\bm{s}_0$. Consider the bits $\bm{y}_0[m]$ and $\bm{y}_0[m+1]$.
  \begin{itemize}
    \item If $\bm{y}_0[m] \neq \bm{y}_0[m+1]$, the deletion did not occur in the data part. In this case, $\bm{s}_0$ equals the first $m$ bits of $\bm{y}_0$. 
    \item Otherwise, if $\bm{y}_0[m] = \bm{y}_0[m+1]$, the deletion occurred in the data part. This implies that the non-data bits $p'_0p'_0 \bm{z}_0$ of $\bm{y}_0$ are intact.
    Together with the non-data bits $p'_1p'_1 \bm{z}_1$ of $\bm{y}_1$, we can reconstruct $p'p' \bm{z}$.
    Next, compute $p \equiv (p-1)\mod3$.
    Let $\bm{s}_0'$ denote the first $m-1$ bits of $\bm{y}_0$.
    Finally, use the decoder of the Tenengolts ternary single-deletion-correcting code to recover $\bm{s}_0 = \mathrm{DEC}(\bm{s}_0' pp \bm{z})$.
  \end{itemize}
\end{IEEEproof}

\begin{corollary}
  For any code length $n$, it holds that
  \begin{equation*}
    |\mathcal{C}_\Romannum{4}| = \frac{3^n}{3^{\lceil \log_3 n \rceil + 3}}.
  \end{equation*}
\end{corollary}

\begin{IEEEproof}
  $\mathcal{C}_\Romannum{4}$ has the same structure, redundancy and cardinality as the Tenengolts ternary single-deletion-correcting code.
\end{IEEEproof}


We now consider the $1$-CDCC case, where a single deletion occurs in exactly one of the two channels.
The affected channel can be identified since only one of the channel outputs has length $n-1$, however, its identity is not known in advance.
We again leverage the nearly optimal VT binary single-deletion-correcting codes \cite{VT} to construct a code that provides a lower bound on the cardinality of $1$-CDCCs.
For $(1, 0)$-CDCC, only the decomposed binary sequence transmitted over the first channel was required to belong to a VT code.
Here, since the deletion may occur in either channel, we require that the concatenation of the decomposed binary sequences belongs to a VT code.

\begin{construction}
  For each $0 \leq a \leq 2n$, let the code $\mathcal{C}_\Romannum{5}(a)$ be defined as
  \begin{equation*}
    \mathcal{C}_\Romannum{5}(a) \triangleq \left\{ \bm{c} \in \Sigma_3^n \ : \ \bm{c}_0  \bm{c}_1 \in VT_a(2n) \right\},
  \end{equation*}
  where $\bm{c}_0, \bm{c}_1$ are the binary decomposed sequences of $\bm{c}$, that is, $\mathcal{D}(\bm{c}) = (\bm{c}_0, \bm{c}_1)$.
\end{construction}

\begin{theorem}
  For any $0 \leq a \leq 2n$, the code $\mathcal{C}_\Romannum{5}(a)$ is a $1$-CDCC.
\end{theorem}

\begin{IEEEproof}
  Fix some $0 \leq a \leq 2n$.
  Let $\bm{c} \in \mathcal{C}_\Romannum{5}(a)$ be the transmitted sequence and let $\bm{c}_0, \bm{c}_1$ denote the decomposed binary sequences of $\bm{c}$, that is, $\mathcal{D}(\bm{c}) = (\bm{c}_0, \bm{c}_1)$.
  Let $\bm{y}_0, \bm{y}_1$ be the outputs of the first and second channel, respectively. The concatenation of the sequences $\bm{y}_0  \bm{y}_1$ 
  is a subsequence of $\bm{c}_0  \bm{c}_1$ that has suffered a single deletion error, and can therefore be corrected by $VT_a(2n)$.
  The corrected sequence is then partitioned into two halves, which correspond to $\bm{c}_0$ and $\bm{c}_1$.
  Finally, reconstruct $\bm{c} = \mathcal{R}(\bm{c}_0, \bm{c}_1)$.
\end{IEEEproof}

\begin{corollary}
  For any code length $n$, there exists $0 \leq a \leq 2n$ such that
  \begin{equation*}
    \mathcal{S}_\mathsf{D}\left(n; 1\right) \geq |\mathcal{C}_\Romannum{5}(a)| \geq \frac{3^n}{2n+1}.
  \end{equation*}
\end{corollary}

\begin{IEEEproof}
  The $VT_a(2n)$ codes partition the space of composite binary sequences of length $n$ into $2n+1$ cosets.
  By the pigeonhole principle, there exists at least one coset of size no smaller than $\frac{3^n}{2n+1}$.
\end{IEEEproof}

The codes $\mathcal{C}_\Romannum{5}(a)$ are also neither constructive nor systematic.
As for $(1, 0)$-CDCC, a systematic $1$-CDCC can be obtained by slightly increasing the redundancy, using the Tenengolts ternary single-deletion-correcting code of length $2n$
on the concatenation of the decomposed binary sequences.
However, its application here requires a modification to ensure that the markers correctly separate the data and redundancy parts in each of the decomposed sequences.

\begin{construction}
  Let $\bm{s} \in \Sigma_3^m$ be a composite binary sequence of length $m$.
  Let $\bm{s}_0$ and $\bm{s}_1$ be the decomposed binary sequences of $\bm{s}$, so $\mathcal{D}(\bm{s}) = (\bm{s}_0, \bm{s}_1)$.
  Let $\mathrm{ENC}(\bm{s}_0\bm{s}_1) = \bm{s}_0 \bm{s}_1 pp \bm{z}$ be the codeword obtained by encoding the concatenation $\bm{s}_0\bm{s}_1$ with the Tenengolts ternary single-deletion-correcting code of length $2n$.
  Define the code
  \begin{equation*}
    \mathcal{C}_\Romannum{6} \triangleq \left\{ \bm{s} p'p' 02 \bm{z} \ : \ \bm{s} \in \Sigma_3^m \right\},
  \end{equation*}
  where
  \begin{equation*}
    p' = \begin{cases}
    2 & \text{if } \bm{s}[m] = 0 \\
    1 & \text{if } \bm{s}[m] = 1 \\
    0 & \text{if } \bm{s}[m] = 2
  \end{cases}.
  \end{equation*}  
\end{construction}

\begin{theorem}
  The code $\mathcal{C}_\Romannum{6}$ is a systematic $1$-CDCC.
\end{theorem}

\begin{IEEEproof}
  It is immediate by the definition of $\mathcal{C}_\Romannum{6}$ that the code is systematic.
  We show that it is a $1$-CDCC.
  Denote the transmitted codeword $\bm{c} = \bm{s} p'p' 02 \bm{z}$, for some message $\bm{s} \in \Sigma_3^m$. 
  Let $\bm{c}_0, \bm{c}_1$ be the decomposed binary sequences of $\bm{c}$, so that, $\mathcal{D}(\bm{c}) = (\bm{c}_0, \bm{c}_1)$.
  These sequences are transmitted on the deletion channels.
  Assume that $\mathcal{D}(p'p'02\bm{z}) = (p'_0p'_0 01\bm{z}_0, p'_1p'_1 01 \bm{z}_1)$, then $\bm{c}_0 = \bm{s}_0 p'_0p'_0 01 \bm{z}_0$ and $\bm{c}_1 = \bm{s}_1 p'_1p'_1 01 \bm{z}_1$.
  Let $\bm{y}_0, \bm{y}_1$ denote the received sequences. Based on the length of the received sequences, we can determine in which channel the deletion occurred.

  \emph{Case 1: The deletion occurred in the first channel.}
  In this case, $\bm{y}_1 = \bm{c}_1$, and $\bm{s}_1$ is known. Thus it suffices to recover $\bm{s}_0$. Let us consider the bits $\bm{y}_1[m]$ and $\bm{y}_1[m+1]$.
  \begin{itemize}
    \item 00 - This case is impossible. If $\bm{y}_1[m] = 0$, then $\bm{s}[m] = 0$, and as such $p' = 2$, yielding $p'_1 = 1$.

    \item 01 - If $\bm{y}_1[m] = 0$, then $\bm{s}[m] = 0$ and $\bm{c}_0[m] = \bm{s}_0[m] = 0$. In turn $p' = 2$, yielding $\bm{c}_0[m+1] = p'_0 = 1$.
    By considering the bits $\bm{y}_0[m]$ and $\bm{y}_0[m+1]$, we can discover if the deletion occurred in the data bits or the redundancy bits of $\bm{c}_0$.
    \begin{itemize}
      \item If $\bm{y}_0[m] \neq \bm{y}_0[m+1]$, the deletion did not occur in the data bits. In this case, $\bm{s}_0$ equals the first $m$ bits of $\bm{y}_0$.
      \item Otherwise, if $\bm{y}_0[m] = \bm{y}_0[m+1]$, the deletion occurred in the data bits. This implies that the non-data bits $\bm{z}_0$ in $\bm{y}_0$ are intact.
      Together with the non-data bits $\bm{z}_1$ in $\bm{y}_1$, we can reconstruct $\bm{z}$. Since $\bm{s}[m] = 0$, then $p = \bm{s}_1[m] + 1 \mod{3} = 1$.
      Let $\bm{s}_0'$ denote the first $m-1$ bits of $\bm{y}_0$.
      Finally, use the decoder of the Tenengolts ternary single-deletion code of length $2n$ to recover $\bm{s}_0\bm{s}_1 = \mathrm{DEC}(\bm{s}_0' \bm{s}_1 pp \bm{z})$.
    \end{itemize}
    
    \item 10 - This case happens when $\bm{s}[m] = 2$, and as such $p' = 0$. Then $\bm{c}_0[m] = 1$ and $\bm{c}_0[m+1] = 0$. We can recover $\bm{s}_0$ similarly to the previous scenario.

    \item 11 - This case is slightly more complex and is the reason we need the two extra symbols in the code. It happens when $\bm{s}[m] = 1$, and as such $p' = 1$.
    Then $\bm{c}_0[m] = 0$ and $\bm{c}_0[m+1] = 0$. We use the bit $\bm{y}_0[m+3]$ to determine if the error occurred before or after the $m+3$-th bit.
    Remember that $\bm{c}_0[m+3] = 0$ and $\bm{c}_0[m+4] = 1$.
    \begin{itemize}
      \item If $\bm{y}_0[m+3] = 0$, then the deletion occurred after this bit. This implies that the data bits $\bm{s}_0$ are intact.
      \item Else, if $\bm{y}_0[m+3] = 1$, then the deletion occurred before this bit. As such the non-data bits $\bm{z}_0$ are intact, and we can recover
      $\bm{s}_0$ similarly to the previous scenarios.
    \end{itemize}
  \end{itemize}

  \emph{Case 2: The deletion occurred in the second channel.}
  In this case, $\bm{y}_0 = \bm{c}_0$, and $\bm{s}_0$ is known. Thus it suffices to recover $\bm{s}_1$. Let us consider the bits $\bm{y}_0[m]$ and $\bm{y}_0[m+1]$.
  \begin{itemize}
    \item 00 - This case happens when $\bm{s}[m]=1$ and as such $p' = 1$. Then $\bm{c}_1[m] = 1$ and $\bm{c}_0[m+1] = 1$.
    We use the bit $\bm{y}_1[m+2]$ to determine if the error occurred before or after the $m+2$-th bit.
    Remember that $\bm{c}_1[m+2] = 1$ and $\bm{c}_1[m+3] = 0$.
    \begin{itemize}
      \item If $\bm{y}_1[m+2] = 1$, then the deletion occurred after this bit. This implies that the data bits $\bm{s}_1$ are intact.
      \item Else, if $\bm{y}_1[m+2] = 0$, then the deletion occurred before this bit. As such the non-data bits $\bm{z}_1$ are intact, and we can recover
      $\bm{s}_1$ similarly to the previous cases.
    \end{itemize}

    \item 01 - If $\bm{y}_0[m+1] = 1$, then $p' = 2$ and $\bm{c}_1[m+1] = p'_1 = 1$. In turn, $\bm{s}[m] = 0$, yielding $\bm{c}_1[m] = \bm{s}_1[m] = 0$.
    In this case, we can recover $\bm{s}_1$ similarly to 01 case in Case 1.

    \item 10 - If $\bm{y}_0[m] = 1$, then $\bm{s}[m] = 2$ and $\bm{c}_1[m] = \bm{s}_1[m] = 1$. In turn, $p' = 0$, yielding $\bm{c}_1[m+1] = p'_1 = 0$.
    In this case, we can recover $\bm{s}_1$ similarly to 01 case in Case 1.

    \item 11 - This case is impossible. If $\bm{y}_0[m] = 1$, then $\bm{s}[m] = 2$, and as such $p' = 0$, yielding $p'_0 = 0$.
  \end{itemize}

\end{IEEEproof}

\begin{corollary}
  For any code length $n$, it holds that
  \begin{equation*}
    |\mathcal{C}_\Romannum{6}| = \frac{3^n}{3^{\lceil \log_3 2n \rceil + 5}}.
  \end{equation*}
\end{corollary}

\begin{IEEEproof}
The code $\mathcal{C}_\Romannum{6}$ has a structure similar to the Tenengolts ternary single-deletion-correcting code of length $2n$.
In addition to the redundancy symbols of the Tenengolts code, each codeword in $\mathcal{C}_\Romannum{6}$ contains two extra symbols, $02$.
Hence, the total redundancy is $\lceil \log_3 2n \rceil + 5$.
\end{IEEEproof}


\section{Conclusion and Future Work}\label{sec:conclusion}

In this work, we introduced the ordered composite DNA channel for composite letters with arbitrary resolution $k \in \mathbb{N}$ over an alphabet of arbitrary size $q$, although our analysis focused on the case $q = 2$.
We defined two families of substitution error-correcting codes for this channel, one where the number of errors in each channel is bounded, and another where the total number of errors is bounded without any restriction on their distribution across the channels.

For a single substitution error and arbitrary resolution $k$, we established lower and upper bounds on the cardinality of the codes for both families.
These bounds are summarized in Table~\ref{tab:bounds-k-substitution}.
We additionally obtained both lower and upper bounds on the cardinality of the codes for both families when the resolution is $k = 2$ and the number of errors is arbitrary.
For up to two errors, we applied the generalized sphere packing bound approach to obtain nontrivial, non-asymptotic bounds.
Table~\ref{tab:bounds-k2-substitution} summarizes these bounds, where $\star$ denotes asymptotic bounds.

In addition, we investigated deletion errors in the ordered composite DNA channel for the case $k = 2$.
We derived lower and upper bounds for the case of a single deletion error, considering both the known-channel and unknown-channel scenarios.
We also presented systematic code constructions for both cases.
Table~\ref{tab:bounds-k2-deletion} summarizes these results.

{
\renewcommand{\arraystretch}{2} 
\begin{table}[ht]
  \caption{Cardinality bounds of composite error correcting codes for arbitrary resolution.}
  \label{tab:bounds-k-substitution}
  \centering
  \begin{tabular}{c c c}
  \toprule
  \textbf{Code Family}                     &  \textbf{Lower Bound}             & \textbf{Upper Bound } \\
  \midrule
  $\mathcal{S}_k\left(n; (1, 0, \ldots, 0)\right)$ & $(k-1)^{n} \sum_{\ell=0}^{n} \binom{n}{\ell} \left(\frac{1}{k-1}\right)^{\ell} 2^{\ell - \lceil\log_2(\ell + 1)\rceil}$                               & $\frac{(k+1)^{n+1} - (k-1)^{n+1}}{2(n+1)}$ \\
  $\mathcal{S}_k\left(n; 1\right)$               & $\frac{(k+1)^n}{\left(k+1\right)^{\lceil \log_{k+1} (2n +1) \rceil}}$, (even $k$)  & $\frac{(k+1)^n}{\frac{2kn}{k+1}-1}$   \\
  \bottomrule
  \end{tabular}
\end{table}
}

{
\renewcommand{\arraystretch}{2} 
\begin{table}[ht]
  \caption{Cardinality bounds of composite error correcting codes for resolution $k=2$.}
  \label{tab:bounds-k2-substitution}
  \centering
  \begin{tabular}{c c c}
  \toprule
  \textbf{Code Family}                             &  \textbf{Lower Bound}                                                               & \textbf{Upper Bound } \\
  \midrule
  $\mathcal{S}_{2}\left(n; (1, 1)\right)$     & $\frac{3^n}{2 ^ { 2 \lceil \log_2(n+1) \rceil}}$                                    & $\frac{3^n}{\frac{(n-3)^2}{6}}$   \\
  $\mathcal{S}_{2}\left(n; 2\right)$          & $\frac{3^n}{3^{2 \lceil\log_3 (n+1) \rceil + 1}}$                           & $\frac{\sqrt{\frac{8n}{6}}}{\sqrt{\frac{8n}{6}}-1} \cdot \frac{3^n}{\frac{8n^2}{9}-\frac{2n (\sqrt{\frac{8n}{6}})}{3}}$ \\
  $\mathcal{S}_{2}\left(n; e\right)$          & $\frac{3^n}{3^{\lceil\log_3 (n+1) \rceil \cdot \lceil \frac{4e-2}{3} \rceil + 1}}$  & $\frac{3^n}{(\frac{4n}{3e})^e} \star$ \\
  $\mathcal{S}_{2}\left(n; (e_0, e_1)\right)$ & $\frac{3^n }{2 ^ { \lceil \log_2(n+1) \rceil \cdot (e_0 + e_1)}}$                    & $\frac{3^n e_0^{e_0} e_1^{e_1}}{(\frac{n}{3})^{e_0+e_1}} \star $  \\
  \bottomrule
  \end{tabular}
\end{table}
}

{
\renewcommand{\arraystretch}{2} 
\begin{table}[ht]
  \centering
  \caption{Cardinality bounds of composite deletion correcting codes for resolution $k=2$.}
  \label{tab:bounds-k2-deletion}
  \begin{tabular}{c c c}
    \toprule
    \textbf{Code Family} & \textbf{Lower Bound} & \textbf{Upper Bound} \\
    \midrule
    $\mathcal{S}_{\mathsf{D}}\left(n; (1, 0)\right)$ & $\frac{3^n}{n+1}$ & $\sum_{\rho=1} \sum_{w=0}^{n-1} \frac{\mathcal{N}(n-1; \rho; w) \cdot \mathcal{V}(n; w)}{\rho}$ \\
    $\mathcal{S}_{\mathsf{D}}\left(n; 1\right)$      & $\frac{3^n}{2n+1}$ & $\sum_{\rho=1} \sum_{w=0}^{n-1} \frac{\mathcal{N}(n-1; \rho; w) \cdot \mathcal{V}(n; w)}{\rho}$ \\
    \bottomrule
  \end{tabular}
\end{table}
}

The idea of composite letters was originally introduced to enhance information capacity in DNA storage systems \cite{Anavy}, \cite{Choi}.
In the absence of errors, the capacity of the ordered composite DNA channel coincides with that of the regular composite DNA channel and is given by $\log_2 |\Phi_{q, k}|$, 
where $\Phi_{q, k}$ denotes the composite alphabet of size $q$ and resolution $k$.


As a possible direction for future work, it would be of interest to study the capacity of the ordered composite DNA channel for binary ($q = 2$) composite letters with resolution $k = 2$ (see Figure~\ref{fig:channel}), 
under the assumption that the underlying channels are independent binary symmetric channels (BSCs) with transition probability $0 \leq p \leq \tfrac{1}{2}$. 
We denote this channel by $\mathsf{C}$ and its capacity by $\text{cap}(\mathsf{C})$.
The input alphabet is $\mathcal{X} \triangleq \Sigma_{k+1} = \{0, 1, 2\}$ and the output alphabet is $\mathcal{Y} \triangleq \Sigma_{k+1} \cup \{\mathord{?}\} = \{0, 1, 2, \mathord{?}\}$.
Let $\mathsf{X}$ and $\mathsf{Y}$ be the transmitted and received random variables, respectively.
The transition probabilities are given in the following matrix,

\begin{equation*}
\mathbb{P}(\mathsf{Y} | \mathsf{X}) = 
  \mathop{\left[
  \begin{array}{ *{4}{c} }
    \colind{(1-p)^2}{$\mathsf{Y}=0$} & \colind{p(1-p)}{$\mathsf{Y}=1$}  &  \colind{p^2}{$\mathsf{Y}=2$} & \colind{p(1-p)}{$\mathsf{Y}=\mathord{?}$} \\
    p(1-p) &  (1-p)^2 & p(1-p) & p^2 \\
    p^2 & p(1-p) &  (1-p)^2 & p(1-p)
  \end{array}
  \right]}
  \begin{array}{@{}c@{}}
    \rowind{$\mathsf{X}=0$} \\ \rowind{$\mathsf{X}=1$} \\ \rowind{$\mathsf{X}=2$}
  \end{array}^{
  \begin{array}{@{}c@{}}
    \\ \mathstrut
  \end{array}
  }.
\end{equation*}

The key challenge in providing a closed form expression for $\text{cap}(\mathsf{C})$ is determining the input distribution $\mathbb{P}_{\mathsf{X}}$ that maximizes the entropy of the output random variable $H(\mathsf{Y})$, 
since the rows of the matrix $\mathbb{P}(\mathsf{Y} | \mathsf{X})$ are permutations of each other and therefore the conditional entropy $H(\mathsf{Y} | \mathsf{X})$ is independent of $\mathbb{P}_{\mathsf{X}}$.
Due to the symmetry of the channel with respect to the letters $0$ and $2$, this input distribution can be assumed to have the form

\begin{equation*}
  \mathbb{P}(\mathsf{X}=x) = \begin{cases*}
    \alpha & if $x = 0$ \\
    1-2\alpha & if $x = 1$ \\
    \alpha & if $x = 2$
  \end{cases*}.
\end{equation*}

\begin{figure}[htbp]
    \begin{minipage}[htbp]{0.5\textwidth}
      \centering
      \includegraphics[width=1\textwidth]{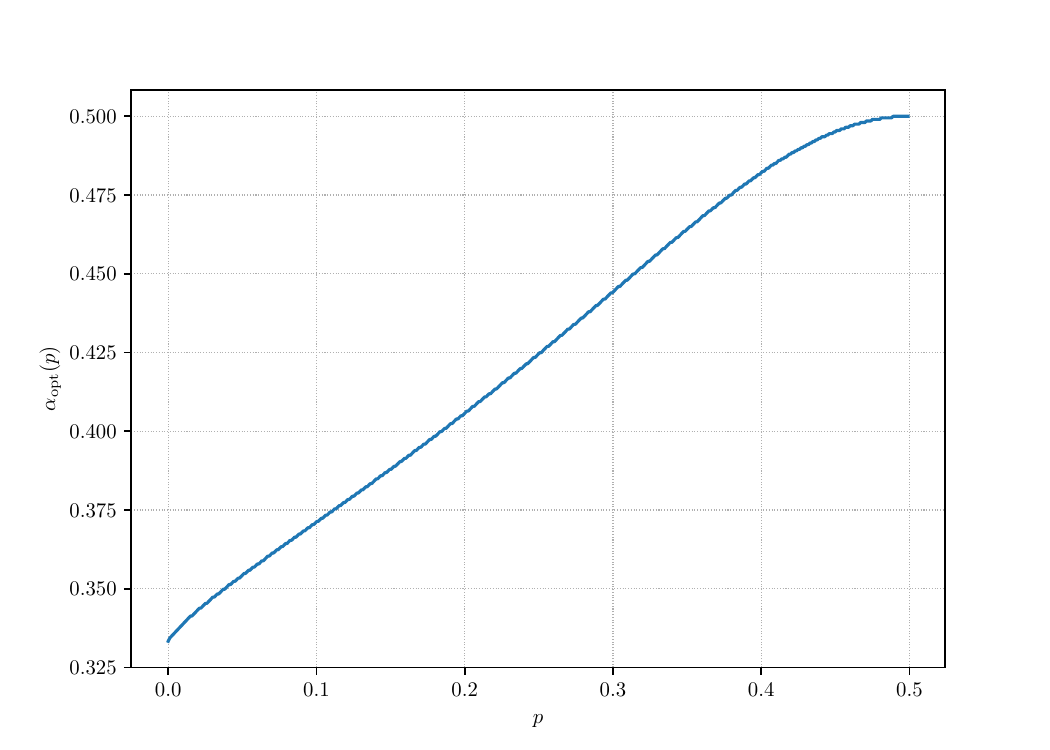}
      \caption{$\alpha_{\rm opt}(p)$ versus $p$.}
      \label{fig:alpha-max}
    \end{minipage}%
    \begin{minipage}[htbp]{0.5\textwidth}
      \centering
      \includegraphics[width=1\textwidth]{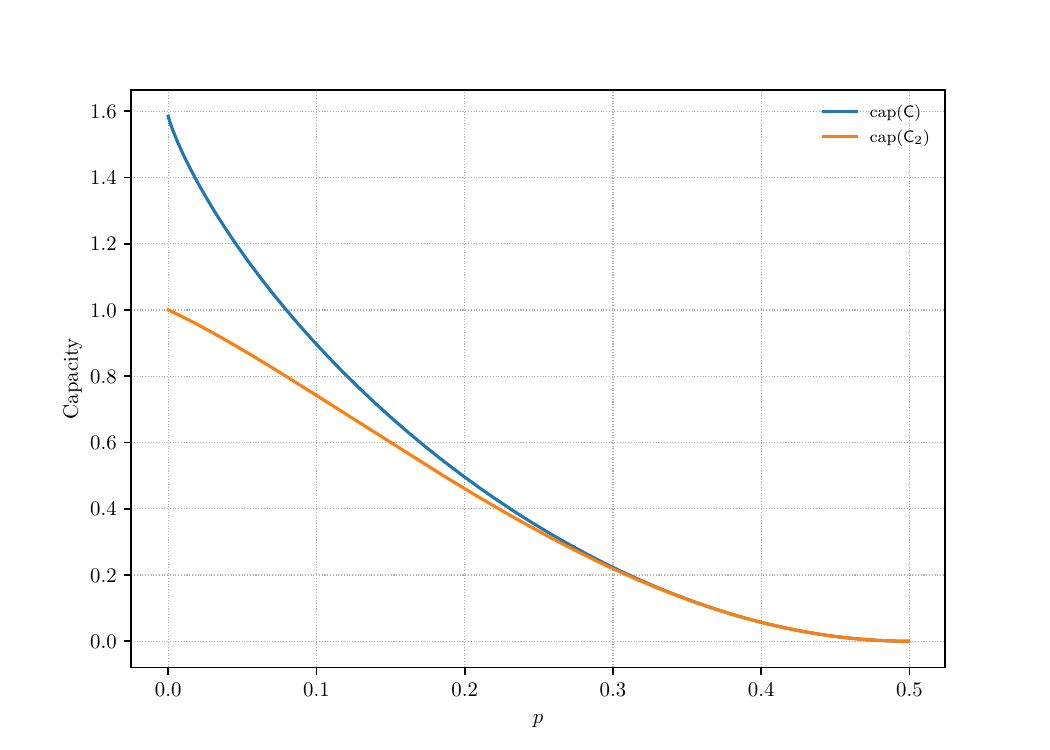}
      \caption{$\text{cap}(\mathsf{C})$ and $\text{cap}(\mathsf{C}_2)$ as a function of $p$.}
      \label{fig:capacities}
    \end{minipage}
\end{figure}

Differentiating $H(\mathsf{Y})$ with respect to $\alpha$ leads to a transcendental equation. We numerically computed the value of $\alpha$ that maximizes the capacity for each crossover probability $p$, denoted by $\alpha_{\rm opt}(p)= \argmax_\alpha \text{cap}(\mathsf{C})$.
The results, shown in Figure~\ref{fig:alpha-max}, suggest that the use of composite letters in this channel is advantageous when the underlying $\text{BSC}(p)$ channels are not too noisy.

To quantify this advantage, we compare the numerically computed $\text{cap}(\mathsf{C})$ with the capacity of a channel composed
of two identical and independent $\text{BSC}(p)$ channels, in which identical copies of a binary sequence are transmitted in each.
This channel model, which we denote by $\mathsf{C}_2$, was studied by Mitzenmacher \cite{Mitzenmacher} who provided an expression for its capacity $\text{cap}(\mathsf{C}_2)$.

Figure~\ref{fig:capacities} illustrates $\text{cap}(\mathsf{C})$ and $\text{cap}(\mathsf{C}_2)$ as functions of the crossover probability $p$.
For the noiseless case ($p=0$), we have $\text{cap}(\mathsf{C}) = \log_2 3$ while $\text{cap}(\mathsf{C}_2) = 1$, demonstrating the advantage of using composite letters.
For $p > 0.3$, the two capacities are nearly identical, indicating that the ordered composite DNA channel is beneficial primarily when $p \leq 0.3$.

\newpage

\appendices


\section{Proofs for Section~\ref{sec:preliminaries}} \label{appendix:preliminaries}

\kAryEcc*

\begin{IEEEproof}
  Let $\mathcal{C}$ be an optimal $(k+1)$-ary $e$-error-correcting code of length $n$, then $|\mathcal{C}| = \mathcal{A}_{k+1}(n; e)$.
  By definition, $\mathcal{C}$ can correct up to $e$ substitution errors in the $k$-resolution composite binary sequence, that is, $\mathcal{C}$ is a $k$-resolution $e$-CECC,
  since each channel error causes at most one error in the $k$-resolution composite binary sequence.
  Therefore, $|\mathcal{C}| = \mathcal{A}_{k+1}(n; e) \leq \mathcal{S}_k(n; e).$
\end{IEEEproof}


\connection*

\begin{IEEEproof}
  Let $\Delta = \sum_{i=0}^{k-1} e_i$.
  Let $\mathcal{C}$ be an optimal $k$-resolution $\Delta$-CECC of length $n$, then $|\mathcal{C}| = \mathcal{S}_k(n; \Delta)$.
  By definition, $\mathcal{C}$ can correct up to $\Delta$ substitution errors introduced collectively by all $k$ channels.
  Therefore, $\mathcal{C}$ can correct up to $e_i$ substitution errors in $\bm{s}_i$ for all $i \in \left\{0, 1, \ldots, k-1\right\}$,
  that is, $\mathcal{C}$ is an $(e_0, e_1, \ldots, e_{k-1})$-CECC. Hence, we have
  \begin{equation*}
  |\mathcal{C}| = \mathcal{S}_k\left(n; \sum_{i=0}^{k-1} e_i\right) \leq \mathcal{S}_{k}\left(n; (e_0, e_1, \ldots, e_{k-1})\right).
  \end{equation*}
\end{IEEEproof}


\reversal*

\begin{IEEEproof}
  Since the error tuples are reversed, the main idea in this proof is to apply a reversal operation to a decomposition output $\begin{bmatrix} 0^{k-\sigma} \ 1^{\sigma} \end{bmatrix}^\intercal$, yielding $\begin{bmatrix} 1^{\sigma} \ 0^{k-\sigma} \end{bmatrix}^\intercal$.
  If for some $\sigma \in \Sigma_{k+1}$ we have $\mathcal{D}(\sigma) = \begin{bmatrix} 0^{k-\sigma} \ 1^{\sigma} \end{bmatrix}^\intercal$, we observe that the reversed vector $\begin{bmatrix} 1^{\sigma} \ 0^{k-\sigma} \end{bmatrix}^\intercal$ is simply the bitwise negation of $\mathcal{D}(k-\sigma)$.

  Formally, let $\mathcal{C}$ be an $(e_0, e_1, \ldots, e_{k-2}, e_{k-1})$-CECC. We construct an $(e_{k-1}, e_{k-2}, \ldots, e_1, e_0)$-CECC $\mathcal{C}'$ of the same size.
  For each codeword $\bm{c} \in \mathcal{C}$ of length $n$, define $\bm{c}' \in \mathcal{C}'$ by $\bm{c}'[i] \triangleq k - \bm{c}[i]$, for all $1 \leq i \leq n$.
  The mapping is a bijection. Let $\mathcal{D}(\bm{c}) = (\bm{c}_0, \bm{c}_1, \ldots, \bm{c}_{k-1})$ and $\mathcal{D}(\bm{c}') = ({\bm{c}'}_0, {\bm{c}'}_1, \ldots, {\bm{c}'}_{k-1})$.
  Then, for every $0 \leq j \leq k-1$, we have $\bar{\bm{c}_j'} = \bm{c}_{k-1-j}$, where $\bar{\cdot}$ denotes bitwise negation.
  Let us show that $\mathcal{C}'$ is indeed an $(e_{k-1}, e_{k-2},  \ldots, e_{1}, e_0)$-CECC.
  Suppose $\bm{c}' \in \mathcal{C}'$ is transmitted, and the $j$-th channel outputs the sequence $\bm{y}'_j$ with at most $e_{k-1-j}$ substitution errors, for all $0 \leq j \leq k-1$.
  Let $\bar{\bm{y}_j'}$ denote the bitwise negation of $\bm{y}'_j$, and define $\bm{y}_{k-1-j} \triangleq \bar{\bm{y}_j'}$.
  Then each $\bm{y}_j$ has at most $e_j$ substitution errors in $\bm{c}_j$, and if we provide $(\bm{y}_0, \bm{y}_1, \ldots, \bm{y}_{k-1})$ to the decoder of $\mathcal{C}$, it can recover $\bm{c}$.
  Finally, we can compute $\bm{c}'$ from $\bm{c}$ using the bijection.

  Applying the construction to an optimal $(e_0, e_1, \ldots, e_{k-2}, e_{k-1})$-CECC and observing that the same construction can be performed on an optimal $(e_{k-1}, e_{k-2}, \ldots, e_1, e_0)$-CECC, we conclude that
  \begin{equation*}
  \mathcal{S}_{k}\left(n; (e_0, e_1, \ldots, e_{k-2}, e_{k-1})\right) = \mathcal{S}_{k}\left(n; (e_{k-1}, e_{k-2}, \ldots, e_{1}, e_0)\right).
  \end{equation*}
\end{IEEEproof}


\swapSingleError*

\begin{IEEEproof}
  It is enough to show for that for any $0 \leq i < k-1$ it holds that $\mathcal{S}_{k}\left(n; \bm{e}_i\right) = \mathcal{S}_{k}\left(n; \bm{e}_{i+1}\right)$, then one can apply the proposition inductively.
  Let us consider the transformation a letter in $ \sigma \in \Sigma_{k+1}$ can undergo if we assume that a single substitution error occurs in the $i$-th channel.
  The letter $\sigma \in \Sigma_{k+1}$ is decomposed into the binary column vector $\bm{v}_{\sigma} = \begin{bmatrix} 0^{k-\sigma} \quad 1^{\sigma}\end{bmatrix}^\intercal$.
  \begin{itemize}
    \item If $\sigma = k-1-i$, then $\bm{v}_{\sigma}[i-1, i, i+1] = 001$. An error in the $i$-th channel will change $\bm{v}_{\sigma}[i-1, i, i+1]$ to $011$, and $k-i-1$ will be transformed to $k-i$.
    \item If $\sigma = k-i$, then $\bm{v}_{\sigma}[i-1, i, i+1] = 011$. An error in the $i$-th channel will change $\bm{v}_{\sigma}[i-1, i, i+1]$ to $001$, and $k-i$ will be transformed to $k-1-i$.
    \item For any other $\sigma$, $\bm{v}_{\sigma}[i-1, i, i+1]$ is either $000$ or $111$. An error in the $i$-th channel will change $\bm{v}_{\sigma}[i-1, i, i+1]$ to either $010$ or $101$, and in both cases $\sigma$ will be transformed to $\mathord{?}$.
  \end{itemize}

  Let $\mathcal{C}$ be an $(\bm{e}_i)$-CECC of length $n$. We will build an $(\bm{e}_{i+1})$-CECC $\mathcal{C}'$ of the same size.
  For each codeword $\bm{c} \in \mathcal{C}$ of length $n$, define $\bm{c}' \in \mathcal{C}'$ by $\bm{c}' \triangleq \bm{c} - \bm{1} \mod (k+1)$, element wise.
  The mapping is a bijection. Let $\bm{c}'$ be the transmitted codeword, and assume a substitution error occurs in the $i+1$-channel.
  Let $\bm{y}' = \mathcal{R}(\bm{y}'_0, \bm{y}'_1, \ldots, \bm{y}'_{k-1})$ be the reconstructed composite binary sequence, where $\bm{y}'_{j}$ is the received sequence in the $j$-th channel.
  Let us show how to decode $\bm{c}'$ given $\bm{y}'$.

  \begin{itemize}
    \item If $\bm{y}'$ is a codeword in $\mathcal{C}'$, then we can decode $\bm{c}' = \bm{y}'$.
    \item Some $\bm{y}'[m] \not \in \left\{k-i-2, k-i-1\right\}$ is transformed to $\mathord{?}$. If $\bm{y}'_i[m] = 0$, then the substitution error is of type $0 \mapsto 1$, otherwise it is of type $1 \mapsto 0$. Fix this error and reconstruct again.
    \item Else, either a letter $k-i-2$ has been transformed to $k-i-1$ or vice-versa.
    In this case, $\bm{y}' \not \in \mathcal{C}'$. Suppose by contradiction $\bm{y}'$ is a codeword in $\mathcal{C}'$.
    Then, by definition both $\bm{c} = \bm{c}' + \bm{1} \mod (k+1)$ and $\bm{y} = \bm{y}' + \bm{1} \mod (k+1)$ are codewords in $\mathcal{C}$.
    A single substitution error in the $i$-th channel on the same position would transform $\bm{c}$ to $\bm{y}$, contradicting the fact that $\mathcal{C}$ is an $(\bm{e}_i)$-CECC.
    Therefore, there exists a unique codeword $\bm{c}' \in \mathcal{C}'$ that differs from $\bm{y}'$ in exactly one letter from the set $\left\{k-i-2, k-i-1\right\}$.
    Decoding can thus be performed by identifying said codeword $\bm{c}'$.
  \end{itemize}

  We have shown that $\mathcal{C}'$ is an $(\bm{e}_{i+1})$-CECC, and since the mapping is a bijection, it holds that $|\mathcal{C}| = |\mathcal{C}'|$.
  Now let $\mathcal{C}$ be an optimal $(\bm{e}_i)$-CECC, and let $\mathcal{C}'$ be the corresponding code obtained by the above transformation.
  Then
  \begin{equation*}
    \mathcal{S}_{k}\left(n; \bm{e}_i\right) = |\mathcal{C}| = |\mathcal{C}'| \leq \mathcal{S}_{k}\left(n; \bm{e}_{i+1}\right).
  \end{equation*}
  Since the equation holds for all $0 \leq i < k-1$, then
  \begin{equation*}
    \mathcal{S}_{k}\left(n; \bm{e}_0\right) \leq \mathcal{S}_{k}\left(n; \bm{e}_1\right) \leq \ldots \leq \mathcal{S}_{k}\left(n; \bm{e}_{k-1}\right).
  \end{equation*}
  However, by Proposition \ref{prop:reversal}, we also have
  \begin{equation*}
    \mathcal{S}_{k}\left(n; \bm{e}_0\right) = \mathcal{S}_{k}\left(n; \bm{e}_{k-1}\right).
  \end{equation*}
  Therefore, we conclude that
  \begin{equation*}
    \mathcal{S}_{k}\left(n; \bm{e}_0\right) = \mathcal{S}_{k}\left(n; \bm{e}_1\right) = \cdots = \mathcal{S}_{k}\left(n; \bm{e}_{k-1}\right).
  \end{equation*}

\end{IEEEproof}


\section{Lemmas and Proofs for Section~\ref{sec:upper-bounds}} \label{appendix:upper}

\begin{proposition} \label{prop:ball-arbitrary}
  Let $\bm{s} \in \mathcal{X}_2^n$ be a $2$-resolution composite binary sequence of length $n$ with $j$ zeroes and $m$ ones. Denote by $r = n-m-j$ the number of twos in $\bm{s}$. Then
  \begin{equation*}
    \begin{split}
      |\mathcal{B}_{2, e}(\bm{s})| &= \sum_{i=0}^{e} \binom{m}{i} 2^i \sum_{\ell=0}^{e-i} \binom{n-m}{\ell} \sum_{p=0}^{\lfloor \frac{e-i-\ell}{2} \rfloor} \binom{n-m-\ell}{p}, \\
      |\mathcal{B}_{2, (e_0, e_1)}(\bm{s})| &= \sum_{\substack{a, b \geq 0\\ a+b \leq j}} \sum_{\substack{c, d \geq 0\\ c+d \leq m}} \sum_{\substack{e, f \geq 0 \\ e+f \leq r}} \mathds{1} \{\substack{b+d+e+f\leq e_0 \\ a+b+c+e\leq e_1}\} \binom{j}{a}\binom{j-a}{b}\binom{m}{c}\binom{m-c}{d}\binom{r}{e}\binom{r-e}{f}.
    \end{split}
  \end{equation*}
\end{proposition}

\begin{IEEEproof}

  \begin{itemize}
    \item We first compute the size of $\mathcal{B}_{2, e}(\bm{s})$.
    Figure \ref{fig:error-channel-e-cecc} shows the transformations that a letter in the reconstructed sequence $\bm{y} = \mathcal{R}(\bm{y}_0, \bm{y}_1)$ can undergo due to errors in the first and second channel outputs $\bm{y}_0$ and $\bm{y}_1$.
    Note that for the dashed edges, the price to pay is 2 errors, since we need both channels to introduce an error at the same position.
    For $\sigma \in \{0, 2\}$, consider the transformations in Figure \ref{fig:error-channel-e-cecc} where
    \begin{itemize}
      \item $i$ represents the number of transformations of type $1 \to \sigma$,
      \item $\ell$ represents the number of transformations of type $\sigma \to 1$, and
      \item $p$ represents the number of transformations of type $\sigma \to \sigma$ (dashed arrows).
    \end{itemize}
    There are $m$ positions where transformations of the first type can occur, and each transformation at a given position can result in one of two possible outputs, namely $\sigma=0$ or $\sigma=2$.
    Thus, the number of ways to introduce $i$ transformations of type $1 \to \sigma$ is $\binom{m}{i} 2^i$.
    There are $n-m$ positions where transformations of the second type can occur. Hence, the number of ways to introduce $\ell$ transformations of type $\sigma \to 1$ is $\binom{n-m}{\ell}$.
    Finally, there are $n-m-\ell$ positions where transformations of the third type can be occur. However, note that $\sigma \to \sigma$ transformations require errors in both channels.
    Therefore, at most $\lfloor \frac{e-i-\ell}{2} \rfloor$ such transformation can be introduced.

    \item Now we compute the size of $\mathcal{B}_{2, (e_0, e_1)}(\bm{s})$.  We choose the following transformations between any pair of letters.
      \begin{itemize}
        \item $a$ transformations of type $0 \to 1$. There are $j$ positions where these transformations can occur.
        \item $b$ transformations of type $0 \to 2$. There are $j-a$ positions where these transformations can occur.
        \item $c$ transformations of type $1 \to 0$. There are $m$ positions where these transformations can occur.
        \item $d$ transformations of type $1 \to 2$. There are $m-c$ positions where these transformations can occur.
        \item $e$ transformations of type $2 \to 0$. There are $r$ positions where these transformations can occur.
        \item $f$ transformations of type $2 \to 1$. There are $r-e$ positions where these transformations can occur.
      \end{itemize}
      The transformations that require errors in the first channel are $0 \to 2$, $1 \to 2$, $2 \to 1$ and $2 \to 2$. As such,
      we require that $b + d + e + f \leq e_0$ via the indicator function.
      The transformations that require errors in the second channel are $0 \to 1$, $0 \to 2$, $1 \to 0$ and $2 \to 0$. As such,
      we require that $a + b + c + e \leq e_1$ via the indicator function.

  \end{itemize}

\end{IEEEproof}


\begin{lemma}\label{lemma:asymp}
  For any positive integer $t$
  \begin{equation*}
      \sum_{i=0}^{\frac{n}{3} - \sqrt{t n \ln n}} \binom{n}{i} 2^{n-i} \lesssim \frac{3^n}{n^{\frac{9t}{4}}} \qquad \text{and} \qquad
      \sum_{i=\frac{n}{3} + \sqrt{t n \ln n}}^{n} \binom{n}{i} 2^{n-i} \lesssim \frac{3^n}{n^{\frac{9t}{4}}}.
  \end{equation*}
\end{lemma}

\begin{IEEEproof}
  Let $t$ be a positive integer. Let $\mathsf{X} \sim B(n, p)$ be a Binomial random variable with parameters $n \in \mathbb{N}$ and $p = \frac{1}{3}$.
  Then
  \begin{equation*}
    \begin{split}
    \mathbb{P}(\mathsf{X} \leq m) & = \sum _{i=0}^{m} \mathbb{P}(\mathsf{X} = i) = \sum _{i=0}^{m} \binom{n}{i} p^i (1-p)^{n-i} \\
    & = \sum _{i=0}^{m} \binom{n}{i} \left( \frac{1}{3} \right)^i \left( \frac{2}{3} \right)^{n-i} 
    = \frac{1}{3^n} \sum _{i=0}^{m} \binom{n}{i} 2^{n-i}.
    \end{split}
  \end{equation*}
  By to the Central Limit Theorem, as $n \to \infty$,
  \begin{equation*}
    \mathsf{Z} = \frac{\mathsf{X} - np}{\sqrt{np(1-p)}} \xrightarrow{d} \mathcal{N}(0, 1),
  \end{equation*} where $\xrightarrow{d}$ denotes convergence in distribution.
  As such it holds that $\mathbb{P}(\mathsf{X} \leq m) \simeq \mathbb{P}(\mathsf{Z} \leq z)$ for $z = \frac{m - np}{\sqrt{np(1-p)}} = \frac{m - \frac{n}{3}}{\sqrt{\frac{n}{3} \cdot \frac{2}{3}}}$.
  Specifically for $m_0 = \frac{n}{3} - \sqrt{t n \ln n}$, $ z_0 $ becomes
  \begin{equation*}
    z_0 = \frac{\frac{n}{3} - \sqrt{t n \ln n} - \frac{n}{3}}{\sqrt{\frac{n}{3} \cdot \frac{2}{3}}} = -3 \cdot \sqrt{\frac{t \ln n}{2}}.
  \end{equation*} Note that $z_0 < 0$.
  Let us compute $\mathbb{P}(\mathsf{Z} \leq z_0)$.
  \begin{equation*}
    \begin{split}
    \mathbb{P}(\mathsf{Z} \leq z_0) &= \mathbb{P}(\mathsf{Z} \geq |z_0|) = \int_{z_0}^{\infty} \frac{1}{\sqrt{2\pi}} e^{-\frac{x^2}{2}} dx 
    \overset{x > z_0}{\leq} \int_{z_0}^{\infty} \frac{x}{z_0} \frac{1}{\sqrt{2\pi}} e^{-\frac{x^2}{2}} dx \\
    & = \frac{1}{\sqrt{2\pi} z_0} \int_{z_0}^{\infty} x e^{-\frac{x^2}{2}} dx  = \frac{e^{\frac{-z_0^2}{2}}}{\sqrt{2\pi} z_0} 
    = \frac{e^{-\frac{9 t \ln n}{4}}}{3\sqrt{\pi t \ln n}} = \frac{n^{-\frac{9t}{4}}}{3\sqrt{\pi t \ln n}} \simeq \frac{1}{n^{\frac{9t}{4}}},
    \end{split}
  \end{equation*}
  meaning that $\mathbb{P}(\mathsf{Z} \leq z_0) \lesssim \frac{1}{n^\frac{9t}{4}}$.
  To recap, we have shown that
  \begin{equation*}
    \frac{1}{3^n} \sum _{i=0}^{\frac{n}{3} - \sqrt{t n \ln n}} \binom{n}{i} 2^{n-i} = \mathbb{P}(\mathsf{X} \leq \frac{n}{3} - \sqrt{t n \ln n}) 
    \simeq \mathbb{P}(\mathsf{Z} \leq -3 \sqrt{\frac{t\ln n}{2}})  \lesssim \frac{1}{n^{\frac{9t}{4}}},
  \end{equation*}
  as required.
  The second part of the lemma can be shown similarly. First,
  \begin{equation*}
    \mathbb{P}(\mathsf{X} \geq m) = \frac{1}{3^n} \sum _{i=m}^{n} \binom{n}{i} 2^{n-i},
  \end{equation*}
  and applying $m_1 = \frac{n}{3} + \sqrt{t n \ln n}$, then
  \begin{equation*}
    \mathbb{P}(\mathsf{X} \geq \frac{n}{3} + \sqrt{t n \ln n}) = \frac{1}{3^n} \sum _{i=\frac{n}{3} + \sqrt{t n \ln n}}^{n} \binom{n}{i} 2^{n-i}.
  \end{equation*}
  For such $m_1$, $z_1$ becomes
  \begin{equation*}
    z_1 = \frac{\frac{n}{3} + \sqrt{t n \ln n} - \frac{n}{3}}{\sqrt{\frac{n}{3} \cdot \frac{2}{3}}} = 3 \cdot \sqrt{\frac{t \ln n}{2}}.
  \end{equation*}
  Note that $z_1 > 0$. Similarly we have that $\mathbb{P}(\mathsf{Z} \geq z_1) \lesssim \frac{1}{n^{\frac{9t}{4}}}$.
  Therefore,
  \begin{equation*}
    \frac{1}{3^n} \sum _{i=\frac{n}{3} + \sqrt{t n \ln n}}^{n} \binom{n}{i} 2^{n-i} = \mathbb{P}(\mathsf{X} \geq \frac{n}{3} + \sqrt{t n \ln n}) 
    \simeq \mathbb{P}(\mathsf{Z} \geq 3 \sqrt{\frac{t\ln n}{2}})  \lesssim \frac{1}{n^{\frac{9t}{4}}}.
  \end{equation*}
  
\end{IEEEproof}


\begin{lemma} \label{lemma:k-resolution-1-cecc}
  For $k \geq 2$ and $n \in \mathbb{N}$
    \begin{equation*}
      \frac{\left( \frac{k+1}{2} \right)^{n}}{\frac{2kn}{k+1}} \leq \sum_{m=0}^n \binom{n}{m} \left(\frac{k-1}{2}\right)^m \frac{1}{n + m} \leq \frac{\left( \frac{k+1}{2} \right)^{n}}{\frac{2kn}{k+1} - 1}.
    \end{equation*}
\end{lemma}

\begin{IEEEproof}
  By the binomial theorem, we have that
  \begin{equation*}
     \sum_{m=0}^n \binom{n}{m} \left(\frac{k-1}{2}\right)^m = \left( \frac{k+1}{2} \right)^{n},
  \end{equation*}
  hence, we need to show that
  \begin{equation*}
    \frac{1}{\frac{2kn}{k+1}} \leq \sum_{m=0}^n \binom{n}{m} \frac{\left(\frac{k-1}{2}\right)^m}{\left( \frac{k+1}{2} \right)^{n}} \frac{1}{n + m} \leq \frac{1}{\frac{2kn}{k+1} - 1}.
  \end{equation*}
  Let $\mathsf{X} \sim B(n, p)$ be a Binomial random variable with parameters $n \in \mathbb{N}$ and $p = \frac{k-1}{k+1}$. Then $q = 1 - p = \frac{2}{k+1}$.
  We first note that
  \begin{equation*}
    \begin{split}
      \mathbb{P}(\mathsf{X} = m) &= \binom{n}{m} p^m q^{n-m} 
      = \binom{n}{m} \left(\frac{k-1}{k+1}\right)^m \left(\frac{2}{k+1}\right)^{n-m} \\
      & = \binom{n}{m} \frac{(k-1)^m}{(k+1)^n} \cdot 2^{n-m}
      = \binom{n}{m} \frac{\left( \frac{k-1}{2} \right)^m}{\left( \frac{k+1}{2} \right)^{n}}.
    \end{split}
  \end{equation*}
  Hence, we want to show that
  \begin{equation*}
    \frac{1}{\frac{2kn}{k+1}} \leq \sum_{m=0}^n \mathbb{P}(\mathsf{X} = m) \frac{1}{n + m} \leq \frac{1}{\frac{2kn}{k+1} - 1}, 
  \end{equation*}
  or equivalently
  \begin{equation*}
    \frac{1}{\frac{2kn}{k+1}} \leq \mathbb{E}[\frac{1}{n + m}] \leq \frac{1}{\frac{2kn}{k+1} - 1}.
  \end{equation*}
  First, we show the left inequality.
  \begin{equation*}
    \mathbb{E}[n +m] = \mathbb{E}[n] + \mathbb{E}[m] = n + n\cdot p = n \cdot (1 + p ) = n \cdot \left(1 + \frac{k-1}{k+1}\right) = \frac{2kn}{k+1}.
  \end{equation*}
  Since $f(x) = \frac{1}{x}$ is a convex function, we can apply Jensen's inequality to obtain
  \begin{equation*}
    \frac{1}{\frac{2kn}{k+1}} = f(\mathbb{E}[n+m]) \leq \mathbb{E}[f(n+m)] = \mathbb{E}[\frac{1}{n+m}].
  \end{equation*}
  To show the right inequality, we use a known bound on the Jensen gap.
  If we assume that $f(x)$ is twice differentiable and there exists $\Lambda$ such that $f''(x) \leq \Lambda$, then we have
  \begin{equation*}
    \mathbb{E}[f(\mathsf{X})] - f(\mathbb{E}[\mathsf{X}]) \leq \frac{\Lambda}{2} \cdot \text{Var}(\mathsf{X}).
  \end{equation*}
  If we apply this to our case, $f(x) = \frac{1}{x}$ is twice differentiable for $x > 0$ and we get
  \begin{equation*}
    \mathbb{E}[\frac{1}{n+m}] = \mathbb{E}[f(n+m)] \leq f(\mathbb{E}[n+m]) + \frac{\Lambda}{2} \cdot \text{Var}(n+m) =  \frac{1}{\mathbb{E}[n+m]} + \frac{\Lambda}{2} \cdot \text{Var}(n+m).
  \end{equation*}
  It holds that
  \begin{equation*}
    \text{Var}(n+m) = \text{Var}(m) = n \cdot p \cdot q = n \cdot \frac{k-1}{k+1} \cdot \frac{2}{k+1} = \frac{2n(k-1)}{(k+1)^2}.
  \end{equation*}
  The second derivative of $f(x) = \frac{1}{x}$ is given by $f''(x) = \frac{2}{x^3}$. Since $0 \leq m \leq n$, then $n \leq n+m \leq 2n$, hence $f''(x) \in [\frac{2}{8n^3}, \frac{2}{n^3}]$.
  As such we can pick $\Lambda = \frac{2}{n^3}$, which gives us
  \begin{equation*}
    \mathbb{E}[\frac{1}{n+m}] \leq \frac{1}{\mathbb{E}[n+m]} + \frac{\Lambda}{2} \cdot \text{Var}(n+m) = \frac{1}{\frac{2kn}{k+1}} + \frac{2(k-1)}{(k+1)^2 \cdot n^2}.
  \end{equation*}
  It remains to show that 
  \begin{equation*}
    \frac{1}{\frac{2kn}{k+1}} + \frac{2(k-1)}{(k+1)^2 \cdot n^2} \leq \frac{1}{\frac{2kn}{k+1} - 1},
  \end{equation*}
  which can be shown that it holds for all $n \in \mathbb{N}, k \geq 2$ by some algebraic manipulations.
\end{IEEEproof}


\begin{lemma} \label{lemma:k-resolution-shift}
  For $k \geq 2$, $j < n -1$ and $j = o(n)$, there exists $M \in \mathbb{N}$ such that for all $n \geq M$ it holds that
    \begin{equation*}
      \sum_{m=0}^n \binom{n}{m} \left(\frac{k-1}{2}\right)^m \frac{1}{n + m - j} \leq  \frac{\left( \frac{k+1}{2} \right)^{n}}{\frac{2kn}{k+1} - (j+1)}.
    \end{equation*}
\end{lemma}

\begin{IEEEproof}
  The only difference between this lemma and \cref{lemma:k-resolution-1-cecc} is that we shift the denominator by $j$.
  The proof is very similar. Let $\mathsf{X}$ be the Binomial random variable defined in the proof of Lemma \ref{lemma:k-resolution-1-cecc}.
  We want to show that
  \begin{equation*}
    \mathbb{E}[\frac{1}{n + m - j}] \leq \frac{1}{\frac{2kn}{k+1} - (j+1)}.
  \end{equation*}
  By the same reasoning on the bound on the Jensen gap, we have
  \begin{equation*}
    \mathbb{E}[\frac{1}{n + m - j}] \leq \frac{1}{\mathbb{E}[n + m - j]} + \frac{\Lambda}{2} \cdot \text{Var}(n + m - j).
  \end{equation*}
  Note that
  \begin{equation*}
    \begin{split}
    \mathbb{E}[n + m - j] &= n + \mathbb{E}[m] - j = n + n \cdot p - j = n \cdot (1 + p) - j = \frac{2kn}{k+1} - j, \\
    \text{Var}(n + m - j) &= \text{Var}(m) = n \cdot p \cdot q = n \cdot \frac{k-1}{k+1} \cdot \frac{2}{k+1} = \frac{2(k-1)n}{(k+1)^2}.
    \end{split}
  \end{equation*}
  Since $0 \leq m \leq n$, then $n-j \leq n+m-j \leq 2n-j$, hence $f''(x) \in [\frac{2}{(2n-j)^3}, \frac{2}{(n-j)^3}]$.
  As such we can pick $\Lambda = \frac{2}{(n-j)^3}$, which gives us
  \begin{equation*}
    \mathbb{E}[\frac{1}{n+m-j}] \leq \frac{1}{\mathbb{E}[n+m-j]} + \frac{\Lambda}{2} \cdot \text{Var}(n+m-j) = \frac{1}{\frac{2kn}{k+1} -j} + \frac{2(k-1)n}{(k+1)^2 \cdot (n-j)^3}.
  \end{equation*}
  Hence, it remains to show that 
  \begin{equation*}
    \frac{1}{\frac{2kn}{k+1} - j} + \frac{2(k-1)n}{(k+1)^2 \cdot (n-j)^3} \leq \frac{1}{\frac{2kn}{k+1} - (j+1)},
  \end{equation*}
  or equivalently
  \begin{equation*}
    \begin{split}
    0 & \leq \frac{1}{\frac{2kn}{k+1} - (j+1)} - \frac{1}{\frac{2kn}{k+1} - j} - \frac{2(k-1)n}{(k+1)^2 \cdot (n-j)^3} \\
    & = \frac{k+1}{2kn - (k+1)(j+1)} - \frac{k+1}{2kn - (k+1)\cdot j} - \frac{2(k-1)n}{(k+1)^2 \cdot (n-j)^3} \\
    & = \frac{(k+1)^4(n-j)^3 - 2(k-1)n\cdot \left(2kn -(k+1)(j+1)\right)\cdot \left(2kn - (k+1)\cdot j\right)}{\left( 2kn - (k+1)(j+1) \right) \cdot \left( 2kn - (k+1)\cdot j\right) \cdot (k+1)^2 \cdot (n-j)^3}.
    \end{split}
  \end{equation*}
  Since $j < n-1$ and $k \geq 2$ then all the terms in the denominator are positive.
  It remains to show that the numerator is non-negative. This certainly holds if 
  \begin{equation*}
    \begin{split}
      & (k+1)^4(n-j)^3 - 2(k-1)n\cdot \left(2kn -(k+1)(j+1)\right)\cdot \left(2kn - (k+1)\cdot j\right) \\
      & = (k+1)^4(n-j)^3 - 2(k-1)n\cdot \left(2kn\right)\cdot \left(2kn\right) \\
      & = (k+1)^4(n-j)^3 - 8(k-1)k^2 n^3 \\
      & \geq 0.
    \end{split}
  \end{equation*}
  Note that for all $k \geq 2$ it holds that $(k+1)^4 > 8 (k-1)k^2$, and since $j = o(n)$, then the term $n^3$ which dominates has a positive coefficient.
  Hence the numerator is non-negative for all $n$ large enough, and the lemma holds.

\end{IEEEproof}


\begin{lemma} \label{lemma:k-resolution-1-1-cecc}
  For $k \geq 2$ and $n \geq 4$
  \begin{equation*}
    \sum_{m=1}^{n} \binom{n}{m}\left( \frac{k-1}{2} \right)^m \frac{1}{m} \leq 
    \sum_{m=1}^{n} \binom{n}{m} \left( \frac{k-1}{2} \right)^m \frac{1}{\frac{(k-1)n}{k+1}-1} 
    = \frac{(\frac{k+1}{2})^n}{\frac{(k-1)n}{k+1}-1}.
  \end{equation*}
\end{lemma}

\begin{IEEEproof}
  The idea of the proof is similar to the one in Lemma \ref{lemma:k-resolution-1-cecc}. 
  Let $\mathsf{X}$ be the Binomial random variable defined in the proof of Lemma \ref{lemma:k-resolution-1-cecc}. 
  We will assume that $\mathsf{X} > 0$ as $\mathsf{X} = 0$ does not contribute to the expected value.
  We want to show that
  \begin{equation*}
    \mathbb{E}[\frac{1}{m}] \leq \frac{1}{\frac{(k-1)n}{k+1}-1}.
  \end{equation*}
  By the same reasoning on the bound on the Jensen gap, we have
  \begin{equation*}
    \mathbb{E}[\frac{1}{m}] \leq \frac{1}{\mathbb{E}[m]} + \frac{\Lambda}{2} \cdot \text{Var}(m).
  \end{equation*}
  Note that
  \begin{equation*}
    \begin{split}
    \mathbb{E}[m] &= n \cdot p = n \cdot \frac{k-1}{k+1} = \frac{(k-1)n}{k+1}, \\
    \text{Var}(m) &= n \cdot p \cdot q = n \cdot \frac{k-1}{k+1} \cdot \frac{2}{k+1} = \frac{2n(k-1)}{(k+1)^2}.
    \end{split}
  \end{equation*}
  and similarly to the proof of Lemma \ref{lemma:k-resolution-1-cecc}, we can take $\Lambda = \frac{2}{n^3}$.
  Hence, it holds that
  \begin{equation*}
    \mathbb{E}[\frac{1}{m}] \leq \frac{1}{\frac{(k-1)n}{k+1}} + \frac{\Lambda}{2} \cdot \text{Var}(m) = \frac{1}{\frac{(k-1)n}{k+1}} + \frac{2(k-1)}{(k+1)^2 \cdot n^2}.
  \end{equation*}
  It remains to show that
  \begin{equation*}
    \frac{1}{\frac{(k-1)n}{k+1}} + \frac{2(k-1)}{(k+1)^2 \cdot n^2} \leq \frac{1}{\frac{(k-1)n}{k+1}-1},
  \end{equation*}
  which can be shown that it holds for all $n \geq 4, k \geq 2$ by some algebraic manipulations.
\end{IEEEproof}


\begin{lemma}\label{lemma:1-1-helper-2}
  \begin{equation*}
    \sum_{m=0}^{n} \binom{n}{m} \sum_{j=0}^{n-m} \binom{n-m}{j} \frac{1}{(j+1)(n-m-j+1)} \leq \frac{3^{n+2}}{(n+1)(n+2)}.
  \end{equation*}
\end{lemma}

\begin{IEEEproof}
  First note that 
  \begin{equation}\label{eq:1-1-helper-3}
    \frac{1}{n-m+2}\left( \frac{1}{j+1} + \frac{1}{n-m-j+1}\right) = \frac{1}{(j+1)(n-m-j+1)}.
  \end{equation}
  Now
  \begin{equation*}
    \begin{split}
      & \sum_{m=0}^{n} \binom{n}{m} \sum_{j=0}^{n-m} \binom{n-m}{j} \frac{1}{(j+1)(n-m-j+1)} \\
      & \overset{(\ref{eq:1-1-helper-3})}{=} \sum_{m=0}^{n} \frac{1}{n-m+2} \binom{n}{m} \sum_{j=0}^{n-m} \binom{n-m}{j} \left( \frac{1}{j+1} + \frac{1}{n-m-j+1}\right) \\
      & = \sum_{m=0}^{n} \frac{1}{n-m+2} \binom{n}{m} \left( \sum_{j=0}^{n-m} \binom{n-m}{j} \frac{1}{j+1} + \sum_{j=0}^{n-m} \binom{n-m}{j} \frac{1}{n-m-j+1} \right) \\
      & \overset{\text{(BI)}}{=} \sum_{m=0}^{n} \frac{1}{n-m+2} \binom{n}{m} \left(  \frac{2^{n-m+1}-1}{n-m+1} + \frac{2^{n-m+1}-1}{n-m+1} \right) \\
      & = \sum_{m=0}^{n} \binom{n}{m} \frac{2^{n-m+2}-2}{(n-m+2)(n-m+1)} \\
      & \leq \sum_{m=0}^{n} \binom{n}{m} \frac{2^{n-m+2}}{(n-m+2)(n-m+1)} \\
      & = \sum_{m=0}^{n} \binom{n}{n-m} \frac{2^{n-m+2}}{(n-m+2)(n-m+1)} \\
      & \overset{\ell \triangleq n-m}{=} \sum_{\ell=0}^{n} \binom{n}{\ell} \frac{2^{\ell+2}}{(\ell+2)(\ell+1)} \\
      & \overset{\text{(BI)}}{=} \frac{3^{n+2} - 2(n+2) -1}{(n+1)(n+2)} \leq \frac{3^{n+2}}{(n+1)(n+2)},
    \end{split}
  \end{equation*}
  where $\overset{\text{(BI)}}{=}$ indicates an application of the binomial identities listed in Appendix~\ref{appendix:binom-identities}.
\end{IEEEproof}


\ballOneOne*

\begin{IEEEproof}
  In addition to the errors we considered in the scenario of $2$-resolution single-CECC codes, we now add the cases where both channels introduce exactly one error.
  We categorize these new errors in the following cases.
  \begin{itemize}
    \item The errors occur at different positions where $\bm{s}$ takes values $ \sigma \in \{0, 2\}$. Note that the only transformations possible are $0 \to 1$ and $2 \to 1$.
    The former requires an error in the second channel, while the latter an error in the first. Therefore, one transformation must be $0 \to 1$, while the other $2 \to 1$, yielding 
    $ j \cdot (n-m-j)$ such combinations of positions.
    \item The errors occur at different positions where $\bm{s}$ takes values $\sigma = 1$. There are $\binom{m}{2}$ such combinations of positions.
    Note that in one of the positions we can pick the channel that introduced the error, therefore we have $2 \binom{m}{2}$ such combinations.
    \item Next suppose that one error occurs at a position where $\bm{s}$ takes the value $\sigma = 1$ and the other at a position where $\bm{s}$ takes the value $\sigma \in \{0, 2\}$. There are $m(n-m)$ such combinations of positions.
    The letter $\sigma \in \{0, 2\}$ automatically determines which channel introduces the error, leaving no option for the other letter.
    \item Finally, both errors occur at the same position. In the case the errors occurred at a position where $\bm{s}$ takes the value $\sigma = 1$ this yields an invalid sequence.
    This leaves us with $n-m$ valid combinations of positions.
  \end{itemize}
  To summarize, we have
  \begin{equation*}
    \begin{split}
      |\mathcal{B}_{2, (1, 1)}(\bm{s})| & = |\mathcal{B}_{2, 1}(\bm{s})| + j (n-m-j) + 2\binom{m}{2} + m (n-m) + (n-m) \\
      & = 2n + 1 + m (n-1) + j (n-m-j) \\
      & = 2n + 1 + m (n-1) + j (n-m-1) - (j^2 - j).
    \end{split}
  \end{equation*}
\end{IEEEproof}


\gspbOneOne*

\begin{IEEEproof}
  We iterate over the fractional transversal weights $w_i$ based on the number of ones $m$ and the number of zeroes $j$ in the $2$-resolution composite binary sequence $\bm{s} \in \mathcal{X}$.
  \begin{equation*}
    \begin{split}
      \sum _{i=1}^N w_i & \leq \sum_{m=0}^n \binom{n}{m} \sum_{j=0}^{n-m} \binom{n-m}{j} \frac{1}{m(n-1) + (j+1) (n - m - j + 1)} \\
      & \overset{(a)}{\leq} \sum_{m=1}^n \binom{n}{m} \sum_{j=0}^{n-m} \binom{n-m}{j} \frac{1}{2} \left( \frac{1}{m(n-1)} + \frac{1}{(j+1) (n - m - j + 1)}\right) \\
      & = \frac{1}{2} \left( \sum_{m=1}^n \binom{n}{m} \frac{2^{n-m}}{m(n-1)} + \sum_{m=0}^n \binom{n}{m} \sum_{j=0}^{n-m} \binom{n-m}{j} \frac{1}{(j+1) (n - m - j + 1)} \right) \\
      & \overset{(b)}{\leq} \frac{1}{2} \left( \frac{2^n}{n-1} \sum_{m=1}^n \binom{n}{m} \left( \frac{1}{2} \right)^m \frac{1}{m} + \frac{3^{n+2}}{(n+1)(n+2)} \right) \\
      & \overset{(c)}{\leq} \frac {1}{2} \left( \frac{3^{n+1}}{(n-1)(n-3)} + \frac{3^{n+2}}{(n+1)(n+2)} \right) \\
      & \overset{(d)}{\leq} = \frac{3^n}{2} \left( \frac{3}{(n-3)^2} + \frac{9}{(n-3)^2} \right) \\
      & = \frac{3^n}{\frac{(n-3)^2}{6}},
    \end{split}
  \end{equation*}
  where 
  \begin{itemize}
    \item $(a)$ uses the well-known identity $\frac{1}{x + y} \leq \frac{1}{2}\left(\frac{1}{x} + \frac{1}{y}\right)$ for $x, y > 0$,
    \item $(b)$ utilizes the result of \cref{lemma:1-1-helper-2}, 
    \item $(c)$ utilizes the result from \cref{lemma:k-resolution-1-1-cecc} for resolution parameter $k=2$, and
    \item $(d)$ uses the fact that for all $ n \geq 4$ it holds that $(n-1)(n-3) \geq (n-3)^2$ and $(n+1)(n+2) \geq (n-3)^2$.
  \end{itemize}
\end{IEEEproof}


\ballTwo*

\begin{IEEEproof}
  The errors can occur in any distribution between the two channels. We categorize these new errors in the following cases.
  \begin{itemize}
  \item The errors occur at different positions where $\bm{s}$ takes values $ \sigma \in \{0, 2\}$. In this case only one $\bm{y} \in \mathcal{B}_{2, 2}(\bm{s})$
  can be received, since both 0 and 2 can only be converted to 1. There are $\binom{n-m}{2}$ such combinations of positions.
  \item The errors occur at different positions where $\bm{s}$ takes value $\sigma = 1$. There are $\binom{m}{2}$ such combinations of positions.
  In this case 4 possible $\bm{y} \in \mathcal{B}_{2, 2}(\bm{s})$ can be received, since each such letter can be converted to a 0 or to a 2.
  Therefore $ 4 \binom{m}{2}$ possible $\bm{y} \in \mathcal{B}_{2, 2}(\bm{s})$ can be received.
  \item One error occur at a position where $\bm{s}$ takes the value $\sigma = 1$ and the other at a position where $\bm{s}$ takes a value $\sigma \in \{0, 2\}$. There are $m(n-m)$ such combinations of positions.
  In this case 2 possible $\bm{y} \in \mathcal{B}_{2, 2}(\bm{s})$ can be received, since $\sigma = 1$ can be converted to a 0 or to a 2.
  Therefore $2\cdot m\cdot(n-m)$ possible $\bm{y} \in \mathcal{B}_{2, 2}(\bm{s})$ can be received.
  \item The errors occur at the same position. In the case the errors occurred at a position where $\bm{s}$ takes the value $\sigma = 1$ this yields an invalid sequence.
  This leaves us with $n-m$ valid combinations of positions.
  \end{itemize}

  To summarize, we have
  \begin{equation*}
    \begin{split}
      |\mathcal{B}_{2, 2}(\bm{s})|  & = |\mathcal{B}_{2, 1}(\bm{s})| + \binom{n-m}{2} + 4 \binom{m}{2} + 2(n-m)m + (n-m)  \\
      & = \frac{n^2}{2} + \frac{3n}{2} + 1 + m(n-1) + \frac{m^2-m}{2}.
    \end{split}
  \end{equation*}

\end{IEEEproof}


\gspbTwo*

\begin{IEEEproof}
  We iterate over the fractional transversal weights $w_i$ based on the number of ones $m$ in the $2$-resolution composite binary sequence $\bm{s} \in \mathcal{X}$.
  For $j \geq 7$,
  \begin{equation*}
    \begin{split}
      \sum _{i=1}^N w_i & \leq \sum_{m=0}^n \binom{n}{m} 2^{n-m}  \frac{2}{(n+m)^2 - n - 7m} \\
      & = 2^n \left( \sum_{m=0}^n \binom{n}{m} \left( \frac{1}{2} \right)^{m}  \frac{2}{(n+m)^2 - n - 7m} \right) \\
      & \leq 2^n \left( \sum_{m=0}^n \binom{n}{m} \left( \frac{1}{2} \right)^{m} \frac{2}{(n+m-j)(n+m)} \cdot \frac{j}{j} \right) \\
      & = 2^n \cdot \frac{2}{j} \left(  \sum_{m=0}^n \binom{n}{m} \left( \frac{1}{2} \right)^{m} \left( \frac{1}{n+m-j} - \frac{1}{n+m} \right) \right) \\
      & = 2^n \cdot \frac{2}{j} \left( \sum_{m=0}^n \binom{n}{m} \left( \frac{1}{2} \right)^{m} \frac{1}{n+m-j} - \sum_{m=0}^n \binom{n}{m} \left( \frac{1}{2} \right)^{m} \frac{1}{n+m} \right) \\
      & \overset{(a)}{\leq} 2^n \cdot \frac{2}{j} \left( \frac{(\frac{3}{2})^n}{\frac{4n}{3}-(j+1)}  - \frac{(\frac{3}{2})^n}{\frac{4n}{3}} \right) \\
      & = 3^n \cdot \frac{j+1}{j} \cdot \frac{1}{\frac{8n^2}{9} - \frac{2n(j+1)}{3}},
    \end{split}
  \end{equation*}
  where $(a)$ is the application of \cref{lemma:k-resolution-1-cecc} and \cref{lemma:k-resolution-shift} for resolution parameter $k=2$.
  We derive the result by $j$ and deduce that for $j=a\sqrt{n}-1$ where $a=\frac{\sqrt{8}}{\sqrt{6}}$ the tightest bound is achieved.
  For this $j$ the bound becomes
  \begin{equation*}
    \sum _{i=1}^N w_i \leq \frac{a\sqrt{n}}{a\sqrt{n}-1} \cdot \frac{3^n}{\frac{8n^2}{9}-\frac{2n (a \sqrt{n})}{3}}
    = \frac{\sqrt{\frac{8n}{6}}}{\sqrt{\frac{8n}{6}}-1} \cdot \frac{3^n}{\frac{8n^2}{9}-\frac{2n (\sqrt{\frac{8n}{6}})}{3}}.
  \end{equation*}
  Since we required $7 \leq j = a \sqrt{n} - 1$, then we require $n \geq 48$.
  Additionally, for resolution parameter $k=2$, \cref{lemma:k-resolution-shift} holds only for such $n$ that satisfy
  \begin{equation*}
    \begin{split}
      0 \leq (k+1)^4(n-j)^3 - 8 (k-1) k^2 n^3 = 81 \left( n - \sqrt{\frac{8n}{6}} + 1 \right)^3 - 32n^3.
    \end{split}
  \end{equation*}
  This last inequality is satisfied for all $n \geq 10$, and thus the theorem holds for $n \geq 48$.
\end{IEEEproof}


\averageBallSize*

\begin{IEEEproof}
  Throughout this proof we extensively use the binomial identities in Appendix~\ref{appendix:binom-identities}, marking the relevant equality signs with $\overset{\text{(BI)}}{=}$ whenever such an identity is applied.
  We begin by computing $\bar{\Delta}_{k, (1, 0, \ldots, 0)}$.
  To do so, we iterate over all $k$-resolution composite binary sequences $\bm{s}$ of length $n$, grouping them according to the value $m \triangleq \#_{k-1}(\bm{s}) + \#_{k}(\bm{s})$, as in \cref{prop:ball-1-0-general}. 
  From that proposition, we use the fact that $|\mathcal{B}_{k, (1, 0, \ldots, 0)}(\bm{s})| = 1 + m$.
  \begin{equation*}
    \begin{split}
      \bar{\Delta}_{k, (1, 0, \ldots, 0)} &= \frac{1}{ |\mathcal{X}| } \sum_{\bm{s} \in \mathcal{X}} |\mathcal{B}_{k, (1, 0, \ldots, 0)}(\bm{s})| = 
      \frac{1}{\left( k+1 \right)^n} \sum_{m=0}^n \binom{n}{m} 2^{m} (k-1)^{n-m} (1 + m) \\
      & = \left(\frac{k-1}{k+1}\right)^n  \left( \sum_{m=0}^n \binom{n}{m} \left(\frac{2}{k-1}\right)^{m} + \frac{2}{k-1} \sum_{m=0}^n m \binom{n}{m} \left(\frac{2}{k-1}\right)^{m-1}  \right) \\ 
      & \overset{\text{(BI)}}{=} \left(\frac{k-1}{k+1}\right)^n \left( \left(\frac{k+1}{k-1}\right)^n + \frac{2n}{k-1}\left(\frac{k+1}{k-1}\right)^{n-1} \right) \\
      & = \frac{2n}{k+1} + 1.
    \end{split}
  \end{equation*}

  Next, we compute $\bar{\Delta}_{k, 1}$.
  To do so, we iterate over all $k$-resolution composite binary sequences $\bm{s}$ of length $n$, grouping them according to the value $m \triangleq \sum_{i=1}^{k-1} \#_{i}(\bm{s})$, as in \cref{prop:ball-1-cecc-general}. 
  From that proposition, we use the fact that $|\mathcal{B}_{k, 1}(\bm{s})| = 1 + n + m$.
  \begin{equation*}
    \begin{split}
      \bar{\Delta}_{k, 1} &= \frac{1}{ |\mathcal{X}| } \sum_{\bm{s} \in \mathcal{X}} |\mathcal{B}_{k, 1}(\bm{s})| = \frac{1}{(k+1)^n} \sum_{m=0}^n \binom{n}{m} 2^{n-m} (k-1)^m (1 + n + m) \\
      &= \left( \frac{2}{k+1} \right)^n \left( (n+1)\sum_{m=0}^n \binom{n}{m} \left(\frac{k-1}{2}\right)^{m} + \frac{k-1}{2} \sum_{m=0}^n m \binom{n}{m} \left(\frac{k-1}{2}\right)^{m-1} \right) \\
      & \overset{\text{(BI)}}{=} \left( \frac{2}{k+1} \right)^n \left( (n+1)\cdot \left(\frac{k+1}{2}\right)^{n} + \frac{(k-1) n}{2} \cdot \left(\frac{k+1}{2}\right)^{n-1} \right) \\
      & = n+1 + \frac{k-1}{k+1} \cdot n = \frac{2kn}{k+1} + 1.
    \end{split}
  \end{equation*}

  We now compute $\bar{\Delta}_{2, (1, 1)}$.
  To do so, we iterate over all $2$-resolution composite binary sequences $\bm{s}$ of length $n$, grouping them according to the number of zeroes $j$ and ones $m$ in the sequence, as in \cref{prop:ball-1-1}.
  From that proposition, we use the fact that $|\mathcal{B}_{2, (1, 1)}(\bm{s})| = 2n + 1 + m (n-1) + j (n-m-j) = 2n + 1 + m(n -1) + j (n-m-1) - (j^2-j).$
  \begin{equation*}
    \begin{split}
      \bar{\Delta}_{2, (1, 1)} &= \frac{1}{ |\mathcal{X}| } \sum_{\bm{s} \in \mathcal{X}} |\mathcal{B}_{2, (1, 1)}(\bm{s})| 
      = \frac{1}{3^n} \sum_{m=0}^n \binom{n}{m} \left( \sum_{j=0}^{n-m} \binom{n-m}{j} \left( 2n + 1 + m(n -1) + j (n-m-1) - (j^2-j) \right)\right) \\
      & = \frac{1}{3^n} \sum_{m=0}^{n} \binom{n}{m} \left[ \left( 2n + 1 + m(n -1) \right) \sum_{j=0}^{n-m} \binom{n-m}{j}
      + (n-m-1) \sum_{j=0}^{n-m} j \binom{n-m}{j} - \sum_{j=0}^{n-m} (j^2 - j) \binom{n-m}{j} \right] \\
      & \overset{\text{(BI)}}{=} \frac{1}{3^n} \sum_{m=0}^{n} \binom{n}{m} \left[ \left( 2n + 1 + m(n-1) \right) 2^{n-m} 
      + (n-m-1) (n-m) 2^{n-m-1} - (n-m)(n-m-1) 2^{n-m-2} \right] \\
      & = \frac{2^n}{3^n}\sum_{m=0}^{n} \binom{n}{m} \cdot \left(\frac{1}{2}\right)^m \cdot \left[ \frac{n^2}{4} + \frac{7n}{4} + 1 + \frac{n-1}{2}\cdot m + \frac{m^2-m}{4} \right] \\
      & \overset{\text{(BI)}}{=} \frac{2^n}{3^n} \left[ (\frac{n^2}{4} + \frac{7n}{4} + 1) \cdot \left(\frac{3}{2}\right)^n 
      + \frac{n^2-n}{4} \cdot \left(\frac{3}{2}\right)^{n-1} +  \frac{(n^2-n)}{16}\cdot\left(\frac{3}{2}\right)^{n-2}\right] \\
      & = \frac{4n^2}{9} + \frac{14n}{9} + 1.
    \end{split}
  \end{equation*}

  Lastly, we compute $\bar{\Delta}_{2, 2}$.
  To do so, we iterate over all $2$-resolution composite binary sequences $\bm{s}$ of length $n$, grouping them according to the number of ones $m$ in the sequence, as in \cref{prop:ball-2}.
  From that proposition, we use the fact that $|\mathcal{B}_{2, 2}(\bm{s})| = \frac{n^2}{2} + \frac{3n}{2} + 1 + m(n-1) + \frac{m^2-m}{2}$.
  \begin{equation*}
    \begin{split}
      \bar{\Delta}_{2, 2} &= \frac{1}{ |\mathcal{X}| } \sum_{\bm{s} \in \mathcal{X}} |\mathcal{B}_{2, 2}(\bm{s})| 
      = \frac{1}{3^n} \sum_{m=0}^n \binom{n}{m} 2^{n-m} \left( \frac{n^2}{2} + \frac{3n}{2} + 1 + m(n-1) + \frac{m^2-m}{2} \right) \\
      &= \frac{2^n}{3^n} \left[ 
      \left(\frac{n^2}{2} + \frac{3n}{2} + 1\right) \sum_{m=0}^n \binom{n}{m} \left(\frac{1}{2}\right)^{m}
      + \frac{n-1}{2}\sum_{m=0}^n m \binom{n}{m} \left(\frac{1}{2}\right)^{m-1} + 
      \frac{1}{8} \sum_{m=0}^n (m^2-m) \binom{n}{m} \left(\frac{1}{2}\right)^{m-2} 
      \right] \\
      & \overset{\text{(BI)}}{=} \frac{2^n}{3^n} \left[ 
      \left(\frac{n^2}{2} + \frac{3n}{2} + 1\right) \cdot \left(\frac{3}{2}\right)^n
      + \frac{(n-1)}{2} \cdot n \cdot \left(\frac{3}{2}\right)^{n-1} 
      + \frac{n^2-n}{8} \cdot \left(\frac{3}{2}\right)^{n-2}\right]
      = \frac{8n^2}{9} + \frac{10n}{9} + 1.
    \end{split}
  \end{equation*}

\end{IEEEproof}


\section{Proofs for Section~\ref{sec:lower-bounds}} \label{appendix:lower-bounds}

\lowerBoundBCH*

\begin{IEEEproof}
  For $q=k+1$, $m = \lceil \log_q(n+1)\rceil$ and distance $d = 2e+1$ consider the primitive $q$-ary BCH code with parameters $[q^m-1, q^m - 2 - m \lceil \frac{(d-2)(q-1)}{q}\rceil, d]$, as shown in Problem 8.12 of~\cite{Roth}.
  The distance of the code is $d = 2e + 1$ and therefore it can correct up to $e$ substitution errors. By shortening this code, we obtain a code of length $n$
  that can correct up to $e$ substitution errors, and cardinality
  \begin{equation*}
    \frac{q^n}{q ^ {m \lceil \frac{(q-1)(2e-1)}{q}\rceil + 1}} = \frac{q^n}{q ^ {\lceil\log_q (n+1) \rceil \cdot \lceil \frac{(q-1)(2e-1)}{q}\rceil + 1}} 
    = \frac{\left(k+1\right)^n}{\left(k+1\right) ^ {\lceil\log_{k+1} (n+1) \rceil \cdot \lceil \frac{k(2e-1)}{k+1}\rceil + 1}}.
  \end{equation*}
  Therefore,
  \begin{equation*}
    \mathcal{S}_{k}\left(n; e\right) \geq \mathcal{A}_{k+1}(n; e) \geq \frac{\left(k+1\right)^n}{\left(k+1\right) ^ {\lceil\log_{k+1} (n+1) \rceil \cdot \lceil \frac{k(2e-1)}{k+1}\rceil + 1}}.
  \end{equation*}
\end{IEEEproof}


\lowerBoundCosets*

\begin{IEEEproof}
  For $m = \lceil \log_2(n+1) \rceil$ and odd distance $d > 2$ consider the binary primitive and narrow-sense $\text{BCH}_{m, d}$ code with parameters 
  $[2^m-1, 2^m-1 - m \cdot \frac{d-1}{2}, d]$. 
  By shortening the code $\text{BCH}_{m,d}$, we can obtain a code of length $n$ with the same redundancy.
  The code $\text{BCH}_{m, d}$ partitions the space of $\{0, 1\}^n$ into $2^{m \cdot \frac{d-1}{2}}$ cosets.
  For each $0 \leq i \leq k-1$, consider $\mathcal{C}_i$ to be a coset of the potentially shortened $\text{BCH}_{m, 2e_i+1}$.
  Then, there are $2^{m\cdot e_i}$ such cosets $\mathcal{C}_i$. It holds that
  \begin{equation*}
    \begin{split}
      (k+1)^n & \leq \bigcup_{\mathcal{C}_0, \mathcal{C}_1, \ldots, \mathcal{C}_{k-1}} \left| \mathcal{C}_{\Romannum{1}}\left( \mathcal{C}_0, \ldots, \mathcal{C}_{k-1} \right)\right| \\
      & \leq \sum_{\mathcal{C}_0}\sum_{\mathcal{C}_1} \cdots \sum_{\mathcal{C}_{k-1}} \left| \mathcal{C}_{\Romannum{1}}\left( \mathcal{C}_0, \ldots, \mathcal{C}_{k-1} \right)\right| \\
      & \leq \max_{\mathcal{C}_0, \mathcal{C}_1, \ldots, \mathcal{C}_{k-1}} \left| \mathcal{C}_{\Romannum{1}}\left( \mathcal{C}_0, \ldots, \mathcal{C}_{k-1} \right)\right| \sum_{\mathcal{C}_0}\sum_{\mathcal{C}_1} \cdots \sum_{\mathcal{C}_{k-1}} 1 \\
      & = \max_{\mathcal{C}_0, \mathcal{C}_1, \ldots, \mathcal{C}_{k-1}} \left| \mathcal{C}_{\Romannum{1}}\left( \mathcal{C}_0, \ldots, \mathcal{C}_{k-1} \right)\right| \cdot 2^{m \sum_{i=0}^{k-1} e_i}.
    \end{split}
  \end{equation*}
  Therefore, there exist at least one tuple of cosets $\mathcal{C}_0, \mathcal{C}_1, \ldots, \mathcal{C}_{k-1}$ such that
  \begin{equation*}
    \left| \mathcal{C}_{\Romannum{1}}\left( \mathcal{C}_0, \ldots, \mathcal{C}_{k-1} \right)\right| \geq \frac{(k+1)^n}{2^{m \sum_{i=0}^{k-1} e_i}}= \frac{(k+1)^n}{2^{\lceil \log_2(n+1) \rceil \sum_{i=0}^{k-1} e_i}},
  \end{equation*}
  and hence
  \begin{equation*}
    \mathcal{S}_{k}\left(n; (e_0, e_1, \ldots, e_{k-1})\right) \geq \left| \mathcal{C}_{\Romannum{1}}\left( \mathcal{C}_0, \ldots, \mathcal{C}_{k-1} \right)\right| \geq \frac{(k+1)^n}{2 ^ { \lceil \log_2(n+1) \rceil \cdot \sum_{i=0}^{k-1}e_i}}.
  \end{equation*}
\end{IEEEproof}


\section{Lemmas and Proofs for Section~\ref{sec:deletions}} \label{appendix:deletions}

\valueOfV*

\begin{IEEEproof}
  We begin by proving the proposition for a specific choice of $\bm{y}_0$, namely the sequence $\bm{y}_0 = 0^{n-w-1}1^w$, which facilitates the understanding of the construction.
  We then generalize the argument to all binary sequences $\bm{y}_0 \in \{0,1\}^{n-1}$ of Hamming weight $w$.
  Assume $\bm{y}_0 = 0^{n-w-1}1^w.$
  It suffices to consider insertions of the bit $0$ into $\bm{y}_0$, since each such insertion determines a unique $\bm{s}_0 \in \mathcal{I}_1(\bm{y}_0)$, 
  and inserting a $1$ at the same position would yield the same $\bm{s}_1$ in the final comparison $\bm{s}_0 \leq \bm{s}_1$.
  There are two types of positions into which we can insert the $0$.
  \begin{itemize}
    \item \textbf{Insertion at the beginning:} Inserting a $0$ at the start of $\bm{y}_0$ yields $\bm{s}_0 = 0^{n-w}1^w$.
    In this case, any binary sequence $\bm{s}_1 \in \{0,1\}^n$ satisfying $\bm{s}_1 \geq \bm{s}_0$ must have the form $\bm{s}_1 = \bm{a}1^w$, where $\bm{a} \in \{0,1\}^{n-w}$.
    There are $2^{n-w}$ such sequences $\bm{s}_1$.
    \item \textbf{Insertion after a 1:} The remaining $w$ possible insertions place the $0$ immediately after one of the $w$ ones in $\bm{y}_0$, producing sequences of the form $\bm{s}_0 = 0^{n-w-1}1^{i}0 1^{w-i}, \quad 1 \leq i \leq w$.
    These $w$ resulting sequences $\bm{s}_0$ differ in their final $w+1$ bits, and thus are distinct. They also differ from the tail of the sequences in the first case.
    In each case, to satisfy $\bm{s}_0 \leq \bm{s}_1$, we may choose any binary sequence $\bm{s}_1 \in \{0,1\}^n$ such that the final $w+1$ bits agree with $\bm{s}_0$ and the first $n-w-1$ bits of $\bm{s}_1$ are greater than or equal to the corresponding bits of $\bm{s}_0$, which are all zero.
    Thus, for each such $\bm{s}_0$, we have $2^{n-w-1}$ valid sequences $\bm{s}_1$.
  \end{itemize}
  Since the tails of all the counted sequences $\bm{s}_1$ are distinct across the two cases, we may add their contributions to obtain the total number of valid sequences $\bm{s}_1$, given by
  \begin{equation*}
    \mathcal{V}(n; w) = 2^{n-w} + w \cdot 2^{n-w-1}.
  \end{equation*}
  Now, we generalize the argument to any binary sequence $\bm{y}_0 \in \{0, 1\}^{n-1}$ of Hamming weight $w$.
  As before, we only consider insertions of the bit $0$. 
  There are again two types of positions into which the $0$ can be inserted.
  \begin{itemize}
    \item \textbf{Insertion at the beginning:} Inserting a $0$ at the beginning of $\bm{y}_0$ results in $\bm{s}_0 = 0\bm{y}_0$.
    Since $\bm{y}_0$ has $w$ ones, the number of binary sequences $\bm{s}_1 \in \{0,1\}^n$ satisfying $\bm{s}_1 \geq \bm{s}_0$ is $2^{n-w}$, 
    as we are free to flip the $n - w$ zero entries in $\bm{s}_0$ (including the inserted $0$) to either $0$ or $1$.

    \item \textbf{Insertion after a 1:} The remaining $w$ possible insertions place the $0$ immediately after one of the $w$ ones in $\bm{y}_0$.
    Let $\bm{s}_0$ denote such a sequence, obtained by inserting a $0$ at position $i$. 
    For each such $\bm{s}_0$, the number of sequences $\bm{s}_1 \in \{0,1\}^n$ satisfying $\bm{s}_1 \geq \bm{s}_0$ is $2^{n-w-1}$, since we may flip the $n - w - 1$ zero entries in $\bm{s}_0$ that were present in $\bm{y}_0$ (excluding the inserted $0$).
    These resulting sequences $\bm{s}_1$ are distinct across different insertions because each inserted $0$ immediately follows a $1$, making it the only case where position $i$ in $\bm{s}_1$ remains a $0$; in all other cases, that position would be a $1$.
  \end{itemize}
  Since all the sequences $\bm{s}_1$ counted in both cases are distinct, we may sum the contributions to obtain the total number of valid sequences $\bm{s}_1$, given by
  \begin{equation*}
    \mathcal{V}(n; w) = 2^{n-w} + w \cdot 2^{n-w-1}.
  \end{equation*}
\end{IEEEproof}


\vertexSetSize*

\begin{IEEEproof}
  We iterate over all the Hamming weights $w$ of the binary sequences $\bm{y}_0 \in \{0, 1\}^{n-1}$ and use the result from \cref{prop:value-v}.
  Note that the number of binary sequences $\bm{y}_0 \in \{0, 1\}^{n-1}$ with Hamming weight $w$ is $\binom{n-1}{w}$.
  Each such sequence contributes $\mathcal{V}(n; w)$ vertices to the vertex set.
  \begin{equation*}
    \begin{split}
      | \mathcal{X}_{(1, 0)} |  & = \sum_{w=0}^{n-1} \binom{n-1}{w} \mathcal{V}(n; w)
      \overset{(\ref{prop:value-v})}{=} \sum_{w=0}^{n-1} \binom{n-1}{w} \left( 2^{n-w} + w \cdot 2^{n-w-1} \right) \\
      & = 2^n \cdot \sum_{w=0}^{n-1} \binom{n-1}{w} \left(\frac{1}{2}\right)^w + 2^n \sum_{w=0}^{n-1} w \binom{n-1}{w} \left(\frac{1}{2}\right)^{w+1} \\
      & \overset{\text{(BI)}}{=} 2^n \left( \frac{3}{2} \right)^{n-1} + 2^{n-2}  \sum_{w=0}^{n-1} w \binom{n-1}{w} \left(\frac{1}{2}\right)^{w-1} \\
      & \overset{\text{(BI)}}{=} 2 \cdot 3^{n-1} + 2^{n-2} \cdot (n-1) \cdot \left( \frac{3}{2} \right) ^{n-2} \\
      & = 2 \cdot 3^{n-1} + (n-1) \cdot 3^{n-2},
    \end{split}
  \end{equation*}
  where $\overset{\text{(BI)}}{=}$ indicates an application of the binomial identities listed in Appendix~\ref{appendix:binom-identities}.
\end{IEEEproof}


\numBinarySequences*

\begin{IEEEproof}
  If the binary sequence has a single run, that is, if $\rho = 1$, then the Hamming weight $w$ must be either $0$ or $n$, corresponding to the all-zero or the all-one sequence, respectively.
  Now consider the case $\rho \geq 2$. Let $\rho_0$ and $\rho_1$ denote the number of zero and one runs, respectively. Then $\rho = \rho_0 + \rho_1$, with $\rho_0, \rho_1 \geq 1$.
  To construct a sequence as in the lemma, we must partition the $w$ ones into $\rho_1$ non-empty groups, and $n-w$ zeros into $\rho_0$ non-empty groups.
  Using the standard stars and bars technique, there are $\binom{w-1}{\rho_1 -1} \cdot \binom{n-w-1}{\rho_0 - 1}$ ways to do so.
  Finally, note that if the binary sequence begins with a $1$ then $\rho_1 = \lceil\frac{\rho}{2}\rceil$ and $\rho_0 = \lfloor\frac{\rho}{2}\rfloor$. 
  Conversely, if it begins with a $0$, then $\rho_1 = \lfloor\frac{\rho}{2}\rfloor$ and $\rho_0 = \lceil\frac{\rho}{2}\rceil$. The result follows.
\end{IEEEproof}


\begin{lemma}\label{lemma:deletion-runs}
  For any length $n$ and Hamming weight $w$, it holds that
  \begin{equation*}
  \sum_{\rho=2} \rho \cdot \mathcal{N}(n; \rho; w) = \binom{n}{w} + 2 (n-1) \binom{n-2}{w-1}.
  \end{equation*}
\end{lemma}

\begin{IEEEproof}
The left-hand side counts the total number of runs across all binary sequences of length $n$ and Hamming weight $w$ that have at least two runs.
We provide an alternative combinatorial computation.
Every binary sequence of length $n$ and weight $w$ has at least one run, contributing $\binom{n}{w}$ runs.
Additional runs occur at positions $i = 2, \dots, n$ whenever the bit at position $i$ differs from the bit at position $i-1$.
Each such difference creates exactly one additional run in the corresponding sequence.
Consider the transition at positions $i-1$ and $i$. If it is $0 \to 1$, then position $i$ is a one and the remaining $w-1$ ones can be distributed among the other $n-2$ positions, giving $\binom{n-2}{w-1}$ sequences. 
By symmetry, the transition $1 \to 0$ contributes the same. Summing over all $n-1$ positions gives $2 (n-1) \binom{n-2}{w-1}$ additional runs.
Adding the first run per sequence, the total number of runs is
\begin{equation*}
  \binom{n}{w} + 2 (n-1) \binom{n-2}{w-1},
\end{equation*}
which proves the lemma.
\end{IEEEproof}


\deletionAverageBallSize*

\begin{IEEEproof}
  We start by computing $|\bar{\Delta}_{(1, 0)}^{\mathsf{D}}|$. It holds that
  \begin{equation*}
      \bar{\Delta}_{(1, 0)}^{\mathsf{D}} = \frac{1}{|\mathcal{X}_2^n|} \sum_{\bm{s} \in \mathcal{X}_2^n} |\mathcal{B}_{(1, 0)}^{\mathsf{D}}(\bm{s})| = \left(\frac{1}{3}\right)^n \sum_{\bm{s} \in \mathcal{X}_2^n} \rho(\bm{s}_0).
  \end{equation*}
  We want to iterate over the number of runs $\rho = \rho(\bm{s}_0)$ instead of the composite binary sequence $\bm{s}$.
  Remember that the number of binary sequences $\bm{s}_0$ of length $n$ with $\rho$ runs and Hamming weight $w$ is given by $\mathcal{N}(n; \rho; w)$.
  For each such sequence $\bm{s}_0$, there exist $2^{n-w}$ corresponding binary sequences $\bm{s}_1$ such that $\mathcal{R}(\bm{s}_0, \bm{s}_1)$ defines a unique composite binary sequence $\bm{s}$.
  Therefore, by using the result of \cref{lemma:deletion-runs} and the binomial identities in Appendix~\ref{appendix:binom-identities} (marked as $\overset{\text{(BI)}}{=}$),
  \begin{equation*}
    \begin{split}
      \bar{\Delta}_{(1, 0)}^{\mathsf{D}} &=  \left(\frac{1}{3}\right)^n \sum_{\bm{s} \in \mathcal{X}_2^n} \rho(\bm{s}_0) = \left(\frac{1}{3}\right)^n \sum_{w=0}^{n} \sum_{\rho=1} \rho \cdot \mathcal{N}(n; \rho; w) \cdot 2^{n-w} \\
      & = \left(\frac{1}{3}\right)^n \left( \mathcal{N}(n; 1; 0) \cdot 2^n +  \mathcal{N}(n; 1; n) +  \sum_{w=1}^{n-1} \sum_{\rho=2} \rho \cdot \mathcal{N}(n; \rho; w) \cdot 2^{n-w} \right) \\
      & = \left(\frac{1}{3}\right)^n \left( 2^n + 1  + 2^n \sum_{w=1}^{n-1} \left(\frac{1}{2}\right)^w \sum_{\rho=2} \rho \cdot \mathcal{N}(n; \rho; w) \right) \\
      & \overset{(\ref{lemma:deletion-runs})}{=} \left(\frac{1}{3}\right)^n \left( 2^n + 1 + 2^n \sum_{w=1}^{n-1} \left(\frac{1}{2}\right)^w \left( \binom{n}{w} + 2 (n-1) \binom{n-2}{w-1} \right) \right) \\
      & = \left(\frac{1}{3}\right)^n \left( 2^n + 1 + 2^n \left( \sum_{w=1}^{n-1} \left(\frac{1}{2}\right)^w \binom{n}{w} +  \sum_{w=1}^{n-1} \left(\frac{1}{2}\right)^{w-1} (n-1) \binom{n-2}{w-1} \right) \right) \\
      & \overset{\text{(BI)}}{=} \left(\frac{1}{3}\right)^n \left( 2^n + 1 + 2^n \left( \left(\frac{3}{2}\right)^n - 1 - 2^n  + (n-1) \left(\frac{3}{2}\right)^{n-2} \right) \right) \\
      & = \left(\frac{1}{3}\right)^n \left( 2^n + 1 + 3^n - 2^n - 1  + \frac{4}{9}(n-1) 3^n \right) = 1 + \frac{4}{9} (n-1).
    \end{split}
  \end{equation*}
  In order to compute $\bar{\Delta}_1^{\mathsf{D}}$, we again leverage the symmetry of the problem.
  \begin{equation*}
      \bar{\Delta}_1^{\mathsf{D}} = \frac{1}{|\mathcal{X}_2^n|} \sum_{\bm{s} \in \mathcal{X}_2^n} |\mathcal{B}_1^{\mathsf{D}}(\bm{s})| = \left(\frac{1}{3}\right)^n \sum_{\bm{s} \in \mathcal{X}_2^n} \rho(\bm{s}_0) + \rho(\bm{s}_1) = \left(\frac{1}{3}\right)^n \sum_{\bm{s} \in \mathcal{X}_2^n} \rho(\bm{s}_0) + \left(\frac{1}{3}\right)^n \sum_{\bm{s} \in \mathcal{X}_2^n} \rho(\bm{s}_1).
  \end{equation*}
  The first term is equal to $\bar{\Delta}_{(1, 0)}^{\mathsf{D}}$, which we have already computed. The second term is identical due to symmetry and the fact that $\mathcal{N}(n; \rho; w) = \mathcal{N}(n; \rho; n-w)$.
  Therefore $\bar{\Delta}_1^{\mathsf{D}} = 2 \cdot \bar{\Delta}_{(1, 0)}^{\mathsf{D}} = 2 + \frac{8}{9} (n-1)$.
\end{IEEEproof}


\section{General Identities} \label{appendix:binom-identities}

The following identities are used extensively throughout the paper.
The first is the binomial theorem, and the others are derived from it via differentiation or integration with respect to $x$.
  \begin{equation*}
    \begin{split}
       (1+x)^n &= \sum_{i=0}^n \binom{n}{i} x^i \\
       n(1+x)^{n-1} &= \sum_{i=0}^n i \binom{n}{i} x^{i-1} \\
       (n^2-n)(1+x)^{n-2} &= \sum_{i=0}^n (i^2-i) \binom{n}{i} x^{i-2}  \\
       \frac{(1+x)^{n+1} -1}{n+1} & = \sum_{i=0}^n \binom{n}{i} \frac{x^{i+1}}{i+1} \\
       \frac{(1+x)^{n+2} - (n+2)x - 1}{(n+1)(n+2)} & = \sum_{i=0}^n \binom{n}{i} \frac{x^{i+2}}{(i+1)(i+2)} \\
    \end{split}.
  \end{equation*}


\section*{Acknowledgment}

The authors thank M. Somoza for helpful discussions and the validation of the proposed model in this work by photolithographic DNA synthesis.

\ifCLASSOPTIONcaptionsoff
  \newpage
\fi



%




\end{document}